\documentclass[12pt]{article}
\usepackage{amsmath}
\usepackage{graphicx,psfrag,epsf}
\usepackage{enumerate}

\usepackage{url} % not crucial - just used below for the URL 
\renewcommand{\baselinestretch}{1.40}

\usepackage[hmargin=1.60cm,vmargin=1.60cm]{geometry}

\usepackage{amssymb}
\usepackage{amsthm} 
\usepackage{array}
\usepackage{booktabs}
\usepackage[format=plain,justification=justified,font=footnotesize]{caption,subfig}
\usepackage{enumitem}
\usepackage{mathrsfs}
\usepackage{setspace}
\usepackage{xcolor}
\usepackage{eqnarray}
\usepackage{physics}
\usepackage{chngcntr}

\usepackage{float}

\usepackage[longnamesfirst]{natbib}

\usepackage[capitalise,nameinlink,noabbrev]{cleveref}

\newtheorem{theorem}{Theorem}
\newtheorem{proposition}{Proposition}

\newtheorem{lemma}{Lemma} 
\newtheorem{assumption}{Assumption}
\newtheorem{remark}{Remark}

\newcommand{\E}{\mathbb{E}}
\setlist[description]{leftmargin=\parindent,labelindent=\parindent}
\newcommand{\indep}{\raisebox{0.05em}{\rotatebox[origin=c]{90}{$\models$}}}

\def \E {\mathbb{E}}

\def\be{\begin{equation}}
	\def\ee{\end{equation}}
\def\bea{\begin{eqnarray}}
	\def\eea{\end{eqnarray}}
\def\beAA{\begin{align}}
	\def\eeAA{\end{align}}

\begin{document}

	\def\spacingset#1{\renewcommand{\baselinestretch}%
		{#1}\small\normalsize} \spacingset{1}

	%%%%%%%%%%%%%%%%%%%%%%%%%%%%%%%%%%%%%%%%%%%%%%%%%%%%%%%%%%%%%%%%%%%%%%%%%%%%%% 
	
	\title{\bf Semiparametric Single-Index Estimation for Average Treatment Effects\thanks{We are grateful to the Editor: Esfandiar Maasoumi, Yuya Sasaki, two anonymous referees, Qi Li, Bin Peng and Yundong Tu and the seminar participants in International Association for Applied Econometrics Annual Conferences, Econometric Society Australasia Meeting, Econometric Society Asian Meeting, and Monash University for valuable comments. This paper is based on the second chapter of the first author's doctoral dissertation at Monash University. Huang gratefully acknowledges financial support from the National Natural Science Foundation of China under Grant Numbers: T2293771. Gao gratefully acknowledges financial support from the Australian Government through Australian Research Council under Grant Number: DP200102769. Oka gratefully acknowledges financial support from the Australian Government through the Australian Research Council's Discovery Projects under Grant Number: DP190101152. Any mistakes, errors, or misinterpretations are our alone.}}
	{\small\author{
		{\sc Difang Huang$^{\dagger}$} and {\sc Jiti Gao$^{\ddagger}$} and {\sc Tatsushi Oka$^{\mathsection}$}\\
		$^{\dagger}$Academy of Mathematics and Systems Science, Chinese Academy of Sciences \\
		$^{\ddagger}$Department of Econometrics and Business Statistics, Monash University\\
		$^{\mathsection}$Keio University, Japan
	}}
	
	\date{}
	\maketitle
	
	\thispagestyle{empty}
	
	\begin{abstract} 
		%We propose a semiparametric method to estimate the average treatment effect under the assumption of unconfoundedness given observational data. Our estimation method alleviates misspecification issues of the propensity score function by estimating the single-index link function involved through Hermite polynomials.  Our approach is computationally tractable and allows for moderately large dimension covariates. We provide the large sample properties of the estimator and show its validity. Also, the average treatment effect estimator achieves the parametric rate and asymptotic normality. Our extensive Monte Carlo study shows that the proposed estimator is valid in finite samples. We also provide an empirical analysis on the effect of maternal smoking on babies' birth weight and the effect of job training program on future earnings.
		We propose a semiparametric method to estimate the average treatment effect under the assumption of unconfoundedness given observational data. Our estimation method alleviates misspecification issues of the propensity score function by estimating the single-index link function involved through Hermite polynomials. Our approach is computationally tractable and allows for moderately large dimension covariates. We provide the large sample properties of the estimator and show its validity. Also, the average treatment effect estimator achieves the parametric rate and asymptotic normality. Our extensive Monte Carlo study shows that the proposed estimator is valid in finite samples. Applying our method to maternal smoking and infant health, we find that conventional estimates of smoking's impact on birth weight may be biased due to propensity score misspecification, and our analysis of job training programs reveals earnings effects that are more precisely estimated than in prior work. These applications demonstrate how addressing model misspecification can substantively affect our understanding of key policy-relevant treatment effects.
	\end{abstract}
	
	\noindent%
	\vfill

	\clearpage
	\pagenumbering{arabic}
	\spacingset{1.45} % DON'T change the spacing!
	
	% =============================

	\section{Introduction}
	
	The literature on program evaluation has attracted significant attention and provided essential tools for empirical studies in social science. A frequently used measure for program evaluation is the average treatment effect (ATE), often estimated with restrictive parametric conditions on either the propensity score function or the outcome equation. However, when these parametric assumptions are violated, the estimator may suffer from non-negligible finite-sample bias.

	There are various approaches for estimating the average treatment effects under the assumption of unconfoundedness \citep[see][for an overview]{Abadie2018}. Doubly robust estimators, including the inverse probability weighting estimator augmented with additional terms \citep[e.g.][]{robins1994a, robins1995a} and the regression imputation estimator with the propensity score as an additional regressor \citep[e.g.][]{robins1992estimating,scharfstein1999adjusting}, have been widely used due to their robustness against model misspecifications. The estimators are valid as long as the specification of either the propensity score function or the outcome equation is correct. For example, \cite{bang2005doubly}, \cite{robins2000a}, and \cite{rotnitzky2012improved} combine inverse propensity score weighting and matching methods with imputation and projection to estimate treatment effects, while the efficiency of these estimators depends on the choice of tuning parameters and computational implementation. \cite{Tan2006,Tan2010} propose a nonparametric doubly robust estimation approach and show that the estimators are efficient if the parametric propensity score model is correctly specified. However, \cite{kang2007a} and \cite{vansteelandt2012on} show that the doubly robust estimation approach suffers from non-trivial finite sample bias when one of the two models is misspecified, with the bias being large when both models are misspecified.
	
	To address concerns about bias reduction, a fast-growing literature focuses on improving the robustness of doubly robust estimators. \cite{Wang2010} consider a nonparametric method to estimate the propensity score and treatment effects. However, this approach is not feasible for moderately large dimensional covariates. \cite{Vermeulen2015,Vermeulen2016} propose a novel data-driven method to reduce estimation bias under potential misspecification of both models. \cite{farrell2015a} provides an inference approach on average treatment effects that is robust to model selection errors. \cite{sloczynski2017a} propose a general approach for the doubly robust estimators of average treatment effects under unconfoundedness. \cite{Chernozhukov2018} provide a general method to estimate unknown functions in the nonparametric influence function and apply it to debias the estimators of average treatment effects that are feasible in high-dimensional datasets.

	In this paper, we propose a semiparametric estimation method for the ATE to address potential model misspecification concerns. Our approach imposes a single-index structure on the propensity score function while allowing for a flexible link function approximated through Hermite polynomials. We simultaneously estimate the index parameter and the link function in the infinitely dimensional function space. Our method is computationally feasible and applicable to models with a large class of error distributions. We establish the model identification under specific regularity conditions, which are incorporated in our estimation procedure, and demonstrate that the proposed average treatment effect estimator achieves the parametric convergence rate and asymptotic normality. Through an extension of the propensity score residual approach in \cite{lee2017simple}, we propose an estimator that remains valid even when the propensity score approaches zero or one.
	
	This paper contributes to the existing literature on average treatment effect by introducing a novel estimator that relaxes assumptions on both propensity score function and outcome equation specification. Our methodology builds upon the work of \cite{Dong2018}. In contrast to their least squares approach, we employ maximum likelihood estimation for the propensity score, which yields more efficient estimation of the propensity score function, although estimation results can be obtained using the delta method based on their least squares approach. Our estimator is computationally efficient and readily implementable in empirical research. Moreover, it provides an explicit asymptotic variance form, eliminating the need for bootstrap methods. Additionally, our estimator demonstrates robustness in datasets with limited overlap in covariate distributions and remains stable in the presence of extreme observations. Through the incorporation of a flexible assumption regarding the single index structure of the propensity score function, our estimator exhibits superior performance with numerous covariates compared to nonparametric approaches.
	
	This paper further connects to the literature on binary outcome model estimation. Building on semiparametric single-index models, \cite{Ardakani2018} offer an innovative approach to estimating average treatment effects. \cite{Wang2010} employ a nonparametric kernel method to approximate the propensity score function; however, their estimation results deteriorate with increasing covariates and become computationally intractable for moderately large dimensional covariates due to the curse of dimensionality. Semiparametric approaches with single-index structures can maintain flexible specifications while circumventing the curse of dimensionality. \cite{sun2021estimation} introduce a semiparametric single-index model for the propensity score and utilize the kernel method to estimate its functional form semiparametrically. \cite{Liu2018} establish a relationship between the treatment indicator and the low-dimensional linear structure of the covariates, estimating the propensity score function with a nonparametric link function. Both studies impose boundedness assumptions on the link function and function support, necessitating additional constraints on asymptotic theory and numerical performance.
	
	Through relaxing the parametric assumption on the propensity score model, we estimate the conditional probability utilizing an orthogonal series-based estimation method. In comparison with the approaches of \cite{Liu2018} and \cite{sun2021estimation}, our estimation methodology eliminates the need for both the boundedness of the regression function's support and the boundedness of the regression function itself \citep[see][]{Chen2007,dong2016estimation,dong2020a,LiQi2007}. The method is computationally tractable and accommodates moderately large dimensional covariates. Our propensity score estimator achieves a super convergence rate $O_{p}(1/N)$ along the direction of the true parameter while maintaining a standard convergence rate $O_{p}(1/\sqrt{N})$ along all other directions, where $N$ denotes the sample size. The methodology can be further extended by assuming the conditional probability depends on the covariates vector through several linear combinations \citep[see][among others]{koenker2009parametric,li2016flexible,Ma2012,Ma2013,RACINE2004}.
	
	The rest of this paper is organized as follows. Section \ref{section:identification} introduces treatment effect parameters of interest, and Section \ref{section:estimation} explains our semiparametric estimation method. We provide the large sample properties of the estimators in Section \ref{section:asymptotic} and simulation results in Section \ref{section:simulation}. We present the empirical results regarding the effect of maternal smoking on babies' birth weight and the effect of job training programs on future earnings in Section \ref{section:empirical} and conclude in Section \ref{section:conclusion}. Appendix A presents the proof of the main theorem. Appendix B provides detailed sieve expansions. Appendix C contains supplementary theoretical proofs. Appendices D and E present additional simulation results and empirical findings, respectively.

	\section{Model and Identification}\label{section:identification}

	\subsection{Setup}

	%% setup
	We consider the setup of the binary treatment and 
	adopt the potential outcome framework proposed by \cite{rubin1974a}.
	We use $Y(0)$ and $Y(1)$
	to denote the potential outcome without and with the treatment, respectively.
	Also, $D$ is the treatment indicator taking 1
	if an individual receives the treatment and 0 otherwise.
	For each unit, we observe either $Y(0)$ or $Y(1)$, and the observed outcome is
	given by:
	\begin{eqnarray*}
		\label{eq:outcome}
		Y := D Y(1) + (1-D) Y(0).  
	\end{eqnarray*}
	Suppose that we observe a vector of covariates $X$ with the support of $\mathcal{X} \subset \mathbb{R}^{d}$
	and 
	define
	the propensity score function
	$\pi(X) := \Pr(D=1|X)$.  The covariate vector $X$ does not include a constant term in its specification. The framework accommodates both continuous and discrete or categorical variables as components of $X$, and notably, we do not impose any specific distributional assumptions on the covariate structure.

	%% treatment effect 
	As a measure of treatment effects, we consider the ATE, defined as:
	\begin{eqnarray*}
		\Delta^{ATE} := \E[Y(1) - Y(0)]  .
	\end{eqnarray*}
	
	Alternatively, \cite{Angrist1998ECMA} discusses the variance weighted ATE, defined as:
	\begin{eqnarray*}
		\Delta_{\omega}^{ATE} :=
		\E
		\big [
		\omega(X)
		\E
		\big (
		Y(1) - Y(0) | X
		\big )
		\big ],  
	\end{eqnarray*}
	where
	$\omega(X) := \operatorname{var}(D | X) / \E[\operatorname{var}(D | X)]$.
	Because 
	$\operatorname{var}(D | X) = \pi(X) \big (1 - \pi(X) \big )$, 
	the weight
	$\omega(X)$
	assigns higher weights to a subpopulation if its propensity score is closer to $\frac{1}{2}$.

	The propensity score function can be used to obtain consistent estimators for the average treatment effect $\Delta^{ATE}$. \cite{rosenbaum1983a} show that under the assumption of unconfoundedness given covariates, the confounding bias can be removed using the propensity score function. \cite{hahn1998a} proposes the regression imputation estimator that achieves the semiparametric efficiency bound for estimating average treatment effect $\Delta^{ATE}$. \cite{hirano2003a} show that the inverse probability weighting approach can achieve semiparametric efficiency by estimating propensity score in a nonparametric fashion, while this approach may suffer from the curse of dimensionality in empirical analysis.

	\subsection{Identification of the Treatment Effect}
	
	This subsection
	introduces assumptions and
	presents an identification result.
	The following conditions are introduced.
	
	\begin{assumption}\label{assumption:id1}
		$\big ( Y(1), Y(0), X, D \big)$ have a joint distribution
		satisfying:
		\begin{description}
			\item[(a)]
			$\big ( Y(1), Y(0) \big ) \  \indep \ D | X$,
			\item[(b)]
			$0<\pi(X)<1$.
		\end{description}
	\end{assumption}

	\Cref{assumption:id1}(a), referred as the ``unconfoundedness assumption'' by \cite{rosenbaum1983a}, has been widely used in the studies on the treatments effects and program evaluation \citep[e.g.,][]{dehejia1999a,heckman1998a}. The validity of this assumption can be assessed by nonparametric approach \citep[see][for example]{rosenbaum2002a, ichino2008from}.
	\Cref{assumption:id1}(b) ensures that
	the conditional probability of treatment occurrence for both treated and non-treated units is positive, usually imposed in the inverse propensity weighting approach \citep[][]{hirano2003a,firpo2007a}. \cite{rosenbaum1983a} refer to the combination of these two assumptions as ``strongly ignorable treatment assignment''.
	Since
	we can identify the propensity score $\pi(X)$
	given the data,
	we can examine this assumption in practice.

	%% parameter
	For the estimation of weighted ATE, we consider the following regression model:
	\begin{equation}\label{eq: U1}
		Y = \big ( D-\pi(X) \big ) \beta + U.
	\end{equation}
	
	For the estimation of ATE, we use the following weighted regression model:
	\begin{equation}\label{eq: U2}
		\upsilon(X)^{-1/2} Y =
		\upsilon(X)^{-1/2} \big ( D-\pi(X) \big ) \gamma+ \upsilon(X)^{-1/2} U,
	\end{equation}
	where
	$\upsilon(X):= \pi(X)\{1-\pi(X)\}$
	is the variance of the treatment status $D$
	conditional on $X$. In the above equations, $\beta$ and $\gamma$ are scalar parameters, and $U$ denotes the error term. 
	
	The variance-based weight function $\upsilon(X)$ optimally assigns weights based on estimation precision, with $\gamma$ attaining its maximum at $\pi(X) = 0.5$ and decreasing towards 0 as $\pi(X)$ approaches either 0 or 1, thereby ensuring higher weights for more precise subsamples and lower weights for less precise ones.
	
	\begin{lemma}\label{lemma:indentification1}
		Suppose that \Cref{assumption:id1} holds.
		Then, we have
		\begin{enumerate}[label=(\alph*)]
			\item
			Parameter $\beta$ in Equation (\ref{eq: U1})
			is equivalent to the weighted ATE,
			or 
			$\beta = \Delta_{\omega}^{ATE} $.
			\item
			Parameter $\gamma$
			in Equation (\ref{eq: U2})
			is equivalent to the ATE,
			or $\gamma  = \Delta^{ATE}$.
		\end{enumerate}
	\end{lemma}

	\Cref{lemma:indentification1} demonstrates that, under \Cref{assumption:id1}, both the ATE and weighted ATE are identifiable from the data; the formal proof is presented in Appendix C. The ATE represents the unweighted average of individual treatment effects, whereas the weighted ATE employs differential weighting based on two criteria: estimation precision and policy relevance.\footnote{\cite{lee2017simple} and \cite{Li2018JASA} demonstrate that the variance-weighted ATE corresponds to the OLS estimand in the regression of the outcome on the treatment and the vector of covariates.}

	\section{Estimation}\label{section:estimation}
	
	This section proposes the estimator of the treatment effect parameters explained
	in the previous section.
	Suppose that we have a sample of size equals to $N$ and observe $\{(Y_{i},X_{i},D_{i})\}_{i=1}^{N}$,
	which are independent copies of $(Y,X,D)$.
	From
	Lemma~\ref{lemma:indentification1},
	we can estimate 
	$\Delta^{ATE}$ and $\Delta_{\omega}^{ATE}$,
	applying a least square estimation
	for Equations (\ref{eq: U1}) and (\ref{eq: U2}),
	once the propensity score is obtained. 
	Thus, we shall focus on the estimation of the
	propensity score.
	\Cref{subsec:ps}
	presents a semiparametric estimation method for the propensity
	score function
	and
	\Cref{subsec:ate} explains a practical estimation procedure.

	\subsection{Estimation of the Propensity Score}\label{subsec:ps}

	%% single index model model
	To estimate the propensity score function, 
	we consider a semiparametric approach,
	in which the propensity function is assumed to have a single-index structure.\footnote{The single-index structure implies a constant ratio \(\frac{\partial \pi(x) / \partial x_j}{\partial \pi(x) / \partial x_k} = \frac{\theta_{j, 0}}{\theta_{k, 0}}\), when the single index is linear in regressors. While this might appear restrictive for estimating ATE in empirical applications, our Monte Carlo studies in  Appendix D demonstrate the approach's robustness to non-single index data generating processes. Furthermore, it is important to note that the single-index structure accommodates non-linearity in regressors, while maintaining linearity in index parameters.}
	
	Specifically, the propensity score is given by 
	\begin{equation}\label{eq:ps1}
		\pi(X) = \Lambda \big( g_0(X'\theta_{0}) \big),
	\end{equation}
	where
	$\Lambda: \mathbb{R} \to [0,1]$ is a known function. $g_{0}: \mathbb{R} \to \mathbb{R}$ is an unknown link function
	and
	$\theta_0$ is a $d \times 1$ vector of unknown parameters. The mapping $\Lambda(\cdot)$
	ensures that the propensity score function
	remains within the unit interval. We employ the logistic function
	in our analysis, 
	specifically $\Lambda(z)= e^{z}/(1+e^{z})$
	for $z \in \mathbb{R}$. Our theoretical results generalize to alternative functions. 
	
	The existing literature on single-index models typically imposes restrictive assumptions regarding the boundedness of either the support or the link function. For instance, \cite{Ichimura1993} requires compactness on the parameter space and the support of regressors, which consequently restricts the link function to be defined on a compact set. Similarly, \cite{Xia2007} and \cite{Cai2014} stipulate that the link function be bounded, despite being defined on the entire real line. More recent work by \cite{Dong2015} necessitates that the link function be smooth and integrable, thereby implying its boundedness. 
	
	Our methodology advances the literature by accommodating an unbounded link function with unbounded support. By conceptualizing the link function $g_{0}$ as a point within an infinitely dimensional function space, we facilitate the simultaneous estimation of the index parameter and the link function. Importantly, the Hilbert space $\mathcal{L}^{2}$ encompasses a wide range of functions---including polynomials, power functions, and bounded functions on $\mathbb{R}$---that are prevalent in both practical applications and econometric theory.
	
	We assume that the true parameter vector $\theta_{0}$ is an element of the compact parameter space $\Theta \subset \mathbb{R}^{d}$ and impose the normalization constraint $ \|\theta_{0}\|=1$ along with $\theta_{0,1}>0$ for identification purposes. The constraint $\theta_{0,1}>0$ serves to pin down the direction of $\theta_{0}$, where the estimation achieves super-convergence rate $O_P(N^{-1})$ in the direction of $\theta_{0}$, while maintaining root-N convergence rate along all directions orthogonal to $\theta_{0}$. This normalization restriction does not affect the optimization procedure under the identification constraint when deriving the index parameter estimator.

	The model permits general forms of heteroskedasticity in the function form of the propensity score. 
	The semiparametric approach with a single-index constraint can retain flexible specifications while avoiding the curse of dimensionality. 
	Our method relies on the optimization with the constraint of the identification condition to derive the estimator of parameter $\theta_{0}$. 
	Compared with the semiparametric approach using nonparametric kernel method \citep[see][]{Liu2018,sun2021estimation}, our estimator does not impose restrictions on the link function or its support and establishes the asymptotic theory for inference.

	%% nonparametric 
	We consider the case where 
	the nonparametric function $g_{0}(\cdot)$
	resides in
	the Hilbert space $\mathcal{L}^2$
	with
	an inner product
	$\langle g, h \rangle
	:=
	\int g(w) h(w) \exp  (-w^{2} / 2) d w$
	for $g, h \in \mathcal{L}^{2}$.
	The Hilbert space contains all polynomials, all power functions and all bounded, real-valued functions, which are often encountered in both applications and econometric theory \citep[see][for a review]{Chen2007}.\footnote{We explain the properties of Hilbert space used for this paper in  Appendix B.}
	Let
	$\{h_{j}(\cdot) \}_{j=1}^{\infty}$
	be the orthonormal basis
	for the Hilbert space.
	Then, we have the orthogonal series expansion for
	any function $g \in \mathcal{L}^2$:
	\begin{equation}\label{eq:ps3}
		g(w)= \sum_{j=1}^{\infty} c_{j} h_{j}(w),
	\end{equation} 
	where the inner product 
	$c_{j}:=\langle g, h_{j}\rangle$.
	In practice,
	one has to choose a truncation parameter $k$
	to approximate the infinite series, such that:
	\begin{equation}\label{eq:new}
		g(\omega)
		=
		\widetilde{g}_{k}(\omega)+\epsilon_{k}(\omega),
	\end{equation}
	where
	$
	\widetilde{g}_{k}(\omega) :=\sum_{j=1}^{k} c_{j} h_{j}(\omega)
	$
	and 
	$\epsilon_{k}(\omega) :=\sum_{j=k+1}^{\infty} c_{j} h_{j}(\omega)
	$.

	By virtue of \Cref{eq:new}, the $g_0(X'\theta_{0}) $ in \Cref{eq:ps1} can have the following representation:
	\begin{equation}\label{eq:new2}
		g_0(X'\theta_{0})  = g_{0,k}(X'\theta_{0})  + \epsilon_{0,k}(X'\theta_{0})
	\end{equation}
	 
	We have  the maximum likelihood framework
	and define the log-likelihood function as follows:

	For the estimation of the single-index model,
	we consider the maximum likelihood framework
	and define the log-likelihood function as follows: 
	%[Do we have to divide it by N?]
	\begin{eqnarray*}
		\ell_{N}
		\big (\theta, \{c_{j}\}_{j=1}^{k} \big)
		:=
		N^{-1} 
		\sum_{i=1}^{N}\big[
		D_i\cdot
		\ln \Lambda \big (
		\widetilde g_{k}(X_i^{\prime} \theta)
		\big)
		+
		(1-D_{i})
		\cdot
		\ln
		\big \{
		1- \Lambda \big (
		\widetilde g_{k}(X_i^{\prime} \theta)
		\big\}
		\big ].
	\end{eqnarray*}
	The estimators 
	$\hat{\theta}$  
	and
	$\{\hat{c}_{j}\}_{j=1}^{k}$
	of $\theta$ and $\{c_{j}\}_{j=1}^{k}$
	are defined as the solution to the following maximization problem: 
	\begin{eqnarray}\label{eq:ps4}
		\max_{(\theta, \{c_{j}\}_{j=1}^{k}) \in \Theta \times \mathbb{R}^{k}}
		\ell_{N}
		\big (\theta, \{c_{j}\}_{j=1}^{k} \big), \ \ \  \mathrm{s.t.} \ \ \| \theta \| = 1.
	\end{eqnarray}
	%[I am not sure if $c_{j}$ is an element of $\R$] [A: I follow the definition from Jiti, where in their paper, page 303, $\Omega_{k}=\left\{\left(\boldsymbol{\theta}, \boldsymbol{C}_{k}\right): \boldsymbol{\theta} \in \Theta,\left\|\boldsymbol{C}_{k}\right\| \leqslant B_{1}\right\} \subseteq \mathbb{R}^{d+k}$].
	
	Letting
	$\widehat{g}_{k}(\cdot):= \sum_{j=1}^{k} \hat{c}_{j} h_{j}(\cdot)$,
	we have an estimator of the propensity score function:
	\be
	\widehat{\pi}(X) :=
	\Lambda
	\big (
	\widehat{g}_{k}(X^{\prime}\hat{\theta})
	\big ).
	\label{jiti3.1}
	\ee

	%%%%%%%%%%%%

	%% DGP 2019
 
	Our propensity score estimation method is grounded in the work of \cite{Dong2018}, which explores a semi-parametric single-index model wherein the link function is permitted to be unbounded and possess unbounded support. In this model, the link function is treated as a point in an infinitely-dimensional function space, and both the index parameter and link function can be estimated simultaneously through an optimization process, subject to the identification condition constraint for the index parameter.
	
	While the least squares approach in \cite{Dong2018} is prevalent, it may not consistently produce efficient estimates, particularly when the underlying linear regression assumptions are not met. Our maximum likelihood estimation (MLE) approach offers several benefits, including a more versatile estimation framework suitable for various statistical models, encompassing those with non-normal error distributions. This adaptability enables the accommodation of different propensity score model types. Moreover, the MLE method boasts favorable asymptotic properties, resulting in more efficient propensity score function estimation, especially in large samples. Our approach can also be expanded to include prior knowledge through Bayesian estimation techniques, further enhancing efficiency and robustness.

	Our estimator has nice finite sample properties. First, compared with a parametric approach, such as probit and logit, our propensity score estimator is more flexible and less prone to misspecification. Second, the estimator can be used for cases with multiple regressors and avoids the curse of dimensionality commonly encountered in nonparametric methods \citep[e.g.,][]{hahn1998a,hirano2003a}. Finally, we do not impose any distributional assumptions on the covariates by including continuous, discrete, and categorical variables.
	
	\subsection{Estimation Procedure}\label{subsec:ate}

	This subsection provides an estimation procedure for the nonlinear constrained optimization problem in Equation (\ref{eq:ps4}). Given a truncation parameter $k$, we obtain the estimators
	$\hat{\theta}$  
	and
	$\{\hat{c}_{j}\}_{j=1}^{k}$
	as follows: 
	
	\vspace{0.5cm}
	\noindent 
	\textbf{Algorithm 1.} For a fixed truncation parameter $k$, 
	\medskip
	
	(i) Use a linear regression estimator
	as the initial estimator
	{\small $\check \theta
		:=
		(\sum_{i=1}^{N} X_i X_i^{\prime} )^{-1}
		\sum_{i=1}^{N} X_iD_i.$}
	
	(ii) Obtain the estimator $\{\hat{c}_{j}\}_{j=1}^{k}$
	as the solution to the problem:
	{\small $\max_{\{c_{j}\}_{j=1}^{k}} 
		\ell_{N}\big (\check{\theta}, \{c_{j}\}_{j=1}^{k} \big)$.}
	
	(iii) Given the estimator
	$\{\hat{c}_{j}\}_{j=1}^{k}$,
	obtain the estimator $\hat{\theta}$ as the solution to
	the constrained optimization problem:
	\begin{equation*}
		\max_{\theta, \lambda} \
		\ell_{N}\big (\theta, \{\hat{c}_{j}\}_{j=1}^{k} \big)
		+\lambda(\|\theta\|^{2}-1),
	\end{equation*}
	where $\lambda$ is a scalar Lagrange multiplier.
	\medskip
	
	While existing works examine the choice of optimal truncation parameter for non-parametric models \citep[e.g.][]{Gao2002}, there is limited theoretical research on the optimal choice of $k$ for the single-index model. For addressing the choice of the truncation parameter, we propose a leave-one-out cross-validation algorithm. Notably, the truncation error becomes asymptotically negligible under the regularity conditions \citep{Dong2018}. 
	
	\vspace{0.5cm}
	\noindent 
	\textbf{Algorithm 2.}
	Let $\mathcal{K}$ be the set of candidate values for the truncation parameter.
	\medskip
	
	(i) For each $k \in \mathcal{K}$, we apply Algorithm 1 to the dataset $\{(Y_{i},X_{i},D_{i})\}$ excluding the $j$-th observation and obtain the leave-one-out estimate, denoted by $\widehat{Y}_{k, -j}$. 
	
	(ii) The optimal truncation parameter is determined by minimizing the prediction mean of squares: 
	\begin{equation*}
		\min_{k \in \mathcal{K}} \sum_{j=1}^{N}(\widehat{Y}_{k,-j}-Y_{j})^{2}.
	\end{equation*}
	
	To demonstrate the empirical validity of our approach, we present simulation results examining the finite sample performance of the estimation method with various choices of truncation parameter $k$ selected from a set of suitable candidate values in Appendix D.4.

	\section{Asymptotic Properties}
	\label{section:asymptotic}
	
	To establish asymptotic properties for the average treatment effect estimator, we assume the following conditions and provide justifications for the following assumptions.

	\begin{assumption}\label{assumption as:1}
		\begin{enumerate}[label=(\alph*)]	
			\item
			The observations $\{(Y_{i},X_{i},D_{i})\}_{i=1}^{N}$ are
			independent copies of $(Y,X,D)$.
			\item
			Define the population objective function
			$\ell(\theta, g) :=
			\E \big[
			D\cdot
			\ln \Lambda \big(
			g(X^{\prime} \theta)
			\big)
			+
			(1-D)
			\cdot
			\ln
			\big\{
			1- \Lambda \big(
			g(X^{\prime} \theta)
			\big)
			\big\} 
			\big]$, where $g(w)= \sum_{j=1}^{\infty} c_{j} h_{j}(w)$.
			Let 
			$\mathcal{G}$ be a subset of $\mathcal{L}^2$ such that $g_{0}\in \mathcal{G}$.
			\begin{enumerate}[label=(\roman*)]	
				\item 
				All derivatives $g_{0}^{(j)}(w) \in \mathcal{L}^2
				$ for $j=1,\ldots, r$ and $r\geq 1$.
				\item
				$\ell(\theta, g)$ has the unique maximum at $(\theta_{0}, g_{0})$.      
				\item 
				$\sup_{(\theta, g) \in \Theta \times \mathcal{G}
				} \E \big\| \{g^{(1)}(X^{\prime} \theta)\}^{2} X X^{\prime} \big\| \leq M$, 
				for some $M>0$.
				%			\item
				%			$\sup_{(g_1, g_2) \in \mathcal{G}
					%			} \E \big\| g_{2}(\omega) - g_{1}(\omega) \big\|^{2} \leq M_{2}\|g_{1}-g_{2}\|_{\mathcal{L}^2}$, 
				%			for some $M_{2}>0$.
			\end{enumerate}
			\item The truncation parameter $k$
			satisfies that
			$k \to \infty$
			and  $k / \sqrt{N} \rightarrow 0$ as $N \rightarrow \infty$.
		\end{enumerate}
	\end{assumption}

	\Cref{assumption as:1}(b) covers the conditions that are commonly used in the existing literature on sieve estimation \citep[see][for a review]{Chen2007} and binary choice model \citep[e.g.,][]{coppejans2001estimation,cosslett1983distribution,Ichimura1993,Klein1993} as a special case and includes many commonly used distributions such as normal distribution and those possessing compact support. Recall we impose $\|\theta_0\|=1$ for identification purpose. The estimator is derived from an optimization with the constraints for index parameter, and we do not need to impose additional parameter normalization conditions.\footnote{\cite{ma2016semiparametric} develops a semiparametric single index estimator for the propensity score with a one-step maximum likelihood-type estimator that imposes parameter normalizations. Our approach relax such restrictions on parameters including function form $g$ and parameter $\theta$ as long as the identification condition $\|\theta_0\|=1$ with $\theta_{0,1}>0$ holds.} \Cref{assumption as:1}(c) imposes the smoothness restrictions on the link function and the divergence restrictions on the truncation parameter to guarantee the truncation error is negligible.

	\begin{remark}
		Assumption \ref{assumption as:1}.b.(iii) covers some conditions commonly used in the existing literature as special cases.
		\begin{enumerate}
			\item If Assumption 5.3.1 of \cite{Ichimura1993} holds, i.e., $X$ belongs to a compact set, then we can write
			$$
			\begin{aligned}
				\E \big\| \{g^{(1)}(X^{\prime} \theta)\}^{2} X X^{\prime} \big\|  &  \leq O(1) \mathrm{E}\left|g^{(1)}\left(X^{\prime} \theta\right)\right|^2=O(1) \int\left\{g^{(1)}(\omega)\right\}^2 f_\theta(\omega) d\omega \\
				& \leq O(1) \int\left\{g^{(1)}(\omega)\right\}^2 \exp \left(-\omega^2 / 2\right) \cdot \exp \left(\omega^2 / 2\right) f_\theta(\omega) d\omega \\
				& \leq O(1) \int\left\{g^{(1)}(\omega)\right\}^2 \exp \left(-\omega^2 / 2\right) d\omega
			\end{aligned}
			$$
			where $f_\theta(\omega)$ is the same as that defined in Assumption \ref{assumption as:1}.b. Then in this case, Assumption \ref{assumption as:1}.b.(iii) reduces to requiring $g^{(1)}(\omega) \in \mathcal{L}^2$ for all $g \in \mathcal{G}$.
			
			\item If Condition C2 of \cite{Xia2007} holds, i.e., $g^{(1)}(\omega)$ is bounded on $\mathbb{R}$, then we can write
			$$
			\E \big\| \{g^{(1)}(X^{\prime} \theta)\}^{2} X X^{\prime} \big\|  \leq O(1) \E \| X \|^2
			$$
		\end{enumerate}
		Then we need only to bound the second moment of $X$. Since our link function can potentially be an unbounded function defined on the whole real line, we adopt the current form of Assumption \ref{assumption as:1}.b.(iii).
	\end{remark}
	
	%%%

	For theoretical developments, we need to assume additional regularity conditions. Since these conditions are standard in the literature, 
	we state the conditions as \Cref{assumption as:2}.
	
	\begin{assumption}\label{assumption as:2}
		Let $\zeta$ be a relatively small positive number and $M$, $M_{1}$ and $M_{2}$ be positive constants. Suppose that the following conditions hold:
		\begin{enumerate}[label=(\alph*)]
			\item For $\Omega(\zeta)=\big\{(\theta, g):\big\|(\theta, g)-(\theta_{0}, g_{0})\big\|_{2} \leq \zeta\big\}$,
			{\footnotesize
			\begin{enumerate}[label=(\roman*)]
				\item $$\sup _{(\theta, g) \in \Omega(\zeta)} \E\big\|g^{(2)}(X_{i}^{\prime} \theta) X_{i} X_{i}^{\prime}\big\|^{2} \leq M,$$
				\item $$\sup _{(\theta, g) \in \Omega(\zeta)} \Big\|\frac{1}{N} \sum_{i=1}^{N}
				\upsilon\big(g(X_{i}^{\prime} \theta)\big)\big(g^{(1)}(X_{i}^{\prime} \theta)\big)^{2} X_{i} X_{i}^{\prime}
				-
				\E \big[\upsilon\big(g(X_{i}^{\prime} \theta)\big)\big(g^{(1)}(X_{i}^{\prime} \theta)\big)^{2} X_{i} X_{i}^{\prime} \big]
				\Big\|=o_{P}(1),$$
				\item $$\sup _{(\theta_{1}, g_{1}),(\theta_{2}, g_{2}) \in \Omega(\zeta)}
				\Big\|
				\frac{1}{N} \sum_{i=1}^{N} 
				\Lambda\big(g_{1}(X_{i}^{\prime} \theta_{1})\big) g_{2}^{(2)}(X_{i}^{\prime} \theta_{2}) X_{i} X_{i}^{\prime}
				-
				\E \big[
				\Lambda\big(g_{1}(X_{i}^{\prime} \theta_{1})\big) g_{2}^{(2)}(X_{i}^{\prime} \theta_{2}) X_{i} X_{i}^{\prime} \big]
				\Big\|=o_{P}(1).$$
			\end{enumerate}
			}
			\item Let $\Sigma_{1}(\theta)=\E \big[\dot{H}(X_{i}^{\prime} \theta) \dot{H}(X_{i}^{\prime} \theta)^{\prime} \big]$ and $\Sigma_{2}(\theta_{0})=\E \big[\dot{H}(X_{i}^{\prime} \theta_{0}) \dot{H}(X_{i}^{\prime} \theta_{0})^{\prime}\|X_{i}\|^{2} \big]$, where $$\dot{H}(w)= \big(h_{1}^{(1)}(w), \ldots, h_{k}^{(1)}(w)\big)^{\prime}.$$
			Let
			$
			\sup_{\{\theta:\|\theta-\theta_{0}\| \leq \zeta\}} \lambda_{\max}\big(\Sigma_{1}(\theta)\big) \leq M_{1}
			$
			and
			$
			\lambda_{\max}\big(\Sigma_{2}(\theta_{0})\big) \leq M_{2}
			$,
			where $\lambda_{\min}(A)$ and $\lambda_{\max}(A)$ denote the minimum and maximum eigenvalues of a square matrix $A$, respectively.
			\item Suppose that (i) $\E\big[\epsilon_{0, k}^{(1)}(X_{i}^{\prime} \theta_{0})\big]^{4}=o(1)$, where $\epsilon_{0, k}(X^{\prime}\theta_{0})$ is defined in \Cref{eq:new2};  (ii) $N / k^{r} \rightarrow 0$.
		\end{enumerate}
	\end{assumption}

	\Cref{assumption as:2}(a) covers commonly used conditions in the existing literature and establishes the uniform convergence of $(\theta_{0}, g_{0})$ in a small neighborhood. The uniform convergence of $(\theta_{0}, g_{0})$ in a small neighborhood is required by \cite{yu2002penalized}, which can be derived using the results of Lemma A2 in \cite{Newey2003}. For simplicity, we directly impose conditions (ii) and (iii) in \Cref{assumption as:2}(a). \Cref{assumption as:2}(b) matches Assumption 2 of \cite{newey1997convergence} and Assumption 3 of \cite{Su2012}, and the restrictions on the derivatives of orthogonal Hermite functions can be derived using the results in \cite{belloni2015some}. \Cref{assumption as:2}(c) is the under-smoothing condition \citep[see][]{belloni2015some,chang2015high} that ensures the negligibility of the truncation error terms and the asymptotic normality of the estimator.

	\begin{assumption}\label{assumption as:3}
		Let $\epsilon$ be a relatively small positive number. As $(N, k) \rightarrow(\infty, \infty),$ let $\Phi(N, k) \rightarrow 0$ uniformly in $\theta \in\{\theta:\|\theta-\theta_{0}\|<\epsilon\}$, where
		\begin{equation*}
			\Phi(N, k)=\frac{1}{N} \sum_{m=1}^{k} \sum_{n=1}^{k} \int_{\mathbb{R}} h_{m}^{2}(\omega) h_{n}^{2}(\omega) f_{\theta}(\omega) d \omega,
		\end{equation*}
		in which $f_{\theta}(\omega)$ is the pdf of $\omega=X^{\prime} \theta$ as defined in \Cref{assumption as:1}.
	\end{assumption}

	\Cref{assumption as:3} imposes restrictions on the probability distribution functions of the regressors $h_{m}(\omega)$ and $h_{n}(\omega)$ to make the functions reduce to $k^{2d}/N$ as we use Hermite polynomials to decompose the link function $g_0(\omega)$ in the function space $\mathcal{L}^{2}$.
	
	\begin{remark}
		For $\Phi(N, k)$, there are two ways to simplify the notation:
		\begin{enumerate}
			\item Impose a stronger version of Assumption \ref{assumption as:3} as follows:
			\begin{itemize}
				\item Suppose $\exp(\|X\|^2)f(z) \leqslant M$ uniformly on $\mathbb{R}^d$
			\end{itemize}
			
			Then, we are able to further organize $\Phi(N, k)$:
			$$
			\begin{aligned}
				\Phi(N, k)= & \frac{1}{N} \sum_{m=1}^k \sum_{n=1}^k \int_{\mathbb{R}^d}|h_m(\omega)|^2|h_n(\omega)|^2 \exp(-\|X\|^2) \cdot \exp(\|X\|^2)f(\omega)d\omega \\
				& \leqslant O(1) \frac{1}{N} \sum_{m=1}^k \sum_{n=1}^k \int_{\mathbb{R}^d}|h_m(\omega)|^2|h_n(\omega)|^2 \exp(-\|X\|^2)d\omega \\
				& \leqslant O(1) \frac{k}{N} \sum_{m=1}^k \int_{\mathbb{R}^d}|h_m(\omega)|^2 \exp(-\|X\|^2/2)d\omega = O\left(\frac{k^2}{N}\right) = O\left(\frac{k^{2d}}{N}\right)
			\end{aligned}
			$$
			
			\item We can follow Assumption 3.iii of \cite{Su2012}, i.e., assuming $\E|h_m(\omega)|^4$ is uniformly bounded.
		\end{enumerate}
		
		Either restriction above will make $\Phi(N, k)$ reduce to $k^{2d}/N$ under the restrictive condition on the pdf of $X$. Indeed, as the link function $g_0(\omega)$ is potentially non-integrable and unbounded, one may need to impose some restrictions on the pdf of the regressors.
	\end{remark}

	The subsequent lemma establishes the consistency and asymptotic properties of the estimators $\hat{\theta}$. The proof of \Cref{lemma:indentification3,lemma:indentification2} is presented in Appendix C.
	
	\begin{lemma}\label{lemma:indentification3}
		Define the norm $\big\|(\theta, g)\big\|_{2}:=
		\big(\|\theta\|^{2}+\| g \|^2\big)^{1 / 2}$
		over the space $\Theta \times \mathcal{L}^2$. 
		Suppose that \Cref{assumption as:1} holds. Then, we have, as $N \rightarrow \infty$,
		\begin{equation*}
			\big\|(\hat{\theta}, \widehat{g}_{k})-(\theta_{0}, g_{0})\big\|_{2} \rightarrow_{p} 0.
		\end{equation*}
	\end{lemma}
	
	In \Cref{lemma:indentification2} below, 
	we establish the convergence rate and the limit distribution
	of the estimator $\hat{\theta}$.
	To present the result,  
	let
	$P_{\theta_{0}} := I_{d}-\theta_{0}\theta_{0}^{\prime}$
	be a $d \times d$ matrix. 
	Under the constraint of $\|\theta_{0}\|=1$,
	$P_{\theta_{0}}$ represents the projection matrix
	that maps any $d \times 1$ vector
	into the orthogonal complement space of the true parameter $\theta_0$.
	Let 
	$V:=[v_{1}, \dots, v_{d-1}]$
	be a $d \times (d-1)$
	matrix consisting of eigenvectors 
	$v_{j}$
	associated with the eigenvalue of 1.
	\medskip
	
	\begin{lemma}\label{lemma:indentification2}
		Let  Assumptions 2 and 3 hold. Then,
		as $N \rightarrow \infty$,
		\begin{enumerate}[label=(\alph*)]
			\item 
			$ \sqrt{N} V^{\prime}(\hat{\theta}-\theta_{0}) \rightarrow_{D} N\big(0, (V^{\prime} Q V)^{-1}V^{\prime} W V(V^{\prime} Q V)^{-1}\big),
			$ 
			\item 
			$\theta_{0}^{\prime}(\hat{\theta} - \theta_{0}) = O_{p}(1/N)$,
		\end{enumerate}
		where
		$Q:=
		\E
		\big[
		\big\{g_{0}^{(1)}(X^{\prime} \theta_{0})\big\}^{2} X X^{\prime}
		\big]
		$
		and
		$W:=
		\E
		\big[
		\big(D - \pi(X) \big)^{2}
		\big\{g_{0}^{(1)}(X^{\prime} \theta_{0})\big\}^{2} X X^{\prime}
		\big]
		$.
	\end{lemma}
	
	\Cref{lemma:indentification2}(a) demonstrates the asymptotic normality of $V^{\prime}(\hat{\theta}-\theta_{0})$ through the transformation of $\hat{\theta}-\theta_{0}$ into $\mathbb{R}^{d-1}$, which is necessary since both $\hat{\theta}$ and $\theta_{0}$ are constrained to the unit sphere in $\mathbb{R}^{d}$. \Cref{lemma:indentification2}(b), which follows from the constraint $\|\theta_{0}\|=1$, characterizes the convergence behavior along the radial direction. Specifically, it reveals that the convergence rate along the direction of $\theta_{0}$ is faster ($O_p(1/N)$) compared to the directions orthogonal to $\theta_{0}$ ($O_p(1/\sqrt{N})$).

	%%%%%%%%%%%%%%%%%%%%%%%%%%%%%%%%%%%%%%%%%%%%%%%%%%%%%%%%%%%%%%%%%%%%%%%%%%%
	
	We are now ready to establish the large sample properties of the estimators for ATE ($\widehat{\Delta}^{ATE}$) and variance weighted ATE ($\widehat{\Delta}_w^{ATE}$) defined in \Cref{subsec:ate} where the propensity score is estimated based on the series estimation approach in \Cref{subsec:ps}. The proof of \Cref{theorem:ate1} below is given in Appendix A.

	The variance weighted ATE estimator $\widehat{\Delta}_w^{ATE}$ has the form:
	\begin{equation*}
		\widehat{\Delta}_w^{ATE} = \frac{\sum_{i=1}^N Y_i(D_i - \hat{\pi}(X_i))}{\sum_{i=1}^N (D_i - \hat{\pi}(X_i))^2}.
	\end{equation*}
	
	The ATE estimator $\widehat{\Delta}^{ATE}$ is:
	\begin{equation*}
		\widehat{\Delta}^{ATE} = \frac{\sum_{i=1}^N \hat{\upsilon}(X_i)^{-1}Y_i(D_i - \hat{\pi}(X_i))}{\sum_{i=1}^N \hat{\upsilon}(X_i)^{-1}(D_i - \hat{\pi}(X_i))^2}.
	\end{equation*}

	\begin{theorem}\label{theorem:ate1} 
		Suppose that Assumptions 2--4 hold. Then, as $N \rightarrow \infty$,
		\begin{enumerate}[label=(\alph*)]
			\item
			$\sqrt{N} (\widehat{\Delta}_{\omega}^{ATE} - \Delta_{\omega}^{ATE})
			\rightarrow_{D} N(0, \sigma_{\omega}^2)$,
			\item
			$\sqrt{N} (\widehat{\Delta}^{ATE}- \Delta^{ATE})\rightarrow_{D} N(0, \sigma^2)$,
		\end{enumerate}
		where
		$
		\sigma_{\omega}^{2}
		:=
		\E
		\big( 
		\{D - \pi(X)\}U 
		\big)^2
		/
		\big(
		\E
		\{\upsilon(X)\}
		\big )^{2}
		$
		and 
		$
		\sigma^2:=
		\E
		\big(
		\upsilon(X)^{-1}\{ D - \pi(X)\}U
		\big)^2
		$.
	\end{theorem}	
	%%%%%%%%%%%%%%%%%%%%%%%%%%%%%%%%%%%%%%%%%%%%%%%%%%%%%%%%%%%%%%%%%%%%%%%%%%%
	
	%% Advantages 
	Our proposed estimator has several advantages over other estimators including the inverse propensity score weighting \citep[][]{hirano2003a,imbens2010a}, matching \citep[][]{abadie2006a,abadie2011a,abadie2016a}, regression adjustment \citep[][]{lane1982analysis}, and doubly robust estimators \citep[][]{Tan2010,tsiatis2006semiparametric,Wooldridge2007,Liu2018}. 
	First, our estimator is computationally simple and easy to implement. There is no need to specify the propensity score function form. Second, our estimator has a simple explicit form of asymptotic variance that works well even in small samples,  and we do not need to estimate the asymptotic variance through bootstrap procedures. Third, compared with the inverse propensity score approach, our estimation method works well in the datasets that are limited overlap in the covariate distributions and is not influenced by extreme observations in empirical applications.\footnote{\cite{Wooldridge2007} proposes the inverse probability weighted M-estimation under a general missing data scheme and shows the misspecication of the error distribution is negligible under some regularity conditions, including the existence of pseudo-parameters. Our approach relaxes the requirement and our simulation and empirical results, reported in Sections~\ref{section:simulation} and~\ref{section:empirical}, show some advantages of our approach in finite samples.} Fourth, by adopting a flexible assumption on the single index structure of propensity score function, our estimator performs well when there are many covariates.
	
	For statistical inferences,  
	we need to estimate the asymptotic variances
	$\sigma_{\omega}^2$ and $\sigma^2$.
	Following the sample analog principles,
	we can estimate the asymptotic variances 
	by using the sample moments based on
	the observables:
	$\hat{U}_{i}$,
	$D_{i} - \widehat{\pi}(X_{i})$,
	and 
	$\widehat{\upsilon}(X_{i}):= \widehat{\pi}(X_{i})\{1-\widehat{\pi}(X_{i})\}$. 
	We denote
	by
	$\hat{\sigma}^2_{\omega}$
	and 
	$\hat{\sigma}^2$
	the estimator of $\sigma_{\omega}^2$ and $\sigma^2$, respectively.
	The proposition below shows that those estimators are consistent. The proof of \Cref{theorem:ate2} is given in  Appendix C.

	\begin{proposition}\label{theorem:ate2}
		Let Assumptions 2--4 hold. Then, as $N \rightarrow \infty$,
		
		(a) $\hat{\sigma}^2_{\omega} = \sigma_{\omega}^2 + o_{P}(1)$; and (b) $\hat{\sigma}^{2} = \sigma^2  + o_{P}(1)$. 
		
	\end{proposition}

	\section{Monte Carlo Studies}\label{section:simulation}
	
	In this section, we conducted comprehensive Monte Carlo simulations to evaluate our semiparametric estimator proposed in \Cref{subsec:ps} for the propensity score function and semiparametric estimator proposed in \Cref{subsec:ate} for the average treatment effect. We repeat each experiment 10,000 times with sample size $N$ = 400, 800, or 1600 and set the truncation parameter $k$ equal to $\lfloor N^{1 / 5}\rfloor$.\footnote{We provide an algorithm for calculating truncation parameter and our simulation results are valid and robust to a range of truncation parameters, as demonstrated in  Appendix D.4.}
	
	\subsection{Propensity Score Function}
	
	We conducted Monte Carlo simulations to evaluate our semiparametric estimator  proposed in \Cref{subsec:ps} for the propensity score function, with detailed simulation setup and results in Appendix D.1. Our first simulation examined four functional forms with two-dimensional covariates $X \sim N(0, I_{2})$, considering both cases where $\|\theta_{0}\|=1$ holds and where it is violated. As shown in \Cref{table: simulation_ps_ks}, our estimator demonstrates small biases that decrease with sample size, achieving the theoretical $O(1/N)$ convergence rate when $\|\theta_{0}\|=1$ and $O(1/\sqrt{N})$ otherwise. We then extended to six-dimensional covariates $X \sim N(0, I_{6})$. The results in \Cref{table: simulation_ps_hd_ks,table: simulation_ps_mis_hd_ks} confirm our estimation approach remains valid with higher dimensions, maintaining small biases and decreasing RMSEs as sample size increases. Finally, we examined the estimator's performance for average treatment effects under various challenging scenarios, including normalization condition violations, probit specifications, heavy-tailed errors, and propensity score misspecification. Results in \Cref{table: simulation with ate2} demonstrate the robustness of our approach, with small biases and RMSEs maintained across different sample sizes.
	
	We further compare our semiparametric (SP) estimator with existing approaches including covariate balancing (CB) and high-dimensional selection (HD) methods for propensity score estimation, with detailed simulation setup and results in Appendix D.2. Our simulation study considers four challenging DGPs featuring unbounded link functions and unbounded support for propensity score functions, including both single-index and non-single-index structures. As shown in \Cref{table: simulation_ps_compare}, while all three estimators demonstrate near-zero bias, our SP estimator outperforms CB \citep{Anna2021JAE,imai2014covariate,Zubizarreta} and HD \citep{belloni2017a,Chernozhukov2018,SunTan2021} approaches in terms of standard deviation and RMSE, particularly for non-single-index DGPs.

	\subsection{Average Treatment Effect}
	
	We conduct an extensive simulation study for the finite sample properties of our semiparametric estimator proposed in \Cref{subsec:ate} for the average treatment effect.  
	\medskip
	
	\noindent{\bf Setting 5A}: $g_0(X^{\prime}\theta_0) = \sin(X^{\prime}\theta_0)$, where $\theta_0 = (\theta_{0,1}, \theta_{0,2})^{\prime} = (0.8,-0.6)^{\prime}$ and the vector of covariates $X=(X_{1}, X_{2})^{\prime}$ are generated from independent standard normal distributions $N(0,1)$. The propensity score function is defined as $D = \Lambda \big( g_0(X'\theta_{0}) \big)$ and the outcome model is generated as $Y=\beta_d D + X_{1}+ X_{2} + \epsilon$, where $\epsilon \sim N(0,1)$ and $\beta_d = 1$.
	\smallskip
	
	\noindent{\bf Setting 5B}: $g_0(X^{\prime}\theta_0) = 0.5\{(X^{\prime}\theta_0)^3 - (X^{\prime}\theta_0)\}$,  where $\theta_0 = (\theta_{0,1}, \theta_{0,2})^{\prime} = (0.8,-0.6)^{\prime}$ and the vector of covariates $X=(X_{1}, X_{2})^{\prime}$ are generated from independent standard normal distributions $N(0,1)$. The propensity score function is defined as $D = \Lambda \big( g_0(X'\theta_{0}) \big)$ and the outcome model is generated as $Y=\beta_d D + X_{1}+ X_{2} + \epsilon$, where $\epsilon \sim N(0,1)$ and $\beta_d = 1$.
	\smallskip
	
	\noindent{\bf Setting 5C}: $g_0(X^{\prime}\theta_0) = \sin(X^{\prime}\theta_0)$, \\where $\theta_0 = (\theta_{0,1}, \theta_{0,2}, \theta_{0,3}, \theta_{0,4}, \theta_{0,5}, \theta_{0,6})^{\prime} = (\sqrt{0.2},\sqrt{0.3},\sqrt{0.25},-\sqrt{0.1},\sqrt{0.08},-\sqrt{0.07})^{\prime}$ and the vector of covariates $X=(X_{1}, X_{2},X_{3}, X_{4},X_{5}, X_{6})^{\prime}$ are generated from independent standard normal distributions $N(0,1)$. The propensity score function is defined as $D = \Lambda \big( g_0(X'\theta_{0}) \big)$ and the outcome model is generated as $Y=\beta_d D + X_{1}+ X_{2} + X_{3} + X_{4} + X_{5} + X_{6} + \epsilon$, where $\epsilon \sim N(0,1)$ and $\beta_d = 1$.
	\smallskip
	
	\noindent{\bf Setting 5D}: $g_0(X^{\prime}\theta_0) = 0.5\{(X^{\prime}\theta_0)^3 - (X^{\prime}\theta_0)\}$, \\where $\theta_0 = (\theta_{0,1}, \theta_{0,2}, \theta_{0,3}, \theta_{0,4}, \theta_{0,5}, \theta_{0,6})^{\prime} = (\sqrt{0.2},\sqrt{0.3},\sqrt{0.25},-\sqrt{0.1},\sqrt{0.08},-\sqrt{0.07})^{\prime}$ and the vector of covariates $X=(X_{1}, X_{2},X_{3}, X_{4},X_{5}, X_{6})^{\prime}$ are generated from independent standard normal distributions $N(0,1)$. The propensity score function is defined as $D = \Lambda \big( g_0(X'\theta_{0}) \big)$ and the outcome model is generated as $Y=\beta_d D + X_{1}+ X_{2} + X_{3} + X_{4} + X_{5} + X_{6} + \epsilon$, where $\epsilon \sim N(0,1)$ and $\beta_d = 1$.
	
	\medskip
	
	We summarize the simulation results in \Cref{table: simulation with ate}. In all four simulation settings, our estimators perform well according to bias and standard deviation criteria. For all the sample sizes under consideration, the estimators have relative small biases and standard deviations. The finite--sample simulation results support that the asymptotic normal approximation is accurate and the rate of convergence is consistent with the results in \Cref{theorem:ate1}.
	
	{\footnotesize
		
		\begin{table}[h!]
			\centering
			\caption{\label{table: simulation with ate}Simulation Results the Estimators of Average Treatment Effect of Settings 5A to 5D.}
			\begin{tabular}{lccrcccr}
				\toprule
				& Setting &  $N$    & \multicolumn{1}{c}{$\widehat{\beta_d}$} &  & Setting & $N$    & \multicolumn{1}{c}{$\widehat{\beta_d}$} \\ \hline
				Bias &     &      &         &  &     &      &         \\\hline
				& 5A & 400  & 0.0049  &  & 5C & 400  & -0.0054 \\
				&     & 800  & -0.0003 &  &     & 800  & 0.0002 \\
				&     & 1600 & -0.0001 &  &     & 1600 & -0.0002 \\
				& 5B & 400  & 0.0042  &  & 5D & 400  & 0.0028 \\
				&     & 800  & -0.0005 &  &     & 800  & -0.0011 \\
				&     & 1600 & -0.0001 &  &     & 1600 & -0.0003 \\\hline
				Std  &     &      &         &  &     &      &         \\\hline
				& 5A & 400  & 0.0418  &  & 5C & 400  & 0.0403  \\
				&     & 800  & 0.0289  &  &     & 800  & 0.0277  \\
				&     & 1600 & 0.0225  &  &     & 1600 & 0.0180  \\
				& 5B & 400  & 0.0380  &  & 5D & 400  & 0.0366  \\
				&     & 800  & 0.0274  &  &     & 800  & 0.0243  \\
				&     & 1600 & 0.0201  &  &     & 1600 & 0.0167   \\ \bottomrule
			\end{tabular}
		\end{table}
	}

	We compare our semiparametric (SP) estimator with the locally efficient estimator \citep{Liu2018} and nonparametric cross-validation estimator \citep{sun2021estimation} for treatment effects estimation, with detailed simulation setup and results in Appendix D.3. While these alternative approaches assume bounded link functions and support, we examine performance under four challenging DGPs with unbounded properties. As shown in \Cref{table: simulation_ate_compare}, our SP estimator demonstrates smaller bias when boundedness assumptions are violated, and achieves lower standard deviation and RMSE across all sample sizes compared to both alternatives. The results confirm our method's superior performance for data with unbounded link functions and support.

	We investigate the sensitivity of our estimator to different truncation parameters $k$, with detailed simulation setup and results in Appendix D.4. Results in \Cref{table: simulation_ps_ks_different_k,table: simulation_ps2_ks_different_k} demonstrate the performance of propensity score estimation across truncation parameters ranging from $k = 2$ to $k = 6$. Similarly, \Cref{table: simulation with ate_different_k,table: simulation with ate2_different_k} show the average treatment effect estimation results across the same range of truncation parameters. The results confirm that our estimators maintain validity across different choices of truncation parameters.

	\section{Empirical Study}\label{section:empirical}
	In this section, we apply the proposed semiparametric method to analyze two real data examples. We study the effect of maternal smoking on babies' birth weight in the main context and consider the effects of the job training program on future earnings in  Appendix E.\footnote{As shown in \cite{cattaneo2015a}, the assumption that maternal smoking is exogenous to the babies' birth outcome may not hold. We apply our approach for this empirical study as it is widely used in treatment effects studies as a benchmark. We also provide the effects of the job training program on future earnings in  Appendix E as an additional empirical study.} In both cases, we demonstrate the validity of our estimation and inference method for the datasets with potential misspecification of propensity score and datasets with limited overlap.
	
	We apply our semiparametric method to analyze the ATE of maternal smoking on babies' birth weight. Low birth weight may increase infant mortality rates and economic costs, including late entry into kindergarten, repeated grades, and longer-term labor market outcomes \citep[][]{arcidiacono2011a,Bao2021JFQA,permutt1989simultaneous,rosenzweig1991inequality}. 
	
	Although a large body of research confirms the negative effect of maternal smoking on babies' birth weight, there is no consensus on its exact magnitude \citep[][]{abrevaya2001the,abrevaya2006estimating,chernozhukov2011inference,evans1999can}. We use the dataset of low birth weight initially by \cite{Almond2005}. This is a rich database of singletons in Pennsylvania with 4,642 detailed observations of mothers and their infants' birth information.
	
	The outcome variable $Y$ is infant birth weight measured in grams. The binary treatment variable $D$ is the mother's smoking status ($D = 1$ indicates mother is a smoker, $D = 0$ indicates mother is a non-smoker). The covariates $X$ include mother's age, mother's marital status, an indicator variable for alcohol consumption during pregnancy, an indicator for whether there was a previous birth where the newborn died, mother's education, father's education, number of prenatal care visits, mother's race, an indicator of firstborn baby, and months since last birth \citep[see][]{cattaneo2010a}.  
	
	The critical assumption in our application is that the mother's smoking status is independent of the infant's birth weight conditional on all observed demographic variables, implying that maternal smoking may impact the babies' birth weights only through its effect on observed covariates. We show the summary statistics for all variables in Table \ref{table: summary_baby}, which provides the summary statistics for baby birth weight data, where $N$ denotes sample size. For each variable, we report the sample average (Mean) and sample standard deviation (Std). *, **, and *** indicate the significance level at 10\%, 5\%, and 1\%, respectively. As it reveals, the comparison between these two groups shows that the control group is quite different from the treated group.
	
	\begin{table}[h!]
		\centering
		\caption{\label{table: summary_baby}Summary Statistics for Baby Birth Weight data.}
%		\resizebox{\columnwidth}{!}{%
			\begin{tabular}{lccccc}
				\toprule
				& \multicolumn{2}{c}{Non-smoking Group}              & \multicolumn{2}{c}{Smoking Group}                  & \multicolumn{1}{c}{Two-Sample} \\
				& \multicolumn{2}{c}{$N =  3778$}               & \multicolumn{2}{c}{$N =  864$}                & \multicolumn{1}{c}{Difference} \\ \cline{2-6} 
				& \multicolumn{1}{c}{Mean} & \multicolumn{1}{c}{Std} & \multicolumn{1}{c}{Mean} & \multicolumn{1}{c}{Std} & \multicolumn{1}{c}{}           \\ \hline
				Infant birth weight               & 3412.91                  & 570.69                  & 3137.66                  & 560.89                  & 275.3***                       \\
				Previous births with dead   babys & 0.25                     & 0.43                    & 0.32                     & 0.47                    & -0.0724***                     \\
				Mother's age                      & 26.81                    & 5.65                    & 25.17                    & 5.30                    & 1.644***                       \\
				Mother's education                & 12.93                    & 2.53                    & 11.64                    & 2.17                    & 1.291***                       \\
				Father's education                & 12.67                    & 3.48                    & 10.70                    & 4.10                    & 1.970***                       \\
				Number of prenatal care visits    & 10.96                    & 3.52                    & 9.86                     & 4.21                    & 1.101***                       \\
				Months since last birth           & 21.90                    & 31.50                   & 28.22                    & 36.92                   & -6.322***                      \\
				1 if mother married               & 0.75                     & 0.43                    & 0.47                     & 0.50                    & 0.278***                       \\
				1 if alcohol consumed             & 0.02                     & 0.14                    & 0.09                     & 0.29                    & -0.0726***                     \\
				1 if mother is white              & 0.85                     & 0.36                    & 0.81                     & 0.39                    & 0.0388**                       \\
				1 if first baby                   & 0.45                     & 0.50                    & 0.37                     & 0.48                    & 0.0816***                      \\ \bottomrule
			\end{tabular}
%		}
	\end{table}
	
	\begin{table}[h!]
		\centering
		\caption{\label{table: ate_baby}Average Treatment Effect Estimation for the Low Birth Weight Data.}
		\begin{tabular}{lcccccc}
			\toprule
			& ATE    & Std   & $Z$-statistics & $p$-value & \multicolumn{2}{c}{95\% CI} \\ \midrule
			AIPW      & -231.20 & 27.35 & -8.45        & 0.00    & -284.82      & -177.59      \\
			IPW       & -232.73 & 26.63 & -8.74        & 0.00    & -284.93      & -180.53      \\
			IPW-RA    & -229.69 & 28.58 & -8.04        & 0.00    & -285.70      & -173.67      \\
			MBC       & -219.06 & 32.96 & -6.65        & 0.00    & -283.65      & -154.47      \\
			PSM       & -235.26 & 31.67 & -7.43        & 0.00    & -297.33      & -173.20      \\
			RA        & -234.97 & 25.25 & -9.31        & 0.00    & -284.46      & -185.48      \\
			Efficient & -295.77 & 38.62 & -7.66        & 0.00    & -371.47      & -220.07      \\
			Local     & -306.32 & 54.50 & -5.62        & 0.00    & -413.14      & -199.50      \\
			Logistic  & -352.11 & 46.78 & -7.53        & 0.00    & -443.80      & -260.42      \\
			Our Estimator ($\Delta^{ATE}$)       & -217.90 & 22.82 & -9.55        & 0.00    & -262.63      & -173.16      \\ \bottomrule
		\end{tabular}
	\end{table}

	\Cref{table: ate_baby} shows the estimation of the average treatment effect based on our approach. The naive difference in the weight of babies belonging to non-smoking and smoking mothers is 275.3 grams. Given the potential confounding effects of covariates on the potential outcome, this result may not be a valid estimate of the average treatment effect. We next compare the results of average treatment effect estimation using the augmented inverse propensity weighting (AIPW) estimator \citep[][]{Tan2010,tsiatis2006semiparametric}, inverse propensity weighting (IPW) estimator \citep[][]{hirano2003a,imbens2010a}, inverse propensity weighting with regression adjustment (IPW-RA) estimator \citep[][]{Wooldridge2007}, bias-corrected matching (MBC) estimator \citep[][]{abadie2006a,abadie2011a}, propensity score matching (PSM) estimator \citep[][]{rosenbaum1983a,abadie2016a}, regression adjustment (RA) estimator \citep[][]{lane1982analysis}, doubly robust with  efficient propensity score estimation (Efficient), locally efficient propensity score estimation (Local), or parametric logistic estimation (Logistic) estimators \citep[][]{Liu2018,Ma2013}. We summarize the above estimation results for the average treatment effect of maternal smoking on babies' birth weight in \Cref{table: ate_baby} with the mean, standard deviation, $Z$-statistics, $p$-value, and 95\% confidence interval. 
	
	We note that the doubly robust estimator using the propensity score estimated by logistic regression is different from other estimators, suggesting that the parametric logistic form may not be suitable for the propensity score function in this data example. In comparison, our estimation approach is valid for potential misspecification of the propensity score function and provides a valid estimation and inference for the average treatment effect. 
	
	To further illustrate the potential misspecification of propensity score, we plot the estimated link function $\hat{g}$ of propensity score using the wild bootstrap simulation with 500 bootstrap repetitions. As we use the logistic function $\Lambda$ in \Cref{eq:ps1}, if the true DGP for propensity score in this data example is logistic, then the estimated link function $\hat{g}$ should be the identity function. We calculate the optimal truncation parameter $k$ based on the minimization of the prediction mean of squares. In this data example, the optimal truncation parameter $k = 2$. We estimate the approximated second-order polynomial fitted function using the ordinary least squares estimation approach as follows (with standard errors in brackets):
	\begin{equation*}
		\widehat{g}_{k}(\omega_{i})=\underset{(0.054)}{1.22} 
		+
		\underset{(0.048)}{0.54}\omega_{i}
		+
		\underset{(0.006)}{0.12}\omega_{i}^2,
	\end{equation*}
	where $\omega_{i}$ is the $x_{i}'\hat{\theta}$ for $1\leq i \leq N$.

	{\small
		\begin{figure}[h!]
			\begin{center}
				\caption{Estimated nonparametric link function $g(\omega)$ of the propensity score for the baby birth weight data.}
				{\includegraphics[width=0.70\textwidth]{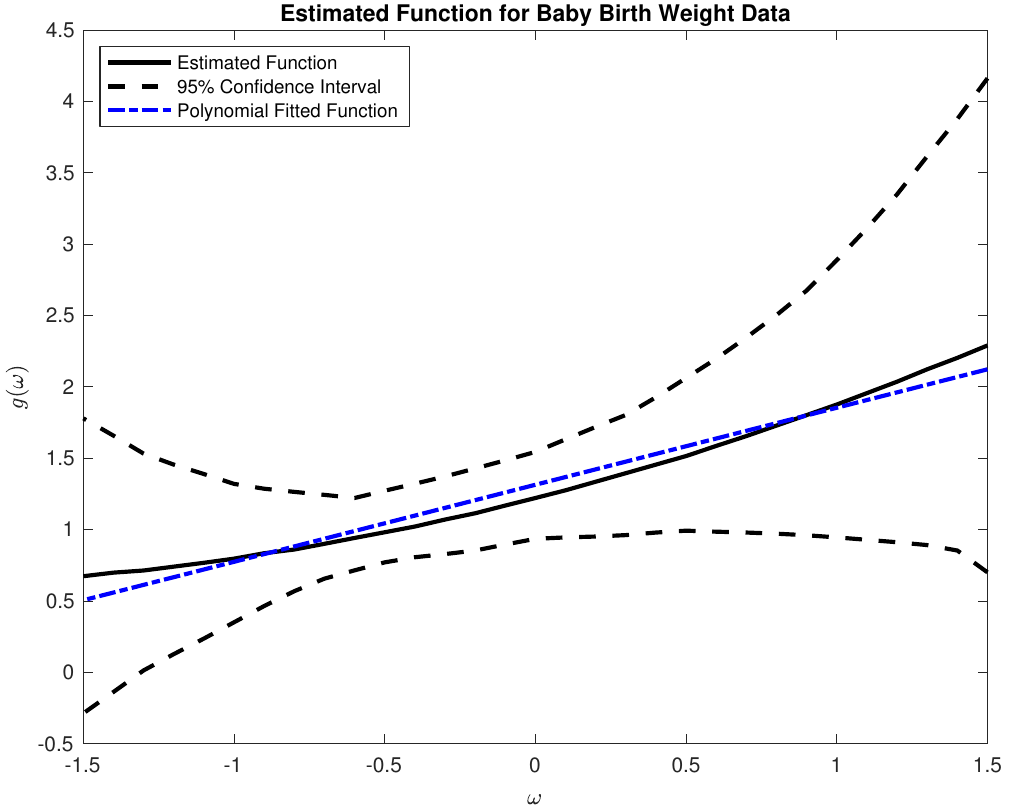}\label{figure: link_baby}}
			\end{center}
		\end{figure}
	}
	
	We show the estimation in \Cref{figure: link_baby} with the estimated link function in solid line and the corresponding 95\% confidence interval in dashed line. We also plot the approximated second-order polynomial fitted function using the ordinary least squares estimation approach as a reference. Our results show that the estimated link function $\hat{g}$ does not seem to be an identity function. Such potential misspecification of the propensity score may explain the differences in treatment effect estimation between our estimator and the doubly robust estimator using the propensity score estimated by logistic regression. Both the estimated link function and the approximated second-order polynomial fitted function are close to each other, demonstrating that the estimation method works well in this empirical application.

	\section{Conclusion}\label{section:conclusion}
	 
	We propose a semiparametric estimation method for determining the average treatment effect. Our approach assumes a single-index structure for the propensity score function, allowing for a flexible link function. We simultaneously estimate the index parameter and link function in an infinitely dimensional function space, and calculate the propensity score through optimization, taking into consideration the constraints of identification conditions. We also establish the large sample properties of our estimator. Our method is empirically applicable and suitable for models encompassing a wide range of error distributions and offers computational flexibility for problems of moderately large dimensions. We conduct an extensive simulation study to assess the finite sample performance of our proposed estimator in finite samples. We present empirical results concerning the impact of maternal smoking on infant birth weight, as well as the effect of job training programs on future earnings. Our approach is less susceptible to misspecification of the propensity score function and remains valid when the propensity score approaches zero or one.
	\bigskip

	\clearpage
	{\small
	
	{\footnotesize
		\bibliographystyle{chicago}
		\bibliography{bibliography}
	}

 	\clearpage
 	
 				% change page number and equation number in appendix
 				\pagenumbering{arabic}		
 				\renewcommand*{\thepage}{IA-\arabic{page}}
 				
 				\begin{center}
 					{\bf \Large Online Appendices}
 				\end{center}
 				
 				Appendix A provides the proof for \Cref{theorem:ate1}. Appendix B gives some preliminary results about Hilbert space, Appendix C gives the proofs of the main results, Appendix D provides some additional numerical results, and Appendix E presents additional empirical application results.
 				\bigskip
 				
 				% Define a command to configure appendix numbering
 				\newcommand{\configureappendix}[1]{%
 					\setcounter{equation}{0}%
 					\renewcommand{\theequation}{#1.\arabic{equation}}%
 					\renewcommand{\thesubsection}{#1.\arabic{subsection}}%
 					\renewcommand{\thefigure}{#1.\arabic{figure}}%
 					\renewcommand{\thetable}{#1.\arabic{table}}%
 					\renewcommand{\thelemma}{#1.\arabic{lemma}}%
 					\renewcommand{\theremark}{#1.\arabic{remark}}%
 					\renewcommand{\thecorollary}{#1.\arabic{corollary}}%
 					\renewcommand{\theassumption}{#1.\arabic{assumption}}%
 					\renewcommand{\theproposition}{#1.\arabic{proposition}}%
 				}
 				
 				% Use the command for each appendix
 				\configureappendix{A}  % For Appendix A
 				\configureappendix{B}  % For Appendix B
 				\configureappendix{C}  % For Appendix C
 				\configureappendix{D}  % For Appendix D
 				\configureappendix{E}  % For Appendix E

				\noindent{\bf \large Appendix A: Proof of \Cref{theorem:ate1}}
				\bigskip

				% change page number and equation number in appendix
				\pagenumbering{arabic}		
				\renewcommand*{\thepage}{A-\arabic{page}}

				\begin{proof}[\textbf{Proof of \Cref{theorem:ate1}}]

					We prove the asymptotic normality of $\widehat{\Delta}_{\omega}^{ATE}$ and $\widehat{\Delta}^{ATE}$ in two steps. In the first step, we take the first-stage errors into account when estimating the propensity score function $\pi(X)$. In the second step, we prove the asymptotic properties of sample analog of treatment effects estimators.
					
					\textbf{(a)} We write
					\bea
					&& \sqrt{N}(\widehat{\Delta}_{\omega}^{ATE}- \Delta_{\omega}^{ATE})
					=
					\Big\{
					\frac{1}{N} \sum_{i=1}^{N} \big(D_{i} - \hat{\pi}(X_i) \big)^{2}
					\Big\}^{-1}
					\nonumber\\
					&& \times 
					\frac{1}{\sqrt{N}}
					\sum_{i=1}^{N} 
					\Big( Y_{i} - \Delta_{\omega}^{ATE} \big(D_{i} - \hat{\pi}(X_i) \big) \Big) 
					\big( D_{i} - \hat{\pi}(X_i) \big)
					\nonumber\\
					&& =
					\Big\{
					\frac{1}{N} \sum_{i=1}^{N} \big(D_{i} - \hat{\pi}(X_i) \big)^{2}
					\Big\}^{-1}
					\frac{1}{\sqrt{N}}
					\sum_{i=1}^{N} 
					\Big( Y_{i} - \Delta_{\omega}^{ATE} \big(D_{i} - \pi(X_i) \big) \Big) 
					\big( D_{i} - \pi(X_i) \big)
					\nonumber\\
					&&
					-
					\Big\{
					\frac{1}{N} \sum_{i=1}^{N} \big(D_{i} - \hat{\pi}(X_i) \big)^{2}
					\Big\}^{-1}
					\frac{1}{\sqrt{N}}
					\sum_{i=1}^{N} 
					\Delta_{\omega}^{ATE} \big( \pi(X_i) - \hat{\pi}(X_i) \big)
					\big( D_{i} - \pi(X_i) \big)
					\nonumber\\
					&&+
					\Big\{
					\frac{1}{N} \sum_{i=1}^{N} \big(D_{i} - \hat{\pi}(X_i) \big)^{2}
					\Big\}^{-1}
					\frac{1}{\sqrt{N}}
					\sum_{i=1}^{N} 
					U_{i}
					\big( \pi(X_i) - \hat{\pi}(X_i) \big)
					\nonumber\\
					&&-
					\Big\{
					\frac{1}{N} \sum_{i=1}^{N} \big(D_{i} - \hat{\pi}(X_i) \big)^{2}
					\Big\}^{-1}
					\frac{1}{\sqrt{N}}
					\sum_{i=1}^{N} 
					\Delta_{\omega}^{ATE} \big( \pi(X_i) - \hat{\pi}(X_i) \big)^{2}
					\nonumber\\
					&& =  A_{1N} + A_{2N} +  A_{3N} +  A_{4N}.
					\nonumber
					\eea
					
					We first consider the asymptotic properties of $\hat{\pi}(X_i) - \pi(X_i)$, by the orthogonality, we write
					\bea
					&&		\big\| \hat{\pi}(X_i) - \pi(X_i) \big\|_{\mathcal{L}^2}^{2}
					=
					\int_{\mathbb{R}} 
					\Big\{ 
					\Lambda\big(\hat{g}_{k}(w)\big)
					-
					\Lambda\big(g_{0}(w)\big) \Big\}^{2} \pi(w) d w 
					\nonumber\\
					&&
					\leq 
					O(1)
					\int_{\mathbb{R}} 
					\Big\{ 
					\big(\hat{g}_{k}(w)\big) 
					-
					\big(g_{0}(w)\big) 
					\Big\}^{2} \pi(w) d w  
					=
					O(1) \int_{\mathbb{R}} \big\{ H(w) (\widehat{C}_{k} - C_{0,k}) + \epsilon_{0, k} \big\}^{2} \pi(w) d w 
					\nonumber\\
					&&
					= O\big(\|\widehat{C}_{k} - C_{0,k}\|^{2} \big) 
					+
					O\big( \| \epsilon_{0, k} \|^{2}_{\mathcal{L}^2} \big),
					\nonumber
					\eea
					where the first inequality follows from the Lipschitz continuity of $\Lambda\big(g(w)\big)$, the third equality follows from the orthogonality of Hermite polynomials.
					
					By Lemma A1 in \cite{Dong2018}, we have $\| \epsilon_{0, k} \|^{2}_{\mathcal{L}^2} = O(k^{-r})$.
					
					By Theorem 2.5 in  \cite{Dong2018}, we have $\|\widehat{C}_{k} - C_{0,k}\|^{2} = O_{P}\big(\frac{k}{N}\big)$.
					
					Therefore, we have $\big\| \hat{\pi}(X_i) - \pi(X_i) \big\|_{\mathcal{L}^2}^{2} =  O_{P}(\frac{k}{N})+O_{P}(k^{-r})$.
					
					We next consider the asymptotic property of $\frac{1}{N} \sum_{i=1}^{N} \big(D_{i} - \hat{\pi}(X_i) \big)^{2}$:
					\bea
					&& \frac{1}{N} \sum_{i=1}^{N} \big(D_{i} - \hat{\pi}(X_i) \big)^{2} 
					= \frac{1}{N} \sum_{i=1}^{N} \big(D_{i} - \pi(X_i) \big)^{2} 
					+ 
					\frac{1}{N} \sum_{i=1}^{N} \big(\hat{\pi}(X_i) - \pi(X_i)\big)^{2}  
					\nonumber\\
					&&+ \frac{2}{N} \sum_{i=1}^{N} \big( D_{i} - \pi(X_i) \big) \big( \hat{\pi}(X_i) - \pi(X_i) \big)
					= H_{1N} + H_{2N} + H_{3N}.
					\nonumber
					\eea
					
					By \Cref{assumption as:1}(a), we have for $H_{1N} = \frac{1}{N} \sum_{i=1}^{N} \big(D_{i} - \pi(X_i) \big)^{2} = \E \big(D_{i} - \pi(X_i) \big)^{2} + o_P(1)$.
					
					Following the proofs as in $K_{11N}$ and  $K_{12N}$, we have $\|H_{2N}\|  = o_P(1)$ and $\|H_{3N}\|  = o_P(1)$.
					
					Therefore, we have $\frac{1}{N} \sum_{i=1}^{N} \big(D_{i} - \hat{\pi}(X_i) \big)^{2}  = \E \big(D_{i} - \pi(X_i) \big)^{2} + o_P(1)$.
					
					For $A_{1N}$, denote the term $\Big( Y_{i} - \Delta_{\omega}^{ATE} \big(D_{i} - \pi(X_i) \big) \Big) 
					\big( D_{i} - \pi(X_i) \big)$ as $P_{i}$. By \Cref{assumption as:1}(a), we have $\E [P_{i}] = 0$ and $\operatorname{Var} [P_{i}] <\infty $. Applying the Lindeberg--Levy CLT, we have:      
					\[
					A_{1N} \rightarrow_{P}  N(0, \sigma_{\omega}^2).
					\]
					
					For $A_{2N}$, we write:
					\bea
					&&\|A_{2 N} \| =
					\Big\{
					\E \big(D_{i} - \pi(X_i) \big)^{2} 
					\Big\}^{-2}
					\E
					\Big\|
					\frac{1}{\sqrt{N}} \sum_{i=1}^{N} [D_{i}-\pi(X_i)] 
					\big\{
					\pi(X_i) - \hat{\pi}(X_i) 
					\big\}
					\Big\| 
					\nonumber\\
					&& =
					O(1)
					\frac{1}{N} \sum_{i=1}^{N} 
					\E
					\Big\|
					\Lambda\big(g_{0}(X_{i}^{\prime} \theta_{0})\big)
					-
					\Lambda\big(\hat{g}_{k}(X_{i}^{\prime} \hat{\theta})\big)
					\Big\| 
					\nonumber\\
					&& = 
					O(1)
					\frac{1}{N} \sum_{i=1}^{N} 
					\E
					\Big\|  
					\Lambda\big(g_{0}(X_{i}^{\prime} \theta_{0})\big)
					-
					\Lambda\big(g_{0}(X_{i}^{\prime} \hat{\theta})\big) 
					+ 
					\Lambda\big(g_{0}(X_{i}^{\prime} \hat{\theta})\big) 
					-
					\Lambda\big(g_{0,k}(X_{i}^{\prime} \hat{\theta})\big) 
					\nonumber\\
					&&
					+
					\Lambda\big(g_{0,k}(X_{i}^{\prime} \hat{\theta})\big) 	
					-		
					\Lambda\big(\hat{g}_{k}(X_{i}^{\prime} \hat{\theta})\big)
					\Big\| 
					\leq
					O(1)
					\frac{1}{N} \sum_{i=1}^{N} 
					\E
					\Big\|  
					\Lambda\big(g_{0}(X_{i}^{\prime} \theta_{0})\big)
					-
					\Lambda\big(g_{0}(X_{i}^{\prime} \hat{\theta})\big) 
					\Big\|
					\nonumber\\
					&&+
					O(1)
					\frac{1}{N} \sum_{i=1}^{N} 
					\E
					\Big\| 
					\Lambda\big(g_{0}(X_{i}^{\prime} \hat{\theta})\big) 
					-
					\Lambda\big(g_{0,k}(X_{i}^{\prime} \hat{\theta})\big) 
					\Big\|+
					O(1)
					\frac{1}{N} \sum_{i=1}^{N} 
					\E
					\Big\| 
					\Lambda\big(g_{0,k}(X_{i}^{\prime} \hat{\theta})\big) 	
					-
					\Lambda\big(\hat{g}_{k}(X_{i}^{\prime} \hat{\theta})\big)
					\Big\|
					\nonumber\\
					&& =
					O(1) A_{21N} + O(1) A_{22N} + O(1) A_{23N},
					\nonumber
					\eea
					where the second equality follows from \Cref{assumption as:1}(a), the fourth inequality follows from the triangular inequality. 
					
					For $A_{21N}$, we write:
					\begin{align*}
						A_{21N} 
						&= 
						\frac{1}{N} \sum_{i=1}^{N} 
						\E
						\Big\|  
						\Lambda\big(g_{0}(X_{i}^{\prime} \theta_{0})\big)
						-
						\Lambda\big(g_{0}(X_{i}^{\prime} \hat{\theta})\big) 
						\Big\|
						\\
						&
						\leq 
						\Big\|
						\E
						\big[
						\Lambda\big(g_{0}(X_{i}^{\prime} \theta_{0})\big)
						-
						\Lambda\big(g_{0}(X_{i}^{\prime} \hat{\theta})\big)
						\big]
						\Big\|+o_{P}(1)
						\\
						& 
						= 
						\Big\|
						\E
						\big[ 
						\Lambda\big(g_{0}(X_{i}^{\prime} \theta^{*})\big) g_{0}^{(1)}(X_{i}^{\prime} \theta^{*})
						(X_{i}^{\prime} \theta_{0}
						-
						X_{i}^{\prime} \hat{\theta})
						\big]
						\Big\| +o_{P}(1) 
						\\
						& 
						\leq
						\|
						\hat{\theta}
						-
						\theta_{0} 
						\|
						\Big\{
						\E
						\big\| 
						\Lambda\big(g_{0}(X_{i}^{\prime} \theta^{*})\big) g_{0}^{(1)}(X_{i}^{\prime} \theta^{*}) X_{i} X_{i}^{\prime}
						\big\|^{2} 
						\Big\}^{1/2}
						+o_{P}(1)  = o_{P}(1),       
					\end{align*}
					where the first inequality follows from \Cref{assumption as:2}(a), the second equality follows from mean value theorem, the third inequality follows from Cauchy-Schwarz inequality, and the last equality follows from \Cref{assumption as:1}(b) and the fact that $\|\hat{\theta} - \theta_{0} \| \rightarrow_{P} 0$ and $\Lambda\big(g(X_{i}^{\prime}\theta)\big) $ is bounded by 1.
					
					For $A_{22N}$, we write:
					\bea
					&& A_{22N} =
					\frac{1}{N} \sum_{i=1}^{N} 
					\E
					\Big\| 
					\Lambda\big(g_{0}(X_{i}^{\prime} \hat{\theta})\big) 
					-
					\Lambda\big(g_{0,k}(X_{i}^{\prime} \hat{\theta})\big) 
					\Big\|
					\leq 
					O(1) 
					\Big\|
					\E
					\big[
					\Lambda\big(g_{0}(X_{i}^{\prime} \hat{\theta})\big)
					-
					\Lambda\big(g_{0, k}(X_{i}^{\prime} \hat{\theta})\big)			
					\big]
					\Big\| + o_{P}(1)
					\nonumber\\
					&& 
					\leq  O(1) 
					\Big\| 
					\E
					\big[\epsilon_{0, k}(X_{i}^{\prime} \hat{\theta})
					\big]
					\Big\| + o_{P}(1)
					\leq O(1)  
					\Big\{ 
					\E 
					\big\| 
					\epsilon_{0, k}(X_{i}^{\prime} \hat{\theta}) 
					\big\|^{2} 
					\Big\}^{1 / 2} + o_{P}(1) = o_{P}(1),
					\nonumber
					\eea
					where the first inequality follows from \Cref{assumption as:2}(a), the second inequality follows from Lipschitz continuity of $\Lambda(g(X_{i}^{\prime} \theta)) $, the third inequality follows from Cauchy-Schwarz inequality, and the last equality follows from \Cref{assumption as:2}(c) and Lemma A1 from \cite{Dong2018}.
					
					For $A_{23N}$, we write
					\bea
					&& A_{23N} =
					\frac{1}{N} \sum_{i=1}^{N} 
					\E
					\Big\| 
					\Lambda\big(g_{0,k}(X_{i}^{\prime} \hat{\theta})\big) 	
					-
					\Lambda\big(\hat{g}_{k}(X_{i}^{\prime} \hat{\theta})\big)
					\Big\|
					\nonumber\\
					&& 
					\leq 
					\Big\| 
					\E 
					\big\{
					\Lambda
					\big(H(X_{i}^{\prime} \hat{\theta})^{\prime} C_{0,k}\big)
					-
					\Lambda
					\big(H(X_{i}^{\prime} \hat{\theta})^{\prime} \widehat{C}_{k}\big)			 
					\big\}
					\Big\| +o_{P}(1) 
					\leq  
					O(1) 
					\Big\| 
					\E
					\big[
					H(X_{i}^{\prime} \hat{\theta})^{\prime} 
					( \widehat{C}_{k}  -  C_{0,k} )
					\big]
					\Big\|  +o_{P}(1) 
					\nonumber\\
					&&
					\leq 
					O(1) 
					\Big\{ 
					(\widehat{C}_{k}  -  C_{0,k})^{\prime} 
					\E
					\big[
					H(X_{i}^{\prime} \hat{\theta}) H(X_{i}^{\prime} \hat{\theta})^{\prime}
					\big]
					(\widehat{C}_{k}  -  C_{0,k}) 
					\Big\}^{1 / 2} 
					\nonumber\\
					&&  \leq O(1)\|\widehat{C}_{k} - C_{0, k} \|+o_{P}(1) 
					\leq O(1)\|\hat{g}_{k}-g_{0}\|_{\mathcal{L}^2}+o_{P}(1)=o_{P}(1),                                    
					\nonumber
					\eea
					where the first inequality follows from \Cref{assumption as:2}(a), the second inequality follows from Lipschitz continuity of $\Lambda\big(g(X_{i}^{\prime} \theta)\big) $, the third inequality follows from Cauchy-Schwarz inequality, the fourth inequality follows from Lemma A4 in \cite{Dong2018}, and the last inequality follows from the definition of $\|\cdot\|_{\mathcal{L}^2}$.

					For $A_{3N}$, we write:
					\begin{align*}
						\|A_{3 N} \| 
						= 
						&
						\Big\{
						\E \big(D_{i} - \pi(X_i) \big)^{2} 
						\Big\}^{-2}
						\E
						\Big\|
						\frac{1}{\sqrt{N}} \sum_{i=1}^{N} 
						U_{i}
						\big\{
						\pi(X_i) - \hat{\pi}(X_i) 
						\big\}
						\Big\| 
						\\
						=
						&
						O(1)
						\frac{1}{N} \sum_{i=1}^{N} 
						\E
						\Big\|
						U_{i}
						\Big\{
						\Lambda\big(g_{0}(X_{i}^{\prime} \theta_{0})\big)
						-
						\Lambda\big(\hat{g}_{k}(X_{i}^{\prime} \hat{\theta})\big)
						\Big\}
						\Big\| 
						\\
						= 
						&
						O(1)
						\frac{1}{N} \sum_{i=1}^{N} 
						\E
						\Big\|  
						U_{i}
						\Big\{
						\Lambda\big(g_{0}(X_{i}^{\prime} \theta_{0})\big)
						-
						\Lambda\big(g_{0}(X_{i}^{\prime} \hat{\theta})\big) 
						\Big\}
						+ 
						U_{i}
						\Big\{
						\Lambda\big(g_{0}(X_{i}^{\prime} \hat{\theta})\big) 
						-
						\Lambda\big(g_{0,k}(X_{i}^{\prime} \hat{\theta})\big) 
						\Big\}
						\\
						+ &
						U_{i}
						\Big\{
						\Lambda\big(g_{0,k}(X_{i}^{\prime} \hat{\theta})\big) 	
						-		
						\Lambda\big(\hat{g}_{k}(X_{i}^{\prime} \hat{\theta})\big)
						\Big\}
						\Big\| 
						=
						o_{P}(1),                                    
					\end{align*}
					where the second equality follows from \Cref{assumption as:1}(a), the fourth inequality follows from the triangular inequality and similar proofs as $A_{2N}$.

					For $A_{4N}$, we write:
					\begin{align*}
						\|
						A_{4N}
						\|
						=
						&
						\Big\{
						\E \big(D_{i} - \pi(X_i) \big)^{2} 
						\Big\}^{-2}
						\E
						\Big\|
						\frac{1}{\sqrt{N}} \sum_{i=1}^{N} 
						\Delta_{\omega}^{ATE} \big( \pi(X_i) - \hat{\pi}(X_i) \big)^{2}
						\Big\| 
						\\
						=
						&
						O(1)
						\frac{1}{N} \sum_{i=1}^{N} 
						\E
						\Big\|
						\big\{
						\Lambda\big(g_{0}(X_{i}^{\prime} \theta_{0})\big)
						-
						\Lambda\big(\hat{g}_{k}(X_{i}^{\prime} \theta_{0})\big)
						\big\}^{2}
						\Big\| 
						\\
						\leq
						& 
						O(1)
						\frac{1}{N} \sum_{i=1}^{N} 
						\E
						\Big\|  
						\big\{
						\Lambda\big(g_{0}(X_{i}^{\prime} \theta_{0})\big)
						-
						\Lambda\big(g_{0}(X_{i}^{\prime} \hat{\theta})\big) 
						\big\}^{2}
						\Big\|
						\\
						& + 
						O(1)
						\frac{1}{N} \sum_{i=1}^{N} 
						\E
						\Big\| 
						\big\{
						\Lambda\big(g_{0}(X_{i}^{\prime} \hat{\theta})\big) 
						-
						\Lambda\big(g_{0,k}(X_{i}^{\prime} \hat{\theta})\big) 
						\big\}^{2}
						\Big\|
						\\
						+ &
						O(1)
						\frac{1}{N} \sum_{i=1}^{N} 
						\E
						\Big\| 
						\big\{
						\Lambda\big(g_{0,k}(X_{i}^{\prime} \hat{\theta})\big) 	
						-
						\Lambda\big(\hat{g}_{k}(X_{i}^{\prime} \hat{\theta})\big)
						\big\}^{2}
						\Big\|
						=
						o_{P}(1),
					\end{align*}
					where the second equality comes from the boundness of $\Delta_{\omega}^{ATE} $, the third inequality comes from triangular inequality, and the fourth equality comes from the similar proofs as $A_{2N}$.
					
					In sum, we have shown the following asymptotic normality:
					\begin{equation*}
						\sqrt{N} (\widehat{\Delta}_{\omega}^{ATE}- \Delta_{\omega}^{ATE})\rightarrow_{D} N(0, \sigma_{\omega}^{2}).
					\end{equation*}
					
					\textbf{(b)} We write:
					\bea 
					&&		\sqrt{N}(\widehat{\Delta}^{ATE}- \Delta^{ATE})
					=
					\Big\{
					\frac{1}{N} \sum_{i=1}^{N}\hat{\upsilon}(X_{i})^{-1} \big(D_{i} - \hat{\pi}(X_i) \big)^{2}
					\Big\}^{-1}
					\nonumber\\
					&& \times 
					\frac{1}{\sqrt{N}}
					\sum_{i=1}^{N} 
					\hat{\upsilon}(X_{i})^{-1}
					\big( Y_{i} - \Delta^{ATE} (D_{i} - \hat{\pi}(X_i) \big) 
					\big( D_{i} - \hat{\pi}(X_i) \big)
					\nonumber\\
					&& =
					\Big\{
					\frac{1}{N} \sum_{i=1}^{N}\hat{\upsilon}(X_{i})^{-1} \big(D_{i} - \hat{\pi}(X_i) \big)^{2}
					\Big\}^{-1}
					\frac{1}{\sqrt{N}}
					\sum_{i=1}^{N} 
					\hat{\upsilon}(X_{i})^{-1}
					\big( Y_{i} - \Delta^{ATE} (D_{i} - \pi(X_i) \big) 
					\big( D_{i} - \pi(X_i) \big)
					\nonumber\\
					&&
					-
					\Big\{
					\frac{1}{N} \sum_{i=1}^{N}\hat{\upsilon}(X_{i})^{-1} \big(D_{i} - \hat{\pi}(X_i) \big)^{2}
					\Big\}^{-1}
					\frac{1}{\sqrt{N}}
					\sum_{i=1}^{N} 
					\hat{\upsilon}(X_{i})^{-1} \Delta^{ATE} \big( \pi(X_i) - \hat{\pi}(X_i) \big)
					\big( D_{i} - \pi(X_i) \big)
					\nonumber\\
					&&+
					\Big\{
					\frac{1}{N} \sum_{i=1}^{N}\hat{\upsilon}(X_{i})^{-1} \big(D_{i} - \hat{\pi}(X_i) \big)^{2}
					\Big\}^{-1}
					\frac{1}{\sqrt{N}}
					\sum_{i=1}^{N} 
					\hat{\upsilon}(X_{i})^{-1} U_{i}
					\big( \pi(X_i) - \hat{\pi}(X_i) \big)
					\nonumber\\
					&&
					-
					\Big\{
					\frac{1}{N} \sum_{i=1}^{N}\hat{\upsilon}(X_{i})^{-1} \big(D_{i} - \hat{\pi}(X_i) \big)^{2}
					\Big\}^{-1}
					\frac{1}{\sqrt{N}}
					\sum_{i=1}^{N} 
					\hat{\upsilon}(X_{i})^{-1}
					\Delta^{ATE} \big( \pi(X_i) - \hat{\pi}(X_i) \big)^{2}
					\nonumber\\
					&& =
					B_{1N} + B_{2N} +  B_{3N} +  B_{4N}.
					\nonumber
					\eea
					
					Following similar derivations to those used for $	\frac{1}{N} \sum_{i=1}^{N}
					\big(D_{i} - \hat{\pi}(X_i) \big)^{2}$, we have
					\[
					\frac{1}{N} \sum_{i=1}^{N} \hat{\upsilon}(X_{i})^{-1} 
					\big(D_{i} - \hat{\pi}(X_i) \big)^{2}
					\rightarrow_{P} 
					\E \Big[ \upsilon(X_{i})^{-1} \big(D_{i} - \pi(X_i) \big)^{2} \Big].
					\]
					
					For $B_{1N}$, denote the term $\Big( Y_{i} - \Delta^{ATE} \big(D_{i} - \pi(X_i) \big) \Big) 
					\big( D_{i} - \pi(X_i) \big)$ as $P_{i}$. By \Cref{assumption as:1}(a), we have $\E [P_{i}] = 0$ and $\operatorname{Var} [P_{i}] <\infty $. Applying the Lindeberg--Levy CLT, we have:      
					\[
					B_{1N} \rightarrow_{P}  N(0, \sigma^2).
					\]
					
					For $B_{2N}$, we write
					\bea
					&&		\|B_{2 N} \| 
					=
					\Big\{
					\E \big[ \upsilon(X_{i})^{-1} \big(D_{i} - \pi(X_i) \big)^{2} \big]
					\Big\}^{-2}
					\nonumber\\
					&&
					\E
					\Big\|
					\frac{1}{\sqrt{N}} \sum_{i=1}^{N} \hat{\upsilon}(X_{i})^{-1} \Delta^{ATE} \big( \pi(X_i) - \hat{\pi}(X_i) \big)
					\big( D_{i} - \pi(X_i) \big)
					\Big\| 
					\nonumber\\
					&& \leq
					O(1)
					\frac{1}{N} \sum_{i=1}^{N} 
					\E
					\Big\|
					\Lambda\big(g_{0}(X_{i}^{\prime} \theta_{0})\big)
					-
					\Lambda\big(\hat{g}_{k}(X_{i}^{\prime} \hat{\theta})\big)
					\Big\| 
					=
					o_{P}(1),
					\nonumber
					\eea
					where the second inequality follows from \Cref{assumption as:1}(a) and boundness of $\upsilon(\cdot)$, and the third equality follows from the similar proofs as $A_{2N}$.
					
					For $B_{3N}$, we write:
					\begin{align*}
						\|B_{3 N} \| 
						= 
						&
						\Big\{
						\E \big[ \upsilon(X_{i})^{-1} \big(D_{i} - \pi(X_i) \big)^{2} \big]
						\Big\}^{-2}
						\E
						\Big\|
						\frac{1}{\sqrt{N}} \sum_{i=1}^{N} 
						\hat{\upsilon}(X_{i})^{-1} U_{i}
						\big\{
						\pi(X_i) - \hat{\pi}(X_i) 
						\big\}
						\Big\| 
						\\
						\leq
						&
						O(1)
						\frac{1}{N} \sum_{i=1}^{N} 
						\E
						\Big\|
						U_{i}
						\big\{
						\Lambda\big(g_{0}(X_{i}^{\prime} \theta_{0})\big)
						-
						\Lambda\big(\hat{g}_{k}(X_{i}^{\prime} \hat{\theta})\big)
						\big\}
						\Big\| 
						=o_{P}(1),
					\end{align*}
					where the second inequality follows from \Cref{assumption as:1}(a) and boundness of $\upsilon(\cdot)$, and the third equality follows from the similar proofs as $A_{3N}$.
					
					For $B_{4N}$, we write:
					\begin{align*}
						\|
						B_{4N}
						\|
						=
						&
						\Big\{
						\E \big[ \upsilon(X_{i})^{-1} \big(D_{i} - \pi(X_i) \big)^{2} \big]
						\Big\}^{-2}
						\E
						\Big\|
						\frac{1}{\sqrt{N}} \sum_{i=1}^{N} 
						\hat{\upsilon}(X_{i})^{-1} \Delta_{\omega}^{ATE} \big( \pi(X_i) - \hat{\pi}(X_i) \big)^{2}
						\Big\| 
						\\
						\leq
						&
						O(1)
						\frac{1}{N} \sum_{i=1}^{N} 
						\E
						\Big\|
						\Lambda\big(g_{0}(X_{i}^{\prime} \theta_{0})\big)
						-
						\Lambda\big(\hat{g}_{k}(X_{i}^{\prime} \hat{\theta})\big)
						\Big\| 
						=o_{P}(1),
					\end{align*}
					where the second inequality follows from \Cref{assumption as:1}(a) and boundedness of $\upsilon(\cdot)$, and the third equality follows from the similar proofs as $A_{4N}$.
					
					In sum, we have:
					\begin{equation*}
						\sqrt{N} (\widehat{\Delta}^{ATE}- \Delta^{ATE})\rightarrow_{D} N(0, \sigma^{2}),
					\end{equation*}
					which completes the proof of \Cref{theorem:ate1}.	
					
				\end{proof}

		\noindent{\bf \large Appendix B: Preliminary Properties for Hilbert Space}
		\bigskip
		
		In this section, we present the properties of the Hilbert Space $\mathcal{L}^{2}$ that includes all polynomials, all power functions, all bounded functions, and even some exponential functions. More importantly, all functions in the space are defined on the entire real line.
		
		An inner product for $g_{1}(w), g_{2}(w) \in \mathcal{L}^{2}$ is given by:
		$$
		\langle g_{1}(w), g_{2}(w)\rangle=\int g_{1}(w) g_{2}(w) \exp(-w^{2} / 2) d w.
		$$
		
		Hermite polynomials form a complete orthogonal sequence in  $\mathcal{L}^{2}$ with each elements defined by:
		\begin{equation*}\label{eq: basis}
			h_{m}(w):=\frac{1}{\sqrt{m !}} \cdot (-1)^{m} \cdot \exp (w^{2} / 2) \cdot \frac{d^{m}}{d w^{m}} \exp (-w^{2} / 2), \ \ m=0,1,2, \cdots.
		\end{equation*} 
		
		The orthogonality of this basis system satisfies $\int h_{m}(w) h_{n}(w) \exp (-w^{2} / 2) d w=\sqrt{2 \pi} \delta_{m n}$, where $\delta_{m n}$ is the Kronecker delta. For $\forall g(w) \in \mathcal{L}^{2}$, we have orthogonal series expansion:
		$$
		g(w)=\sum_{j=1}^{\infty} c_{j} h_{j}(w), \quad c_{j}=\langle g(w), h_{j}(w)\rangle.
		$$

		For any truncation parameter $k \geq 1$, we split the orthogonal series expansion into two parts as following:
		\begin{equation} 
			g(w)=g_{k}(w)+\epsilon_{k}(w),
		\end{equation}
		where $g_{k}(w) :=H(w)^{\prime} C_{k}$, $H(w)=\big(h_{1}(w), \ldots, h_{k}(w)\big)^{\prime}$, $C_{k}=(c_{1}, \ldots, c_{k})^{\prime}$, and the approximation residual is given by $\epsilon_{k}(w)=\sum_{j=k+1}^{\infty} c_{j} h_{j}(w)$. 
		
		The quantity of $\|H(w)\|$ is crucial for the asymptotic properties of our estimators, using the results in the Theorem 1.1 of \cite{levin1992christoffel} and \cite{dong2020a}, we have:
		\begin{equation*}
			\frac{1}{\sqrt{k}} \sum_{j=1}^{k} h_{j}^{2}(w) \asymp \exp (w^{2}) \left(\max\left\{k^{-2 / 3}, 1-\frac{|w|}{2\sqrt{k}}\right\}\right)^{1 / 2},
		\end{equation*}
		uniformly for $k \geq 1$ and $w \in\{u:|u| \leq \sqrt{2 k}(1+L k^{-2 / 3})\},$ where $L>0$ is given.\footnote{In our context, $f \asymp g \Longleftrightarrow \exists C, D>0: C|g| \leq|f| \leq D|g|$.} In addition, we can show that the order of $\|H(w)\|$ is $O(\sqrt{k})$ when the condition for $1-|x| / \sqrt{2 k}>k^{-2 / 3}$ satisfies as long as the truncation parameter $k$ diverge.
		\bigskip

		\setcounter{equation}{0}
		\renewcommand{\theequation}{C.\arabic{equation}}
		\renewcommand{\thesubsection}{C.\arabic{subsection}}
		\renewcommand{\thefigure}{C.\arabic{figure}}
		\renewcommand{\thetable}{C.\arabic{table}}
		\renewcommand{\thelemma}{C.\arabi{lemma}}
		\renewcommand{\theremark}{C.\arabic{remark}}
		\renewcommand{\thecorollary}{C.\arabic{corollary}}
		\renewcommand{\theassumption}{C.\arabic{assumption}}
		\renewcommand{\theproposition}{C.\arabic{proposition}}				
		
		\noindent{\bf \large Appendix C: Proofs of Main Results}
		\bigskip

		\begin{proof}[\textbf{Proof of \Cref{lemma:indentification1}}]
			
			%% (a) %%
			\textbf{(a)}
			We have 
			$\E[\operatorname{var}(D | X)]>0$
			under \Cref{assumption:id1}(b),
			and thus the regression parameter $\beta$ can be written as:
			\begin{eqnarray*}
				\beta
				=
				\frac{
					\E
					\big [
					\big (D-\pi(X) \big )
					Y
					\big]
				}{
					\E[\operatorname{var}(D | X)]
				} .
			\end{eqnarray*}
			Given \Cref{assumption:id1}(a)
			and
			that
			$Y = D Y(1) + (1-D) Y(0) $, we have:
			\begin{eqnarray*}
				\E[D Y|X] = \pi(X)\E[Y(1)|X]
				\ \ \ \mathrm{and}  \ \ \ 
				\E[Y|X] = \pi(X)\E[Y(1)|X] + \big (1- \pi(X) \big ) \E[Y(0)|X]. 
			\end{eqnarray*}
			It follows from the law of iterated expectation that:
			\begin{eqnarray*}
				\beta = 
				\frac{
					\E[
					\pi(X)
					\{1-\pi(X)\}
					\E( Y(1) - Y(0)|X)           
					]
				}{
					\E[\operatorname{var}(D | X)]
				}.
			\end{eqnarray*}
			This leads to the desired result.
			
			%% (b) %%
			\textbf{(b)}
			Under \Cref{assumption:id1}(b) and the law of iterated expectation,
			$
			\E[\upsilon(X)^{-1}\operatorname{var}(D | X) ] = 1 
			$
			and thus the regression parameter $\gamma$ can be written as:
			\begin{equation*}
				\gamma
				= 
				\frac{
					\E[\upsilon(X)^{-1}  \big( D - \pi(X)\big)Y ] 
				}
				{
					\E[ \upsilon(X)^{-1} \operatorname{var}(D | X)]
				}.
			\end{equation*}
			Given the unconfoundedness under \Cref{assumption:id1}(a) and that $ Y=D Y(1)+ (1 - D) Y(0)$, we have:
			\begin{eqnarray*}
				\E
				[
				\upsilon(X)^{-1} D Y|X
				] 
				= 
				\frac{\E[Y(1)|X]}{1-\pi(X)}
				\ \ \ \mathrm{and} \ \ \
				\E[
				\upsilon(X)^{-1}   \pi(X) Y |X
				] 
				= 
				\frac{\pi(X)\E[Y(1)|X]}{1-\pi(X)} + \E[Y(0)|X].   	
			\end{eqnarray*}
			Thus, the desired result follows.
		\end{proof}

		\begin{proof}[\textbf{Proof of \Cref{lemma:indentification3}}]

			Note that the objective function $\ell_{N}\big (\check{\theta}, \{c_{j}\}_{j=1}^{k} \big)$ in Equation (\ref{eq:ps4}) is interchangeable with the sample version of population objective function $\ell_{N}(\theta, g)$.
			
			Our proof is composited of two parts. In the first part, we prove the convergence of sample objective function $\ell_{N}(\theta, g)$ to the population objective $\ell(\theta, g)$ using Lemma A2 in \cite{Newey2003}. In the second part, we prove the consistency of $(\hat{\theta}, \widehat{g}_{k})$ by contradiction.
			
			% (2)
			
			%% (a) %%
			\textbf{(a)} We begin to prove the convergence of sample objective function $\ell_{N}(\theta, g)$ to the population objective $\ell(\theta, g)$: 
			\[
			\max _{\Theta \times G} | \ell_{N}(\theta, g) -\ell(\theta, g)| \rightarrow_{P} 0.
			\]
			
			We proceed by checking whether Conditions (i) to (iii) in Lemma A2 of \cite{Newey2003} hold.
			
			For Condition (i), as we assume that $\Theta$ is a compact subset of the parameter space $\mathbb{R}^{d}$ and $\Theta$ is defined with norm. Therefore, Condition (i) holds.

			For Condition (ii), as we assume that the data generating process is independent and identically distributed in \Cref{assumption as:1}(a), therefore, we apply the Law of Large Numbers (LLN) and show that:
			\[
			\ell_{N}(\theta, g)  = \ell(\theta, g) \big(1+o_{P}(1) \big).
			\]
			Therefore, Condition (ii) holds. 
			
			For Condition (iii), we prove by showing the continuity of $\ell(\theta, g)$ holds, which indicates that the continuity of $\ell_{N}(\theta, g)$ holds with probability approaching 1.

			For any given $(\theta_{1}, g_{1})$ and $(\theta_{2}, g_{2})$ belonging to $\Theta \times G$, we have:
			\[
			\big| \ell(\theta_{1}, g_{1}) - \ell(\theta_{2}, g_{2}) \big| 
			\leq
			\big| \ell(\theta_{1}, g_{1}) - \ell(\theta_{1}, g_{2}) \big|
			+
			\big| \ell(\theta_{1}, g_{2}) - \ell(\theta_{2}, g_{2}) \big| 
			= A_1 + A_2.
			\]

			For $A_{1}$, we have:
			\bea
			&&\big|
			\ell(\theta_{1}, g_{1}) - \ell(\theta_{1}, g_{2}) 
			\big| 
			=  
			\Big| 
			\E
			\big[ 
			D_{i} g_{1}(X_{i}^{\prime} \theta_{1}) - D_{i} g_{2}(X_{i}^{\prime} \theta_{1})  +
			\ln \big( 1+ e^{g_{2}(X_{i}^{\prime} \theta_{1})} \big) 
			- 
			\ln \big( 1+ e^{g_{1}(X_{i}^{\prime} \theta_{1})} \big)
			\big] 
			\Big| 
			\nonumber\\
			&&\leq 
			O(1)
			\big\{ 
			\E
			\big[ 
			g_{1}(X_{i}^{\prime} \theta_{1})
			- 
			g_{2}(X_{i}^{\prime} \theta_{1})
			\big]^{2}  \big\}^{1 / 2} +
			\big\{ 
			\E 
			\big[ 
			\ln \big( 1+ e^{g_{2}(X_{i}^{\prime} \theta_{1})} \big) 
			- 
			\ln \big( 1+ e^{g_{1}(X_{i}^{\prime} \theta_{1})} \big) 
			\big]^{2}  
			\big\}^{1/2}
			\nonumber\\
			&& =   A_{11} +A_{12},
			\nonumber
			\eea
			where the second inequality follows from the fact that $D_{i}$ is a binary variable and the Cauchy-Schwarz inequality.

			%		\begin{align*}
				%			\E
				%			\big[
				%			g_{1}(X_{i}^{\prime} \theta_{1})
				%			-
				%			g_{2}(X_{i}^{\prime} \theta_{1})
				%			\big]^{2} 
				%			&
				%			\leq  M \|g_{1}-g_{2}\|_{\mathcal{L}^2}^{2}.
				%		\end{align*}

			For $A_{11}$, we have:
			\bea
			&&\E[g_{1}(X_{i}^{\prime} \theta_{1})-g_{2}(X_{i}^{\prime} \theta_{1})]^{2} =\int\big(g_{1}(w)-g_{2}(w)\big)^{2} f_{\theta_{1}}(w) d w
			\nonumber\\
			&& =\int\big(g_{1}(w)-g_{2}(w)\big)^{2} \exp (-w^{2} / 2) \cdot \exp (w^{2} / 2) f_{\theta_{1}}(w) dw
			\nonumber\\
			&& \leq  M \int\big(g_{1}(w)-g_{2}(w)\big)^{2} \exp (-w^{2} / 2) dw = M \|g_{1}-g_{2}\|_{\mathcal{L}^2}^{2},
			\nonumber
			\eea
			where the inequality follows from \Cref{assumption as:1}(b) due to $\sup _{\{(\theta, w) \in \Theta \times \mathbb{R}\}} \exp (w^{2} / 2) f_{\theta}(w) \leq M$.

			For $A_{12}$, note that $\forall a,b>0$, we have:
			\[
			\big|\ln(1+a) - \ln(1+b) \big|< \big| \ln(a) - \ln(b) \big|.
			\]
			
			Therefore, we have:
			\begin{align*}
				\E 
				\big[ 
				\ln \big( 1+ e^{g_{2}(X_{i}^{\prime} \theta_{1})} \big) - \ln \big( 1+ e^{g_{1}(X_{i}^{\prime} \theta_{1})} \big) 
				\big]^{2} 
				&\leq 
				\E 
				\big[ 
				g_{2}(X_{i}^{\prime} \theta_{1})  - g_{1}(X_{i}^{\prime} \theta_{1}) 
				\big]^{2}
				\\
				&\leq M \|g_{1}-g_{2}\|_{\mathcal{L}^2}^{2}.
			\end{align*}
			
			For $A_1$, we have shown:
			\begin{equation}\label{eq:proof1}
				\big|
				\ell(\theta_{1}, g_{1}) - \ell(\theta_{1}, g_{2}) 
				\big|  
				\leq 
				O(1)
				\|g_{1}-g_{2}\|_{\mathcal{L}^2}.
			\end{equation}

			For $A_2$, we write:
			\bea
			&&	\big|
			\ell(\theta_{1}, g_{2}) - \ell(\theta_{2}, g_{2}) 
			\big|
			=
			\Big| 
			\E\big[ 
			D_{i} g_{2}(X_{i}^{\prime} \theta_{1}) - D_{i} g_{2}(X_{i}^{\prime} \theta_{2}) 
			+
			\ln \big( 1+ e^{g_{2}(X_{i}^{\prime} \theta_{2})} \big) - \ln \big( 1+ e^{g_{2}(X_{i}^{\prime} \theta_{1})} \big) 
			\big] 
			\Big| 
			\nonumber\\
			&& \leq O(1)
			\big\{ 
			\E
			\big[
			g_{2}(X_{i}^{\prime} \theta_{1}) - g_{2}(X_{i}^{\prime} \theta_{2})
			\big]^{2}  \big\}^{1 / 2} 
			+
			\big\{ 
			\E
			\big[ 
			\ln 
			\big( 1+ e^{g_{2}(X_{i}^{\prime} \theta_{2})} \big)
			- 
			\ln 
			\big( 1+ e^{g_{2}(X_{i}^{\prime} \theta_{1})} \big) 
			\big]^{2}  
			\big\}^{1/2}
			\nonumber\\
			&&= A_{21} +A_{22},
			\nonumber
			\eea
			where the inequality follows from the fact that $D_{i}$ is a binary variable and Cauchy-Schwarz inequality. 
			
			Let $\theta^{*}$ lie between $\theta_{1}$ and $\theta_{2}$, and for $A_{21}$, and write
			\bea
			&&\E
			\big[
			g_{2}(X_{i}^{\prime} \theta_{1}) - g_{2}(X_{i}^{\prime} \theta_{2})
			\big]^{2}  
			=\E 
			\big[ 
			(\theta_{2}-\theta_{1})^{\prime} 
			X_{i} X_{i}^{\prime} (\theta_{2}-\theta_{1}) \{g_{2}^{(1)}(X_{i}^{\prime} \theta^{*})\}^{2} 
			\big]
			\nonumber\\
			&& \leq  
			\| \theta_{2}-\theta_{1} \|^{2} \E
			\big\|
			X_{i} X_{i}^{\prime}\{g_{2}^{(1)}(X_{i}^{\prime} \theta^{*})\}^{2}
			\big\| 
			\leq O(1)\|\theta_{2}-\theta_{1}\|^{2},
			\nonumber
			\eea
			where the second inequality follows from \Cref{assumption as:1}(b).
			
			Note that for $A_{22}$, we write:
			\bea
			&&		\E 
			\big[ 
			\ln \big( 1+ e^{g_{2}(X_{i}^{\prime} \theta_{2})} \big)
			-
			\ln \big( 1+ e^{g_{2}(X_{i}^{\prime} \theta_{1})} \big) 
			\big]^{2}
			\nonumber\\
			&& =
			\E
			\big[
			(\theta_{2}-\theta_{1})^{\prime} X_{i} X_{i}^{\prime}(\theta_{2}-\theta_{1})
			\times \big\{ 
			\Lambda\big(g_{2}(X_{i}^{\prime} \theta^{*})\big) g_{2}^{(1)}(X_{i}^{\prime} \theta^{*})
			\big\}^{2}
			\big] 
			\nonumber\\
			&&	\leq 
			\| \theta_{2} - \theta_{1}\|^{2} 
			\times
			\E
			\big\|
			X_{i} X_{i}^{\prime}\{ \Lambda\big(g_{2}(X_{i}^{\prime} \theta^{*})\big) g_{2}^{(1)}(X_{i}^{\prime} \theta^{*})\}^{2}
			\big\| 
			\leq O(1)\|\theta_{2}-\theta_{1}\|^{2},
			\nonumber
			\eea
			where the second inequality follows from the fact that $\big\| \Lambda \big(g(X_{i}^{\prime} \theta)\big)  \big\| \leq 1$ and \Cref{assumption as:1}(b).
			
			For $A_2$, we have shown:
			\begin{equation}\label{eq:proof2}
				\big|
				\ell(\theta_{1}, g_{2}) - \ell(\theta_{2}, g_{2}) 
				\big|
				\leq 
				O(1) \| \theta_{2} - \theta_{1} \|.
			\end{equation}
			
			Combining \Cref{eq:proof1,eq:proof2}, we obtain:
			\begin{equation*}
				\big|
				\ell(\theta_{1}, g_{1}) - \ell(\theta_{2}, g_{2}) 
				\big| 
				\leq 
				O(1) 
				\big\| (\theta_{1}, g_{1}) - (\theta_{2}, g_{2})
				\big\|_{2},
			\end{equation*}
			which indicates the continuity of $\ell(\theta, g)$. Therefore, Condition (iii) of Lemma A2 of \cite{Newey2003} holds. We then have shown $\max_{\Theta \times G} \big| \ell_{N}(\theta, g) -\ell(\theta, g) \big| \rightarrow_{P} 0$.
			
			\textbf{(b)} We next show the consistency of $(\hat{\theta}, \widehat{g}_{k})$ by contradiction. By definition of $(\hat{\theta}, \widehat{g}_{k})$, regardless of the value of $\lambda$, we have:
			\[
			\ell_{N}(\hat{\theta}, \widehat{g}_{k}) - \ell_{N}(\theta_{0}, g_{0}) > 0.
			\]
			
			By the uniform convergence and the continuity of $\ell(\theta, g)$, if $(\hat{\theta}, \widehat{g}_{k})  \nrightarrow_{P} (\theta_{0}, g_{0})$, we have:
			\[
			\ell_{N}(\theta_{0}, g_{0}) - \ell_{N}(\hat{\theta}, \widehat{g}_{k}) > 0,
			\]
			with probability approaching 1, which contradicts the definition of $(\hat{\theta}, \widehat{g}_{k})$ as the maximizer for the sample version of objective function. 
			
			Therefore, we have $\big\|(\hat{\theta}, \widehat{g}_{k})-(\theta_{0}, g_{0})\big\|_{2} \rightarrow_{P} 0$.
			
		\end{proof}
		
		\begin{proof}[\textbf{Proof of \Cref{lemma:indentification2}}]

			To simplify the notation, we denote the constrained maximization objective function in Equation (\ref{eq:ps4}) with:
			\[
			\mathcal{W}_{N}(\theta, C_{k},\lambda) := 
			\frac{1}{N} \ell_{N} \big(\theta,\left\{c_{j}\right\}_{j=1}^{k}\big)
			+
			\lambda\big(\|\theta\|^{2}-1\big),
			\]
			where $\lambda$ is the Lagrange multiplier. 
			
			We prove the asymptotic normality of $\hat{\theta}$ in two parts. In the first part, we use the routine procedure in \cite{yu2002penalized} to study the objective function $\mathcal{W}_{N}(\theta, C_{k},\lambda) $. In the second part, we analyze the asymptotic property of each function used in this procedure.
			
			%% (1) %%
			\textbf{(1)} 
			In the first part, by the definition of $(\hat{\theta}, \left\{\hat{c}_{j}\right\}_{j=1}^{k})$, we have:
			\be
			\pdv{\mathcal{W}_{N}(\hat{\theta}, \widehat{C}_{k}, \hat{\lambda})}{\theta} = 0, \ \ \pdv{\mathcal{W}_{N}(\hat{\theta}, \widehat{C}_{k}, \hat{\lambda})}{C_{k}} = 0, \ \ \pdv{\mathcal{W}_{N}(\hat{\theta}, \widehat{C}_{k}, \hat{\lambda})}{\lambda} = 0.
			\nonumber
			\ee
			
			By multiplying $\hat{\theta}$ in both sides of $\pdv*{\mathcal{W}_{N}(\hat{\theta}, \widehat{C}_{k}, \hat{\lambda})}{\theta} = 0$, 
			we have:
			\[
			\hat{\lambda}=
			\frac{1}{2N} \sum_{i=1}^{N}
			\big[
			D_{i}
			-
			\Lambda\big(\hat{g}_{k}(X_{i}^{\prime} \hat{\theta})\big)
			\big] 
			\hat{g}_{k}^{(1)}(X_{i}^{\prime} \hat{\theta}) X_{i}^{\prime} \hat{\theta}.
			\]
			
			Multiplying $V^{\prime}$ to both sides of $\pdv*{\mathcal{W}_{N}(\hat{\theta}, \widehat{C}_{k}, \hat{\lambda})}{\theta} = 0$, we have:
			\bea
			&& 0 =
			V^{\prime} \pdv{\mathcal{W}_{N}(\hat{\theta}, \widehat{C}_{k}, \hat{\lambda})}{\theta}
			=
			V^{\prime} \pdv{\mathcal{W}_{N}(\theta_{0}, \widehat{C}_{k}, \hat{\lambda})}{\theta} + 
			V^{\prime} \pdv{\mathcal{W}_{N}(\bar{\theta}, \widehat{C}_{k}, \hat{\lambda})}{\theta}{\theta^{\prime}}  (\hat{\theta} - \theta_{0})
			\nonumber\\
			&&=
			V^{\prime} \pdv{\mathcal{W}_{N}(\theta_{0}, \widehat{C}_{k}, \hat{\lambda})}{\theta} 
			+ 
			V^{\prime} \pdv{\mathcal{W}_{N}(\bar{\theta}, \widehat{C}_{k}, \hat{\lambda})}{\theta}{\theta^{\prime}}  (V V^{\prime}+\theta_{0} \theta_{0}^{\prime}) (\hat{\theta} - \theta_{0})
			\nonumber\\
			&& =
			V^{\prime} \pdv{\mathcal{W}_{N}(\theta_{0}, \widehat{C}_{k}, \hat{\lambda})}{\theta} 
			+ 
			V^{\prime} \pdv{\mathcal{W}_{N}(\bar{\theta}, \widehat{C}_{k}, \hat{\lambda})}{\theta}{\theta^{\prime}} V V^{\prime} (\hat{\theta} - \theta_{0}) - \frac{1}{2} \|\hat{\theta} - \theta_{0} \|^{2} V^{\prime} \pdv{\mathcal{W}_{N}(\bar{\theta}, \widehat{C}_{k}, \hat{\lambda})}{\theta}{\theta^{\prime}} \theta_{0},
			\nonumber
			\eea
			where the second equality follows from Taylor expansion with $\overline{\theta}$ lies between $\theta_{0}$ and $\hat{\theta}$, the third equality follows from $I_{d}=P_{\theta_{0}}+\theta_{0} \theta_{0}^{\prime}=V V^{\prime}+\theta_{0} \theta_{0}^{\prime}$, and the fourth equality follows from $\hat{\theta}^{\prime} \theta_{0}-1 = - \|\hat{\theta}-\theta_{0}\|^{2}+(1-\hat{\theta}^{\prime} \theta_{0})$.
			
			%% (2) %%
			\textbf{(2)}
			In the second part, we analyze the asymptotic property of $\pdv*{\mathcal{W}_{N}(\theta, C_{k},\lambda)}{\theta}{\theta^{\prime}}$ and $\pdv*{\mathcal{W}_{N}(\theta, C_{k}, \lambda)}{\theta}$ used in the above procedure. Similar to the proofs in \Cref{lemma:indentification3}, it is easy to show $\big\|(\bar{\theta}, \widehat{g}_{k})-(\theta_{0}, g_{0})
			\big\|_{2} \rightarrow_{P} 0$. We therefore focus on a sufficiently small neighborhood of $(\theta_{0}, g_{0})$ in the following proof.
			
			In the following proofs, we will show $\pdv{\mathcal{W}_{N}(\bar{\theta}, \widehat{C}_{k}, \hat{\lambda})}{\theta}{\theta^{\prime}} \rightarrow_{P} Q$,
			where $\bar{\theta}$ lies between $\theta_{0}$ and $\hat{\theta}$.  
			
			We are also going to show $\sqrt{N}V^{\prime}\pdv{\mathcal{W}_{N}(\theta_{0}, \widehat{C}_{k}, \hat{\lambda})}{\theta} \rightarrow_{D} N(0, V^{\prime} W V )$.
			
			\textbf{(2.1)}
			We first focus on the function $\pdv*{\mathcal{W}_{N}(\theta, C_{k},\lambda)}{\theta}{\theta^{\prime}}$ and note that:
			\bea
			&& \pdv{\mathcal{W}_{N}(\theta, C_{k},\lambda)}{\theta}{\theta^{\prime}} 
			=
			\frac{1}{N}\sum_{i=1}^{N}
			\big(\Lambda\big(g_{k}(X_{i}^{\prime} \theta)\big) - D_{i} \big) g_{k}^{(2)}(X_{i}^{\prime} \theta) X_{i} X_{i}^{\prime}
			\nonumber\\
			&&+
			\frac{1}{N} \sum_{i=1}^{N}
			\Big( 
			\Lambda\big(g_{k}(X_{i}^{\prime} \theta)\big) 
			- 
			\Lambda\big(g_{k}(X_{i}^{\prime} \theta)\big)^2 
			\Big) 
			\big\{
			g_{k}^{(1)}(X_{i}^{\prime} \theta)
			\big\}^2 X_{i} X_{i}^{\prime} +2 \lambda I_{d}
			=
			K_{1N} + K_{2N} + 2 \lambda I_{d}.
			\nonumber
			\eea
			
			For $K_{1N} $, we write:
			\bea
			&& K_{1N}  =
			\frac{1}{N} \sum_{i=1}^{N} 
			\Big[ \Lambda\big(g_{0}(X_{i}^{\prime} \theta_{0})\big) 
			- 
			D_{i} \Big] 
			g_{k}^{(2)}(X_{i}^{\prime} \theta) X_{i} X_{i}^{\prime}
			\nonumber\\
			&& + 
			\frac{1}{N} \sum_{i=1}^{N} 
			\Big[ \Lambda\big(g_{k}(X_{i}^{\prime} \theta)\big) 
			- 
			\Lambda\big(g_{0}(X_{i}^{\prime} \theta_{0})\big) \Big] g_{k}^{(2)}(X_{i}^{\prime} \theta) X_{i} X_{i}^{\prime} 
			=  K_{11N}  + K_{12N}.
			\nonumber
			\eea                                                                 
			
			Following Lemma A2 in \cite{Dong2018} and \Cref{assumption as:1}(a), we write:
			\[
			\frac{1}{N} \sum_{i=1}^{N} 
			\Big[ \Lambda\big(g_{0}(X_{i}^{\prime} \theta_{0})\big) 
			- 
			D_{i} \Big] 
			g_{k}^{(2)}(X_{i}^{\prime} \theta) X_{i} X_{i}^{\prime}
			=
			\E \Big\{
			\Big[ \Lambda\big(g_{0}(X_{i}^{\prime} \theta_{0})\big) 
			- 
			D_{i} \Big] 
			g_{k}^{(2)}(X_{i}^{\prime} \theta) X_{i} X_{i}^{\prime}
			\Big\}
			+ o_{P}(1).
			\]
			
			Therefore, we have the term $K_{11N}=O_{P}(\frac{1}{\sqrt{N}})$ uniformly. 
			
			For $K_{12N}$, we write:
			\begin{align*} 
				K_{12N}
				=
				&
				\frac{1}{N}\sum_{i=1}^{N} \Lambda\big(g_{k}(X_{i}^{\prime} \theta)\big) g_{k}^{(2)}(X_{i}^{\prime} \theta) X_{i} X_{i}^{\prime} 
				-
				\frac{1}{N}\sum_{i=1}^{N} \Lambda\big(g_{0, k}(X_{i}^{\prime} \theta)\big) g_{k}^{(2)}(X_{i}^{\prime} \theta) X_{i} X_{i}^{\prime}
				\\ 
				&
				+
				\frac{1}{N}\sum_{i=1}^{N} \Lambda\big(g_{0, k}(X_{i}^{\prime} \theta)\big) g_{k}^{(2)}(X_{i}^{\prime} \theta) X_{i} X_{i}^{\prime} 
				-
				\frac{1}{N}\sum_{i=1}^{N} \Lambda\big(g_{0}(X_{i}^{\prime} \theta)\big) g_{k}^{(2)}(X_{i}^{\prime} \theta) X_{i} X_{i}^{\prime}
				\\ 
				&
				+
				\frac{1}{N}\sum_{i=1}^{N} \Lambda\big(g_{0}(X_{i}^{\prime} \theta)\big) g_{k}^{(2)}(X_{i}^{\prime} \theta) X_{i} X_{i}^{\prime} 
				-
				\frac{1}{N}\sum_{i=1}^{N} \Lambda\big(g_{0}(X_{i}^{\prime} \theta_{0})\big) g_{k}^{(2)}(X_{i}^{\prime} \theta) X_{i} X_{i}^{\prime} 
				\\ 
				=
				& K_{121N} + K_{122N} + K_{123N}.
			\end{align*}

			For $K_{121N}$, we have:
			\bea
			&&		\|K_{121N}\|_{(\theta, C_{k})=(\bar{\theta}, \widehat{C}_{k})} 
			=
			\Big\|
			\frac{1}{N} \sum_{i=1}^{N} \Lambda\big( \hat{g}_{k}(X_{i}^{\prime} \bar{\theta})\big) \hat{g}_{k}^{(2)}(X_{i}^{\prime} \bar{\theta}) X_{i} X_{i}^{\prime}
			\nonumber\\
			&&-
			\frac{1}{N} \sum_{i=1}^{N}  \Lambda\big( g_{0, k}(X_{i}^{\prime} \bar{\theta})\big) \hat{g}_{k}^{(2)}(X_{i}^{\prime} \bar{\theta}) X_{i} X_{i}^{\prime}
			\Big\|
			\nonumber\\
			&&
			\leq 
			\Big\| 
			\E 
			\big[ 
			\big\{
			\Lambda
			\big(H(X_{i}^{\prime} \bar{\theta})^{\prime} \widehat{C}_{k}\big)
			-
			\Lambda
			\big(H(X_{i}^{\prime} \bar{\theta})^{\prime} C_{0,k}\big) 
			\big\} 
			g_{k}^{(2)}(X_{i}^{\prime} \bar{\theta}) X_{i} X_{i}^{\prime} 
			\big] 
			\Big\| 
			+o_{P}(1) 
			\nonumber\\
			&& 
			\leq  
			O(1) 
			\Big\| 
			\E
			\big[
			H(X_{i}^{\prime} \bar{\theta})^{\prime} 
			( \widehat{C}_{k}  -  C_{0,k} )
			g_{k}^{(2)}(X_{i}^{\prime} \bar{\theta}) X_{i} X_{i}^{\prime}
			\big]
			\Big\|  +o_{P}(1) 
			\nonumber\\
			&&
			\leq 
			O(1) 
			\Big\{ 
			(\widehat{C}_{k}  -  C_{0,k})^{\prime} 
			\E
			\big[
			H(X_{i}^{\prime} \bar{\theta}) H(X_{i}^{\prime} \bar{\theta})^{\prime}
			\big]
			(\widehat{C}_{k}  -  C_{0,k}) 
			\times
			\E 
			\big\|
			g_{k}^{(2)}(X_{i}^{\prime} \bar{\theta}) X_{i} X_{i}^{\prime}
			\big\|^{2}
			\Big\}^{1 / 2} 
			\nonumber\\ 
			&& \leq O(1)\|\widehat{C}_{k} - C_{0, k} \|+o_{P}(1) 
			\leq O(1)\|\hat{g}_{k}-g_{0}\|_{\mathcal{L}^2}+o_{P}(1)=o_{P}(1),       
			\nonumber                                              
			\eea
			where the first inequality follows from \Cref{assumption as:2}(a), the second inequality follows from Lipschitz continuity of $\Lambda\big(g(X_{i}^{\prime} \theta)\big) $, the third inequality follows from Cauchy-Schwarz inequality, the fourth inequality follows from Lemma A4 in \cite{Dong2018}, and the last inequality follows from the definition of $\|\cdot\|_{\mathcal{L}^2}$.
			
			For $K_{122N}$, we write:
			\bea
			&&		K_{122N}
			\|_{(\theta, C_{k})=(\bar{\theta}, \widehat{C}_{k})} 
			=
			\Big\|
			\frac{1}{N} \sum_{i=1}^{N} \Lambda\big(g_{0, k}(X_{i}^{\prime} \bar{\theta})\big) \hat{g}_{k}^{(2)}(X_{i}^{\prime} \theta) X_{i} X_{i}^{\prime}
			\nonumber\\
			&& -
			\frac{1}{N} \sum_{i=1}^{N} \Lambda\big(g_{0}(X_{i}^{\prime} \bar{\theta})\big) \hat{g}_{k}^{(2)}(X_{i}^{\prime} \bar{\theta}) X_{i} X_{i}^{\prime}
			\Big\|
			\nonumber\\
			&& 
			\leq 
			\Big\|
			\E
			\big[
			\Lambda\big(g_{0, k}(X_{i}^{\prime} \bar{\theta})\big) \hat{g}_{k}^{(2)}(X_{i}^{\prime} \bar{\theta}) X_{i} X_{i}^{\prime}
			-
			\Lambda\big(g_{0}(X_{i}^{\prime} \bar{\theta})\big) \hat{g}_{k}^{(2)}(X_{i}^{\prime} \bar{\theta}) X_{i} X_{i}^{\prime}
			\big]
			\Big\| 
			+ o_{P}(1)
			\nonumber\\
			&& 
			\leq  O(1) 
			\Big\| 
			\E
			\big[\epsilon_{0, k}(X_{i}^{\prime} \bar{\theta}) \hat{g}_{k}^{(2)}(X_{i}^{\prime} \bar{\theta}) X_{i} X_{i}^{\prime}
			\big]
			\Big\| + o_{P}(1)
			\nonumber\\
			&& 
			\leq O(1)  
			\Big\{ 
			\E 
			\big\| 
			\epsilon_{0, k}(X_{i}^{\prime} \bar{\theta}) 
			\big\|^{2} 
			\E 
			\big\| 
			\hat{g}_{k}^{(2)}(X_{i}^{\prime} \bar{\theta}) X_{i} X_{i}^{\prime}
			\big\|^{2}
			\Big\}^{1 / 2} + o_{P}(1) 
			= o_{P}(1),
			\nonumber
			\eea
			where the first inequality follows from \Cref{assumption as:2}(a), the second inequality follows from Lipschitz continuity of $\Lambda(g(X_{i}^{\prime} \theta)) $, the third inequality follows from Cauchy-Schwarz inequality, and the last equality follows from \Cref{assumption as:2}(c) and Lemma A1 from \cite{Dong2018}
			
			For $K_{123N}$, we write:
			\bea
			&&\|K_{123N}\|_{(\theta, C_{k})=(\bar{\theta}, \widehat{C}_{k})} 
			=
			\Big
			\|
			\frac{1}{N} \sum_{i=1}^{N}  \Lambda\big(g_{0}(X_{i}^{\prime} \bar{\theta})\big) \hat{g}_{k}^{(2)}(X_{i}^{\prime} \bar{\theta}) X_{i} X_{i}^{\prime}
			\nonumber\\
			&& -
			\frac{1}{N} \sum_{i=1}^{N}  \Lambda\big(g_{0}(X_{i}^{\prime} \theta_{0})\big) 
			\hat{g}_{k}^{(2)}(X_{i}^{\prime} \bar{\theta}) X_{i} X_{i}^{\prime}
			\Big\|
			\nonumber\\
			&&
			\leq 
			\Big\|
			\E
			\big[
			\Lambda\big(g_{0}(X_{i}^{\prime} \bar{\theta})\big) \hat{g}_{k}^{(2)}(X_{i}^{\prime} \bar{\theta}) X_{i} X_{i}^{\prime}
			-
			\Lambda\big(g_{0}(X_{i}^{\prime} \theta_{0})\big) \hat{g}_{k}^{(2)}(X_{i}^{\prime} \bar{\theta}) X_{i} X_{i}^{\prime}
			\big]
			\Big\|
			+o_{P}(1)
			\nonumber\\
			&& 
			= 
			\Big\|
			\E
			\big[ 
			\Lambda\big(g_{0}(X_{i}^{\prime} \theta^{*})\big) g_{0}^{(1)}(X_{i}^{\prime} \theta^{*})(X_{i}^{\prime} \bar{\theta}
			-
			X_{i}^{\prime} \theta_{0}) \hat{g}_{k}^{(2)}(X_{i}^{\prime} \bar{\theta}) X_{i} X_{i}^{\prime}
			\big]
			\Big\| 
			+o_{P}(1) 
			\nonumber\\
			&& 
			\leq
			\|
			\bar{\theta}
			-
			\theta_{0} 
			\|
			\Big\{
			\E
			\big\| 
			\Lambda\big(g_{0}(X_{i}^{\prime} \theta^{*})\big) g_{0}^{(1)}(X_{i}^{\prime} \theta^{*}) X_{i} X_{i}^{\prime}
			\big\|^{2} 
			\E
			\big\|
			\hat{g}_{k}^{(2)}(X_{i}^{\prime} \bar{\theta}) X_{i}
			\big\|^{2}
			\Big\}^{1 / 2}
			+o_{P}(1) 
			= o_{P}(1),    
			\nonumber                                          
			\eea
			where $\theta^{*}$ lies between $\theta$ and $\theta_{0},$ the first inequality follows from \Cref{assumption as:2}(a), the second equality follows from mean value theorem, the third inequality follows from Cauchy-Schwarz inequality, and the last equality follows from \Cref{assumption as:1}(b) and the fact that $\|\bar{\theta} - \theta_{0} \| \rightarrow_{P} 0$ and $\Lambda\big(g(X_{i}^{\prime}\theta)\big) $ is bounded by 1.

			For $K_{2N}$, we write:
			\bea
			&&	K_{2N} = \frac{1}{N} \sum_{i=1}^{N}  \upsilon\big(g_{k}(X_{i}^{\prime} \theta)\big) g_{k}^{(1)}(X_{i}^{\prime} \theta)^{2} X_{i} X_{i}^{\prime} =
			\frac{1}{N} \sum_{i=1}^{N} \upsilon\big(g_{0}(X_{i}^{\prime} \theta_{0})\big)g_{0}^{(1)}(X_{i}^{\prime} \theta_{0})^{2} X_{i} X_{i}^{\prime} 
			\nonumber\\
			&& 
			+
			\frac{1}{N} \sum_{i=1}^{N} \upsilon\big(g_{k}(X_{i}^{\prime} \theta)\big)g_{k}^{(1)}(X_{i}^{\prime} \theta)^{2} X_{i} X_{i}^{\prime}
			-
			\frac{1}{N} \sum_{i=1}^{N} \upsilon(g_{0, k}(X_{i}^{\prime} \theta))g_{0, k}^{(1)}(X_{i}^{\prime} \theta)^{2} X_{i} X_{i}^{\prime} 
			\nonumber\\ 
			&& +\frac{1}{N} \sum_{i=1}^{N} \upsilon\big(g_{0, k}(X_{i}^{\prime} \theta)\big)g_{0, k}^{(1)}(X_{i}^{\prime} \theta)^{2} X_{i} X_{i}^{\prime}-\frac{1}{N} \sum_{i=1}^{N} \upsilon\big(g_{0}(X_{i}^{\prime} \theta)\big)g_{0}^{(1)}(X_{i}^{\prime} \theta)^{2} X_{i} X_{i}^{\prime}  
			\nonumber\\
			&& +\frac{1}{N} \sum_{i=1}^{N} \upsilon\big(g_{0}(X_{i}^{\prime} \theta)\big) g_{0}^{(1)}(X_{i}^{\prime} \theta)^{2} X_{i} X_{i}^{\prime}-\frac{1}{N} \sum_{i=1}^{N} \upsilon\big(g_{0}(X_{i}^{\prime} \theta_{0})\big) g_{0}^{(1)}(X_{i}^{\prime} \theta_{0})^{2} X_{i} X_{i}^{\prime}
			\nonumber\\
			&& = 
			\frac{1}{N} \sum_{i=1}^{N} \upsilon\big(g_{0}(X_{i}^{\prime} \theta_{0})\big)g_{0}^{(1)}(X_{i}^{\prime} \theta_{0})^{2} X_{i} X_{i}^{\prime}+K_{21N}+K_{22N}+K_{23N}. 
			\nonumber
			\eea
			
			By \Cref{assumption as:2}, we have:
			\begin{align*}
				\frac{1}{N} \sum_{i=1}^{N} \upsilon\big(g_{0}(X_{i}^{\prime} \theta_{0})\big)g_{0}^{(1)}(X_{i}^{\prime} \theta_{0})^{2} X_{i} X_{i}^{\prime}  
				&
				= 
				\E
				\Big[
				\upsilon\big(g_{0}(X_{i}^{\prime} \theta_{0})\big)g_{0}^{(1)}(X_{i}^{\prime} \theta_{0})^{2} X_{i} X_{i}^{\prime}
				\Big] + o_{P}(1) 
				:= Q + o_{P}(1).
			\end{align*}
			
			For $K_{21N},$ we write:
			\bea
			&& \|K_{21N}\|_{(\theta, C_{k})=(\bar{\theta}, \widehat{C}_{k})} 
			=
			\Big\|
			\frac{1}{N} \sum_{i=1}^{N} 
			\upsilon\big(g_{k}(X_{i}^{\prime} \bar{\theta})\big) 
			\big(
			\dot{H}(X_{i}^{\prime} \bar{\theta})^{\prime} \widehat{C}_{k}
			\big)^{2} 
			X_{i} X_{i}^{\prime} 
			\nonumber\\
			&& -
			\frac{1}{N} \sum_{i=1}^{N} \upsilon\big(g_{0, k}(X_{i}^{\prime} \bar{\theta})\big) \big(\dot{H}(X_{i}^{\prime} \theta)^{\prime} C_{0, k}\big)^{2} 
			X_{i} X_{i}^{\prime} 	
			\Big\|
			\nonumber\\ 
			&& \leq 
			O(1) 
			\Big\| 
			\E
			\big[
			\big(\dot{H}(X_{i}^{\prime} \bar{\theta})^{\prime} \widehat{C}_{k}\big)^{2} X_{i} X_{i}^{\prime}
			-
			\big(\dot{H}(X_{i}^{\prime} \bar{\theta})^{\prime} C_{0, k}\big)^{2} X_{i} X_{i}^{\prime}
			\big]
			\Big\| +o_{P}(1) 
			\nonumber\\
			&& =O(1)
			\Big\|
			\E
			\big[
			(\widehat{C}_{k} - C_{0, k})^{\prime} \dot{H}(X_{i}^{\prime} \bar{\theta}) \dot{H}(X_{i}^{\prime} \bar{\theta})^{\prime}(\widehat{C}_{k} + C_{0, k}) X_{i} X_{i}^{\prime}
			\big]
			\Big\| +o_{P}(1) 
			\nonumber\\ 
			&& \leq 
			O(1)
			\Big\{
			\E
			\big\|
			(\widehat{C}_{k} - C_{0, k})^{\prime} \dot{H}(X_{i}^{\prime} \bar{\theta})
			\big\|^{2} 
			\E
			\big\|\dot{H}(X_{i}^{\prime} \bar{\theta})^{\prime}(\widehat{C}_{k}+C_{0, k}) X_{i} X_{i}^{\prime}
			\big\|^{2}
			\Big\}^{1 / 2}
			+o_{P}(1) 
			\nonumber\\ 
			&& \leq 
			O(1) 
			\Big\{
			(\widehat{C}_{k} - C_{0, k})^{\prime} 
			\E
			\big[\dot{H}(X_{i}^{\prime} \theta) \dot{H}(X_{i}^{\prime} \bar{\theta})^{\prime}
			\big](\widehat{C}_{k} - C_{0, k})
			\Big\}^{1 / 2} 
			\nonumber\\ 
			&&
			\cdot
			\Big\{
			2 \E\big\|g_{k}^{(1)}(X_{i}^{\prime} \bar{\theta}) X_{i} X_{i}^{\prime}\big\|^{2}
			+
			2 \E\big\|g_{0, k}^{(1)}(X_{i}^{\prime} \bar{\theta}) X_{i} X_{i}^{\prime}\big\|^{2}
			\Big\}^{1 / 2} +o_{P}(1) 
			\nonumber\\ 
			&& \leq  O(1)\|\widehat{C}_{k}-C_{0, k}\|+o_{P}(1) 
			\leq   O(1)\|\hat{g}_{k} - g_{0}\|_{\mathcal{L}^2}+o_{P}(1) = o_{P}(1), 
			\nonumber
			\eea
			where the first inequality follows from \Cref{assumption as:2}(a) and the boundedness of function $\upsilon(\cdot)$, the second inequality follows from Cauchy-Schwarz inequality, the
			fourth inequality follows from \Cref{assumption as:1}(b) and \Cref{assumption as:2}(b), and the last equality follows from the definition of $\|\cdot\|_{\mathcal{L}^2}$ and \Cref{lemma:indentification3}.
			
			For $K_{22N}$, we write:
			\begin{align*}
				\|K_{22N}\|_{\theta=\bar{\theta}}  
				= 
				&
				\Big\|
				\frac{1}{N} \sum_{i=1}^{N} \upsilon\big(g_{0, k}(X_{i}^{\prime} \bar{\theta})\big) g_{0, k}^{(1)}(X_{i}^{\prime} \bar{\theta})^{2} X_{i} X_{i}^{\prime}-\frac{1}{N} \sum_{i=1}^{N} \upsilon\big(g_{0}(X_{i}^{\prime} \bar{\theta})\big) g_{0}^{(1)}(X_{i}^{\prime} \bar{\theta})^{2} X_{i} X_{i}^{\prime}
				\Big\|
				\\
				\leq 
				& 
				O(1) 
				\Big\| 
				\E
				\big[
				\{ g_{0, k}^{(1)}(X_{i}^{\prime} \bar{\theta})^{2}
				-
				g_{0}^{(1)}(X_{i}^{\prime} \bar{\theta})^{2}\} X_{i} X_{i}^{\prime}
				\big]
				\Big\| + o_{P}(1) 
				\\
				=
				&
				O(1)
				\Big\| 
				\E
				\big[
				\big(g_{0, k}^{(1)}(X_{i}^{\prime} \bar{\theta})
				-
				g_{0}^{(1)}(X_{i}^{\prime} \bar{\theta})\big)\big(g_{0, k}^{(1)}(X_{i}^{\prime} \bar{\theta})+g_{0}^{(1)}(X_{i}^{\prime} \bar{\theta})\big) X_{i} X_{i}^{\prime}
				\big]
				\Big\|  +o_{P}(1) 
				\\
				=
				&
				O(1)
				\Big\|
				\E
				\big[
				\epsilon_{0, k}^{(1)}(X_{i}^{\prime} \bar{\theta})\big(g_{0, k}^{(1)}(X_{i}^{\prime} \bar{\theta})
				+
				g_{0}^{(1)}(X_{i}^{\prime} \bar{\theta})\big) X_{i} X_{i}^{\prime}
				\big]
				\Big\|+o_{P}(1) 
				\\ 
				\leq 
				& 
				O(1) 
				\E 
				\big\|
				\epsilon_{0, k}^{(1)}(X_{i}^{\prime} \bar{\theta}) g_{0, k}^{(1)}(X_{i}^{\prime} \bar{\theta}) X_{i} X_{i}^{\prime}
				\big\| 
				+
				O(1) 
				\E 
				\big\|
				\epsilon_{0, k}^{(1)}(X_{i}^{\prime} \bar{\theta}) g_{0}^{(1)}(X_{i}^{\prime} \bar{\theta}) X_{i} X_{i}^{\prime}
				\big\| +o_{P}(1) 
				\\ 
				\leq 
				&
				O(1)
				\Big\{
				\E 
				\big\|
				\epsilon_{0, k}^{(1)}(X_{i}^{\prime} \bar{\theta}) 
				\big\|^{2} 
				\E 
				\big\|
				g_{0, k}^{(1)}(X_{i}^{\prime} \bar{\theta}) X_{i} X_{i}^{\prime}
				\big\|^{2}
				\Big\}^{1/ 2}
				\\ 
				&
				+
				O(1)
				\Big\{
				\E 
				\big\|
				\epsilon_{0, k}^{(1)}(X_{i}^{\prime} \bar{\theta})
				\big\|^{2} 
				\E 
				\big\|
				g_{0}^{(1)}(X_{i}^{\prime} \bar{\theta}) X_{i} X_{i}^{\prime}
				\big\|^{2}
				\Big\}^{1/ 2} + o_{P}(1)
				= o_{P}(1), 
			\end{align*}
			where the first inequality follows from the boundedness of function $\upsilon(\cdot)$ and \Cref{assumption as:2}(a), the third inequality
			follows from Cauchy-Schwarz inequality, and the last equality follows from \Cref{assumption as:1}(b) and Lemma A1 in \cite{Dong2018}.
			
			For $K_{23N}$, we write:
			\begin{align*}
				\|K_{23N}\|_{\theta=\bar{\theta}} 
				=
				& 
				\Big\|
				\frac{1}{N} \sum_{i=1}^{N} \upsilon\big(g_{0}(X_{i}^{\prime} \bar{\theta})\big) g_{0}^{(1)}(X_{i}^{\prime} \bar{\theta})^{2} X_{i} X_{i}^{\prime}
				-
				\frac{1}{N} \sum_{i=1}^{N}\upsilon\big(g_{0}(X_{i}^{\prime} \theta_{0})\big)g_{0}^{(1)}(X_{i}^{\prime} \theta_{0})^{2} X_{i} X_{i}^{\prime}
				\Big\| 
				\\ 
				\leq 
				& 
				O(1)
				\Big\|
				\E
				\big[g_{0}^{(1)} (X_{i}^{\prime} \bar{\theta})^{2} X_{i} X_{i}^{\prime}
				-
				g_{0}^{(1)} (X_{i}^{\prime} \theta_{0})^{2} X_{i} X_{i}^{\prime}
				\big]
				\Big\|+o_{P}(1) \\
				=
				&
				O(1)
				\Big\|
				\E
				\big[g_{0}^{(2)}(X_{i}^{\prime} \theta^{*})(X_{i}^{\prime} \bar{\theta}-X_{i}^{\prime} \theta_{0})\big(g_{0}^{(1)}(X_{i}^{\prime} \bar{\theta})+g_{0}^{(1)}(X_{i}^{\prime} \theta_{0})\big) X_{i} X_{i}^{\prime}
				\big]
				\Big\| +o_{P}(1)
				\\
				\leq 
				&
				O(1)
				\|\bar{\theta} - \theta_{0} \| 
				\E
				\Big[
				\big\|g_{0}^{(2)}(X_{i}^{\prime} \theta^{*}) X_{i}\big\|  \big(g_{0}^{(1)}(X_{i}^{\prime} \bar{\theta})+g_{0}^{(1)}(X_{i}^{\prime} \theta_{0}) \big) X_{i} X_{i}^{\prime}
				\Big] 
				+o_{P}(1) 
				\\ 
				\leq
				& 
				O(1)\|\bar{\theta}-\theta_{0}\|
				\Big\{ 
				\E
				\big\|g_{0}^{(2)}(X_{i}^{\prime} \theta^{*}) X_{i}
				\big\|^{2} 
				\E
				\big\|g_{0}^{(1)}(X_{i}^{\prime} \bar{\theta}) X_{i} X_{i}^{\prime}
				\big\|^{2}
				\Big\}^{1 / 2}\\
				+ &
				O(1)
				\|\bar{\theta}-\theta_{0}\|
				\Big\{ 
				\E\big\|g_{0}^{(2)}(X_{i}^{\prime} \theta^{*}) X_{i}\big\|^{2} \E\big\|g_{0}^{(1)}(X_{i}^{\prime} \theta_{0}) X_{i} X_{i}^{\prime}\big\|^{2}
				\Big\}^{1 / 2} +o_{P}(1) 
				= 
				o_{P}(1),
			\end{align*}
			where $\theta^{*}$ lies between $\theta$ and $\theta_{0}$, the first inequality follows from the boundedness of function $\upsilon(\cdot)$ and \Cref{assumption as:2}(a), the second equality follows from mean value theorem, the third inequality follows from triangular inequality and Cauchy-Schwarz inequality, and the last equality follows from \Cref{assumption as:1}(b) and $\|\bar{\theta}-\theta_{0}\| \rightarrow_{P} 0$.

			For $\hat{\lambda}$, we write $K_{3N}$:
			\begin{align*}
				K_{3N} 
				= 
				& 
				\frac{1}{N} \sum_{i=1}^{N} \Big[
				D_{i} 
				-
				\Lambda\big(g_{0}(X_{i}^{\prime} \theta_{0})\big)
				\Big] 
				g_{k}^{(1)}(X_{i}^{\prime} \theta) X_{i}^{\prime}\theta
				\\
				+ &
				\frac{1}{N} \sum_{i=1}^{N} \Big[ 
				\Lambda\big(g_{0}(X_{i}^{\prime} \theta_{0})\big)  
				- 
				\Lambda\big(g_{k}(X_{i}^{\prime} \theta)\big) \Big] 
				g_{k}^{(1)}(X_{i}^{\prime} \theta) X_{i}^{\prime}\theta
				=
				K_{31N}  + K_{32N} .                                                                   
			\end{align*}
			
			Similar to the proofs for the term $K_{11N}$, $K_{31N}=O_{P}(\frac{1}{\sqrt{N}})$ uniformly. 
			
			For $K_{32N}$, we write:
			\begin{align*} 
				K_{32N}
				=
				&
				\frac{1}{N}\sum_{i=1}^{N} \Lambda\big(g_{0}(X_{i}^{\prime} \theta_{0})\big) g_{k}^{(1)}(X_{i}^{\prime} \theta) X_{i}^{\prime} \theta
				-
				\frac{1}{N}\sum_{i=1}^{N} \Lambda\big(g_{0}(X_{i}^{\prime} \theta)\big) g_{k}^{(1)}(X_{i}^{\prime} \theta) X_{i}^{\prime} \theta		
				\\ 
				&
				+
				\frac{1}{N}\sum_{i=1}^{N} \Lambda\big(g_{0}(X_{i}^{\prime} \theta)\big) g_{k}^{(1)}(X_{i}^{\prime} \theta) X_{i}^{\prime} \theta
				-
				\frac{1}{N}\sum_{i=1}^{N} \Lambda\big(g_{0, k}(X_{i}^{\prime} \theta)\big) g_{k}^{(1)}(X_{i}^{\prime} \theta) X_{i}^{\prime} \theta			
				\\ 
				&
				+
				\frac{1}{N}\sum_{i=1}^{N} \Lambda\big(g_{0, k}(X_{i}^{\prime} \theta)\big) g_{k}^{(1)}(X_{i}^{\prime} \theta) X_{i}^{\prime} \theta
				-
				\frac{1}{N}\sum_{i=1}^{N} \Lambda\big(g_{k}(X_{i}^{\prime} \theta)\big) g_{k}^{(1)}(X_{i}^{\prime} \theta) X_{i}^{\prime} \theta	
				\\ 
				=
				& K_{321N} + K_{322N} + K_{323N}.
			\end{align*}
			
			For $K_{321N}$, we write:
			\bea
			&&			\|K_{321N}\|_{(\theta, C_{k})=(\hat{\theta}, \widehat{C}_{k})} 
			=
			\Big\|
			\frac{1}{N} \sum_{i=1}^{N}  \Lambda\big(g_{0}(X_{i}^{\prime} \theta_{0})\big) 
			\hat{g}_{k}^{(1)}(X_{i}^{\prime} \hat{\theta}) X_{i}^{\prime} \hat{\theta}			
			-
			\frac{1}{N} \sum_{i=1}^{N}
			\Lambda\big(g_{0}(X_{i}^{\prime} \hat{\theta})\big) \hat{g}_{k}^{(1)}(X_{i}^{\prime} \hat{\theta}) X_{i}^{\prime} \hat{\theta}
			\Big\|
			\nonumber\\
			&&
			\leq 
			\Big\|
			\E
			\big[
			\Lambda\big(g_{0}(X_{i}^{\prime} \theta_{0})\big) \hat{g}_{k}^{(1)}(X_{i}^{\prime} \hat{\theta}) X_{i}^{\prime} \hat{\theta}
			-
			\Lambda\big(g_{0}(X_{i}^{\prime} \hat{\theta})\big) \hat{g}_{k}^{(1)}(X_{i}^{\prime} \hat{\theta}) X_{i}^{\prime} \hat{\theta}
			\big]
			\Big\|
			+o_{P}(1)
			\nonumber\\
			&&= 
			\Big\|
			\E
			\big[ 
			\Lambda\big(g_{0}(X_{i}^{\prime} \theta^{*})\big) g_{0}^{(1)}(X_{i}^{\prime} \theta^{*})
			(
			X_{i}^{\prime} \theta_{0}
			-
			X_{i}^{\prime} \hat{\theta}
			) 
			\hat{g}_{k}^{(1)}(X_{i}^{\prime} \hat{\theta}) X_{i}^{\prime} \hat{\theta}
			\big]
			\Big\| 
			+o_{P}(1) 
			\nonumber\\
			&& 
			\leq
			\|
			\hat{\theta}
			-
			\theta_{0} 
			\|
			\Big\{
			\E
			\big\| 
			\Lambda\big(g_{0}(X_{i}^{\prime} \theta^{*})\big) g_{0}^{(1)}(X_{i}^{\prime} \theta^{*}) X_{i} X_{i}^{\prime}
			\big\|^{2} 
			\E
			\big\|
			\hat{g}_{k}^{(1)}(X_{i}^{\prime} \hat{\theta}) \hat{\theta}
			\big\|^{2}
			\Big\}^{1 / 2}
			+o_{P}(1) 
			= o_{P}(1),     
			\nonumber
			\eea                                         
			where $\theta^{*}$ lies between $\hat{\theta}$ and $\theta_{0},$ the first inequality follows from \Cref{assumption as:2}(a), the second equality follows from mean value theorem, the third inequality follows from Cauchy-Schwarz inequality, and the last equality follows from \Cref{assumption as:1}(b) and the fact that $\|\hat{\theta} - \theta_{0} \| \rightarrow_{P} 0$ and $\Lambda\big(g(X_{i}^{\prime}\theta)\big) $ is bounded by 1.
			
			For $K_{322N}$, we write:
			\bea
			&& \|
			K_{322N}
			\|_{(\theta, C_{k})=(\hat{\theta}, \widehat{C}_{k})} 
			=
			\Big\|
			\frac{1}{N} \sum_{i=1}^{N} \Lambda\big(g_{0}(X_{i}^{\prime} \hat{\theta})\big) \hat{g}_{k}^{(1)}(X_{i}^{\prime} \hat{\theta}) X_{i}^{\prime} \hat{\theta}
			-
			\frac{1}{N} \sum_{i=1}^{N} \Lambda\big(g_{0, k}(X_{i}^{\prime} \hat{\theta})\big) \hat{g}_{k}^{(1)}(X_{i}^{\prime} \theta) 
			X_{i}^{\prime} \hat{\theta}
			\Big\|
			\nonumber\\
			&& 
			\leq 
			\Big\|
			\E
			\big[
			\Lambda\big(g_{0}(X_{i}^{\prime} \hat{\theta})\big) \hat{g}_{k}^{(1)}(X_{i}^{\prime} \hat{\theta}) X_{i}^{\prime} \hat{\theta}
			-
			\Lambda\big(g_{0, k}(X_{i}^{\prime} \hat{\theta})\big) \hat{g}_{k}^{(1)}(X_{i}^{\prime} \hat{\theta}) X_{i}^{\prime} \hat{\theta}
			\big]
			\Big\| 
			+ o_{P}(1)			
			\nonumber\\
			&& 
			\leq  O(1) 
			\Big\| 
			\E
			\big[\epsilon_{0, k}(X_{i}^{\prime} \hat{\theta}) \hat{g}_{k}^{(1)}(X_{i}^{\prime} \hat{\theta}) X_{i}^{\prime} \hat{\theta}
			\big]
			\Big\| 
			+ o_{P}(1)
			\nonumber\\
			&& 
			\leq O(1)  
			\Big\{ 
			\E 
			\big\| \epsilon_{0, k}(X_{i}^{\prime} \hat{\theta}) \big\|^{2} 
			\E 
			\big\| \hat{g}_{k}^{(1)}(X_{i}^{\prime} \hat{\theta}) X_{i}^{\prime} \hat{\theta}
			\big\|^{2}
			\Big\}^{1 / 2} + o_{P}(1) 
			= o_{P}(1),
			\nonumber
			\eea
			where the first inequality follows from \Cref{assumption as:2}(a), the second inequality follows from Lipschitz continuity of $\Lambda(g(X_{i}^{\prime} \theta)) $, the third inequality follows from Cauchy-Schwarz inequality, and the last equality follows from \Cref{assumption as:2}(c) and Lemma A1 from \cite{Dong2018}.

			For $K_{323N}$, we have:
			\bea
			&&\|K_{323N}\|_{(\theta, C_{k})=(\hat{\theta}, \widehat{C}_{k})} =
			\Big
			\|
			\frac{1}{N} \sum_{i=1}^{N}  \Lambda\big( g_{0, k}(X_{i}^{\prime} \hat{\theta})\big) \hat{g}_{k}^{(1)}(X_{i}^{\prime} \hat{\theta}) X_{i}^{\prime} \hat{\theta}
			-
			\frac{1}{N} \sum_{i=1}^{N} \Lambda\big( \hat{g}_{k}(X_{i}^{\prime} \hat{\theta})\big) \hat{g}_{k}^{(1)}(X_{i}^{\prime} \hat{\theta}) X_{i}^{\prime} \hat{\theta}
			\Big\|
			\nonumber\\
			&& 
			\leq 
			\Big\| 
			\E 
			\big[ 
			\big\{
			\Lambda\big(H(X_{i}^{\prime} \hat{\theta})^{\prime} C_{0,k}\big) 
			-
			\Lambda\big(H(X_{i}^{\prime} \hat{\theta})^{\prime} \widehat{C}_{k}\big)			
			\big\} 
			g_{k}^{(1)}(X_{i}^{\prime} \hat{\theta}) X_{i}^{\prime} \hat{\theta} 
			\big] 
			\Big\| +o_{P}(1) 
			\nonumber\\
			&& 
			\leq  
			O(1) 
			\Big\| 
			\E
			\big[
			H(X_{i}^{\prime} \hat{\theta})^{\prime} 
			( \widehat{C}_{k}  -  C_{0,k} )
			g_{k}^{(1)}(X_{i}^{\prime} \hat{\theta}) X_{i}^{\prime} \hat{\theta}
			\big]
			\Big\|  +o_{P}(1) 
			\nonumber\\
			&&
			\leq 
			O(1) 
			\Big\{ 
			(\widehat{C}_{k}  -  C_{0,k})^{\prime} 
			\E
			\big[
			H(X_{i}^{\prime} \hat{\theta}) H(X_{i}^{\prime} \hat{\theta})^{\prime}
			\big]
			(\widehat{C}_{k}  -  C_{0,k}) 
			\E 
			\big\|
			g_{k}^{(1)}(X_{i}^{\prime} \hat{\theta}) X_{i}^{\prime} \hat{\theta}
			\big\|^{2}
			\Big\}^{1 / 2} 
			\nonumber\\ 
			&& \leq O(1)\|\widehat{C}_{k} - C_{0, k} \|+o_{P}(1) 
			\leq O(1)\|\hat{g}_{k}-g_{0}\|_{\mathcal{L}^2}+o_{P}(1)=o_{P}(1),    
			\nonumber                                                 
			\eea
			where the first inequality follows from \Cref{assumption as:2}(a), the second inequality follows from Lipschitz continuity of $\Lambda\big(g(X_{i}^{\prime} \theta)\big) $, the third inequality follows from Cauchy-Schwarz inequality, the fourth inequality follows from Lemma A4 in \cite{Dong2018}, and the last inequality follows from the definition of $\|\cdot\|_{\mathcal{L}^2}$.

			Combing the above results, we conclude that: $\pdv{\mathcal{W}_{N}(\bar{\theta}, \widehat{C}_{k}, \hat{\lambda})}{\theta}{\theta^{\prime}} \rightarrow_{P} Q$.
			
			\textbf{(2.2)}
			We next focus on the function $\pdv*{\mathcal{W}_{N}(\theta, C_{k}, \lambda)}{\theta}$ and note that:
			\begin{align*}
				\sqrt{N}\pdv{\mathcal{W}_{N}(\theta_{0}, \widehat{C}_{k}, \hat{\lambda})}{\theta}  
				=
				&
				-\frac{1}{\sqrt{N}} 
				\sum_{i=1}^{N} 
				\Big[D_{i}-\Lambda\big(\hat{g}_{k}(X_{i}^{\prime} \theta_{0})\big)\Big] \hat{g}_{k}^{(1)}(X_{i}^{\prime} \theta_{0}) X_{i}+2 \sqrt{N} \hat{\lambda} \theta_{0} 
				\\
				=
				&
				(-I_{d}+\theta_{0} \hat{\theta}^{\prime}) \frac{1}{\sqrt{N}} \sum_{i=1}^{N} \Big[D_{i}-\Lambda\big(g_{0}(X_{i}^{\prime} \theta_{0})\big)\Big] g_{0}^{(1)}(X_{i}^{\prime} \theta_{0}) X_{i}  
				\\
				&
				-\frac{1}{\sqrt{N}} \sum_{i=1}^{N} \Big[D_{i}-\Lambda\big(g_{0}(X_{i}^{\prime} \theta_{0})\big)\Big] \big(\hat{g}_{k}^{(1)}(X_{i}^{\prime} \theta_{0})-g_{0, k}^{(1)}(X_{i}^{\prime} \theta_{0})\big) X_{i}
				\\
				&+\frac{1}{\sqrt{N}} \sum_{i=1}^{N} \Big[D_{i}-\Lambda\big(g_{0}(X_{i}^{\prime} \theta_{0})\big)\Big] \epsilon_{0, k}^{(1)}(X_{i}^{\prime} \theta_{0}) X_{i} 
				\\ 
				&-\frac{1}{\sqrt{N}} \sum_{i=1}^{N}\Big[\Lambda\big(g_{0}(X_{i}^{\prime} \theta_{0})\big) - \Lambda\big(\hat{g}_{k}(X_{i}^{\prime} \theta_{0})\big)\Big] \hat{g}_{k}^{(1)}(X_{i}^{\prime} \theta_{0}) X_{i}
				\\
				&
				+
				\theta_{0} \hat{\theta}^{\prime} \frac{1}{\sqrt{N}} \sum_{i=1}^{N} \Big[D_{i}-\Lambda\big(g_{0}(X_{i}^{\prime} \theta_{0})\big)\Big] 
				\big( 
				\hat{g}_{k}^{(1)}(X_{i}^{\prime} \hat{\theta})-g_{0}^{(1)}(X_{i}^{\prime} \theta_{0})
				\big) 
				X_{i}
				\\
				&
				+\theta_{0} \hat{\theta}^{\prime} \frac{1}{\sqrt{N}} \sum_{i=1}^{N}\Big[ \Lambda\big(g_{0}(X_{i}^{\prime} \theta_{0})\big)- \Lambda\big(\hat{g}_{k}(X_{i}^{\prime} \hat{\theta})\big)\Big] \hat{g}_{k}^{(1)}(X_{i}^{\prime} \hat{\theta}) X_{i}
				\\
				=
				& (-I_{d}+\theta_{0} \hat{\theta}^{\prime}) J_{1 N}- J_{2 N}+ J_{3 N}-  J_{4 N} + \theta_{0} \hat{\theta}^{\prime} J_{5 N}+ \theta_{0} \hat{\theta}^{\prime} J_{6 N}.
			\end{align*}
			
			For $J_{1N}$, denote the term $\Big[D_{i}-\Lambda\big(g_{0}(X_{i}^{\prime} \theta_{0})\big)\Big] g_{0}^{(1)}(X_{i}^{\prime} \theta_{0}) X_{i}$ as $Z_{i}$. By \Cref{assumption as:1}(a), we have $\E [Z_{i}] = 0$ and $\operatorname{Var} [Z_{i}]=W<\infty $. Applying the Lindeberg--Levy CLT, we have
			\begin{equation*}
				J_{1 N} \rightarrow_{D} N(0, W).
			\end{equation*}

			%% complex terms: J_{2N} J_{5N} %%
			
			For $J_{2 N}$, we write:
			\bea
			&& \|J_{2 N} \| = 
			\E
			\Big\|
			\frac{1}{\sqrt{N}} \sum_{i=1}^{N} \Big[D_{i}-\Lambda\big(g_{0}(X_{i}^{\prime} \theta_{0})\big)\Big] \big(g_{k}^{(1)}(X_{i}^{\prime} \theta_{0})-g_{0, k}^{(1)}(X_{i}^{\prime} \theta_{0})\big) X_{i}
			\Big\|^{2} 
			\nonumber\\
			&&
			= \frac{1}{N} \sum_{i=1}^{N} 
			\E
			\Big\|
			\big[D_{i}-\Lambda\big(g_{0}(X_{i}^{\prime} \theta_{0})\big)\big]
			\dot{H}(X_{i}^{\prime} \theta_{0})^{\prime}(C_{k}-C_{0,k}) X_{i}
			\Big\|^{2} 
			\nonumber\\
			&&
			= \frac{1}{N} \sum_{i=1}^{N} (C_{k}-C_{0,k})^{\prime} 
			\E
			\Big\{
			\big[D_{i}-\Lambda\big(g_{0}(X_{i}^{\prime} \theta_{0})\big)\big]
			\dot{H}(X_{i}^{\prime} \theta_{0}) \dot{H}(X_{i}^{\prime} \theta_{0})^{\prime}\|X_{i}\|^{2}
			\Big\}
			(C_{k}-C_{0,k})
			\nonumber\\
			&&
			\leq O(1) \| C_{k} - C_{0,k} \|^{2}
			\leq O(1) \| g_{k} - g_{0} \|^{2}_{\mathcal{L}^2},
			\nonumber
			\eea
			where the first inequality follows from \Cref{assumption as:2}(c).  In connection with similar proofs in \Cref{lemma:indentification3}, we obtain $J_{2 N}=o_{P}(1)$. Similarly, we have $J_{5 N}=o_{P}(1)$.
			
			For $J_{3 N}$, we write:
			\bea
			&&\|J_{3 N} \| 
			= 
			\E
			\Big\|
			\frac{1}{\sqrt{N}} \sum_{i=1}^{N} \Big[D_{i}-\Lambda\big(g_{0}(X_{i}^{\prime} \theta_{0})\big)\Big] \epsilon_{0, k}^{(1)}(X_{i}^{\prime} \theta_{0}) X_{i}
			\Big\|^{2} 
			\leq O(1)
			\Big\{
			\E 
			\big\| \epsilon_{0, k}^{(1)}(X_{i}^{\prime} \theta_{0}) 
			\big\|^{4} 
			\E 
			\|X_{i}\|^{4}
			\Big\}^{1 / 2} 
			\nonumber\\
			&& = O(1)
			\Big\{
			\E \big\|\epsilon_{0, k}^{(1)}(X_{i}^{\prime} \theta_{0}) \big\|^{4}
			\Big\}^{1 / 2} 
			= o(1),
			\nonumber
			\eea
			where the first inequality follows from Cauchy-Schwarz inequality, and the second equality follows from \Cref{assumption as:2}(c).
			
			For $J_{4N}$, we have:
			\bea
			&&\E
			\|
			\sqrt{N} V^{\prime} J_{4N}
			\|^{2}  =
			\frac{1}{N} 
			\E
			\Big\|
			\sum_{i=1}^{N} 
			\Big[
			\Lambda\big(g_{0}(X_{i}^{\prime} \theta_{0})\big)
			-
			\Lambda\big(\hat{g}_{k}(X_{i}^{\prime} \theta_{0})\big)
			\Big] 
			\hat{g}_{k}^{(1)}(X_{i}^{\prime} \theta_{0}) V^{\prime} X_{i} 
			\Big\|^{2} 
			\nonumber\\
			&&\leq 
			\frac{1}{N}
			\sum_{i=1}^{N}
			\E
			\Big[
			\big\{
			\Lambda\big(g_{0}(X_{i}^{\prime} \theta_{0})\big)
			-
			\Lambda\big(\hat{g}_{k}(X_{i}^{\prime} \theta_{0})\big)
			\big\} 
			g^{(1)}(X_{i}^{\prime} \theta_{0}) 
			\Big]^2 
			\E\| V^{\prime} X_{i} \|^{2} 
			\nonumber\\
			&&= O(1) \int 
			\Big[
			\big\{
			\Lambda\big(g_{0}(\omega)\big)
			-
			\Lambda\big(\hat{g}_{k}(\omega)\big)
			\big\}
			g^{(1)}(\omega)
			\Big]^2 
			f_{\theta_{0}}(\omega) d \omega 
			=o(1),
			\nonumber
			\eea
			where the first inequality comes from Cauchy-Schwarz inequality and the third equality follows from the boundness of function $\Lambda(\cdot)$ and \Cref{assumption as:1}(b).

			For $J_{6 N}$, we write:
			\bea
			&&J_{6 N} =
			\frac{1}{N} \sum_{i=1}^{N}
			\Big[ 
			\Lambda\big(g_{0}(X_{i}^{\prime} \theta_{0})\big)-\Lambda\big(\hat{g}_{k}(X_{i}^{\prime} \hat{\theta})\big)
			\Big] 
			\hat{g}_{k}^{(1)}(X_{i}^{\prime} \hat{\theta}) X_{i} 
			\nonumber\\
			&& =  \frac{1}{N} \sum_{i=1}^{N}
			\Big[ 
			\Lambda\big(g_{0}(X_{i}^{\prime} \theta_{0})\big)-\Lambda\big(\hat{g}_{k}(X_{i}^{\prime} \hat{\theta})\big)
			\Big]
			\Big[
			\hat{g}_{k}^{(1)}(X_{i}^{\prime} \theta_{0})+\hat{g}_{k}^{(1)}(X_{i}^{\prime} \hat{\theta})-\hat{g}_{k}^{(1)}(X_{i}^{\prime} \theta_{0})
			\Big] X_{i} 
			\nonumber\\
			&&=
			\frac{1}{N} \sum_{i=1}^{N}\Big[ \Lambda\big(g_{0}(X_{i}^{\prime} \theta_{0})\big)-\Lambda\big(\hat{g}_{k}(X_{i}^{\prime} \hat{\theta})\big)\Big] \hat{g}_{k}^{(1)}(X_{i}^{\prime} \theta_{0}) X_{i}
			\nonumber\\
			&&
			+
			\frac{1}{N} \sum_{i=1}^{N}\Big[ 
			\Lambda\big(g_{0}(X_{i}^{\prime} \theta_{0})\big)
			-
			\Lambda\big(\hat{g}_{k}(X_{i}^{\prime} \hat{\theta})\big)\Big] \hat{g}_{k}^{(2)}(X_{i}^{\prime} \theta^{*}) X_{i} X_{i}^{\prime}(\hat{\theta}-\theta_{0}) 
			\nonumber\\
			&& =
			\frac{1}{N} \sum_{i=1}^{N}\Big[  \Lambda\big(g_{0}(X_{i}^{\prime} \theta_{0})\big)-\Lambda\big(\hat{g}_{k}(X_{i}^{\prime} \theta_{0})\big) + \Lambda\big(\hat{g}_{k}(X_{i}^{\prime} \theta_{0})\big) -\Lambda\big(\hat{g}_{k}(X_{i}^{\prime} \hat{\theta})\big)\Big] \hat{g}_{k}^{(1)}(X_{i}^{\prime} \theta_{0}) X_{i}
			+o_{P}(\|\theta_{0}-\hat{\theta}\|) 
			\nonumber\\
			&& =
			\frac{1}{N} \sum_{i=1}^{N}\Big[\Lambda\big(g_{0}(X_{i}^{\prime} \theta_{0})\big)-\Lambda\big(\hat{g}_{k}(X_{i}^{\prime} \theta_{0})\big)\Big] \hat{g}_{k}^{(1)}(X_{i}^{\prime} \theta_{0}) X_{i} 
			+\frac{1}{N} \sum_{i=1}^{N} \upsilon\big(\hat{g}_{k}(X_{i}^{\prime} \tilde{\theta})\big) \hat{g}_{k}^{(1)}(X_{i}^{\prime} \theta_{0}) X_{i} X_{i}^{\prime}(\theta_{0}-\hat{\theta})
			\nonumber\\
			&& 	 +o_{P}(\|\theta_{0}-\hat{\theta}\|) =
			J_{4N} + \frac{1}{N} \sum_{i=1}^{N} \upsilon\big(\hat{g}_{k}(X_{i}^{\prime} \tilde{\theta})\big) \hat{g}_{k}^{(1)}(X_{i}^{\prime} \theta_{0}) X_{i} X_{i}^{\prime}(\theta_{0}-\hat{\theta})	 +o_{P}(\|\theta_{0}-\hat{\theta}\|),
			\nonumber
			\eea
			where $\theta^{*}$ and $\tilde{\theta}$ both lie between $\hat{\theta}$ and $\theta_{0}$, and the second and fifth equalities follow from mean value theorem.

			Note that $I_{d}-\theta_{0} \tilde{\theta}' \rightarrow_{P}  I_{d}-\theta_{0} \theta_{0}^{\prime}$, where $I_{d}-\theta_{0} \theta_{0}^{\prime}$ has eigenvalues $0,1, \ldots, 1$ and for eigenvalue 0 with eigenvector $\theta_{0}$. Therefore, we rotate the function $\pdv*{\mathcal{W}_{N}}{\theta}$   using the matrix $P_1$ in order to have non-singular asymptotic variance covariance matrix.
			
			Moreover, note that $V^{\prime}\theta_{0} = 0$, we have 
			$\sqrt{N}V^{\prime}\theta_{0} \hat{\theta}^{\prime} \frac{1}{N} \sum_{i=1}^{N} \upsilon\big(\hat{g}_{k}(X_{i}^{\prime} \tilde{\theta})\big) \hat{g}_{k}^{(1)}(X_{i}^{\prime} \theta_{0}) X_{i} X_{i}^{\prime} = 0$.
			
			In sum, we have $\sqrt{N}V^{\prime}\pdv{\mathcal{W}_{N}(\theta_{0}, \widehat{C}_{k}, \hat{\lambda})}{\theta}  \rightarrow_{D} N(0, V^{\prime} W V )$.
			
			Given the asymptotic properties of $\pdv*{\mathcal{W}_{N}}{\theta}$ and $\pdv*{\mathcal{W}_{N}}{\theta}{\theta^{\prime}}$, we have:
			\[
			\sqrt{N}V^{\prime}(\hat{\theta}-\theta_{0}) 
			\rightarrow_{D} 
			N\big(0, (V^{\prime} Q V)^{-1} V^{\prime} W V (V^{\prime} Q V)^{-1}\big).
			\]
			
			\textbf{(b)} 
			Using
			\bea
			\hat{\theta}^{\prime} \theta_{0}-1 
			&= & (\hat{\theta}-\theta_{0})^{\prime} \theta_{0} =(\hat{\theta}-\theta_{0})^{\prime}(\theta_{0} - \hat{\theta}+\hat{\theta}) 
			\nonumber\\
			&= & -\|\hat{\theta}-\theta_{0}\|^{2}+(\hat{\theta}-\theta_{0})^{\prime} \hat{\theta} = - \|\hat{\theta}-\theta_{0}\|^{2}+(1-\hat{\theta}^{\prime} \theta_{0}). 
			\nonumber
			\eea
			we have $1-\hat{\theta}^{\prime} \theta_{0} = \frac{1}{2} \|\hat{\theta}-\theta_{0}\|^{2}$, where $1-\hat{\theta}^{\prime} \theta_{0}$ measures the distance from the coordinate to the surface of the unit ball and shows that the convergence speed of $\hat{\theta}$ is faster in the direction of $\theta_{0}$ compared with all other directions orthogonal to $\theta_{0}$. 
			
			Meanwhile, we also have $\hat{\theta}-\theta_0 = (VV^{\prime} + \theta_{0}\theta_{0}^{\prime})(\hat{\theta}-\theta_{0})=V V^{\prime}(\hat{\theta}-\theta_{0}) + \theta_{0} (\theta_{0}^{\prime}\hat{\theta} - 1)$, which gives $\| \hat{\theta}-\theta_{0} \|^{2} = \big\| V^{\prime}(\hat{\theta}-\theta_{0}) \big\|^{2}  + (\theta_{0}^{\prime}\hat{\theta} - 1)^{2}$.
			
			In sum, we have:
			\[
			1 - \theta_{0}^{\prime}\hat{\theta} = 
			\frac{1}{1 + \theta_{0}^{\prime}\hat{\theta}} 
			\big\| V^{\prime}(\hat{\theta}-\theta_{0}) \big\|^{2} = \frac{1}{2} \big\| V^{\prime}(\hat{\theta}-\theta_{0}) \big\|^{2} \big(1+ o_{P}(1)\big) = O_{P}(1/N).
			\]

		\end{proof}
		
		\begin{proof}[\textbf{Proof of \Cref{theorem:ate2}}]
			
			\textbf{(a)}
			By the definition of $\hat{\sigma}_{\omega}^{2}$, we have:
			\begin{equation*}
				\hat{\sigma}_{\omega}^{2}
				= 
				\left(\frac{1}{N} \sum_{i=1}^{N}
				\hat{\upsilon}(X_{i})
				\right)^{-2} 
				\frac{1}{N} \sum_{i=1}^{N}
				\Big[
				\big(D_i - \hat{\pi}(X_{i})\big)\widehat{U}_i
				\Big]^{2}.
			\end{equation*}
			
			Observe that
			\bea
			&& \hat{\sigma}_{\omega}^{2} - \sigma_{\omega}^{2} =
			\left(\frac{1}{N} \sum_{i=1}^{N}
			\hat{\upsilon}(X_{i})
			\right)^{-2} 
			\frac{1}{N} \sum_{i=1}^{N}
			\Big[
			\big(D_i - \hat{\pi}(X_{i})\big)\widehat{U}_i
			\Big]^{2}
			-
			\sigma_{\omega}^{2}
			\\
			&& = 
			\left(\frac{1}{N} \sum_{i=1}^{N}
			\hat{\upsilon}(X_{i})
			\right)^{-2} 
			\frac{1}{N} \sum_{i=1}^{N}
			\Big[
			\big(D_i - \hat{\pi}(X_{i})\big)\widehat{U}_i
			\Big]^{2}-
			\big(\frac{1}{N} \sum_{i=1}^{N}
			\hat{\upsilon}(X_{i})
			\big)^{-2} 
			\frac{1}{N} \sum_{i=1}^{N}
			\Big[\big(D_i - \pi(X_{i})\big)U_i\Big]^{2}
			\nonumber\\
			&&
			+
			\left(\frac{1}{N} \sum_{i=1}^{N}
			\hat{\upsilon}(X_{i})
			\right)^{-2} 
			\frac{1}{N} \sum_{i=1}^{N}
			\Big[\big(D_i - \pi(X_{i})\big)U_i\Big]^{2}			-	
			\big(\frac{1}{N} \sum_{i=1}^{N}
			\upsilon(X_{i})
			\big)^{-2} 
			\frac{1}{N} \sum_{i=1}^{N}
			\Big[\big(D_i - \pi(X_{i})\big)U_i\Big]^{2}		
			\nonumber\\
			&&+
			\left(\frac{1}{N} \sum_{i=1}^{N}
			\hat{\upsilon}(X_{i})
			\right)^{-2} 
			\frac{1}{N} \sum_{i=1}^{N}
			\Big[\big(D_i - \pi(X_{i})\big)U_i\Big]^{2}		
			-
			\sigma_{\omega}^{2} 
			=
			C_{1N} + C_{2N} + C_{3N}.
			\nonumber
			\eea
			
			For $C_{1N}$, we have:
			\begin{align*}
				C_{1N}
				=
				&
				O(1)
				\frac{1}{N} \sum_{i=1}^{N}
				\Big[\big(D_i - \hat{\pi}(X_{i})\big)\widehat{U}_i\Big]^{2}
				-
				O(1)
				\frac{1}{N} \sum_{i=1}^{N}
				\Big[\big(D_i - \pi(X_{i})\big)U_i\Big]^{2}
				\\
				\leq
				&
				O(1) 
				\frac{1}{N} \sum_{i=1}^{N} 
				\Big\{
				\Big[\big(D_i - \hat{\pi}(X_{i})\big)\widehat{U}_i\Big]^{2}
				-
				\Big[\big(D_i - \hat{\pi}(X_{i})\big)U_i\Big]^{2}
				\Big\}
				\\
				+ &
				O(1) 
				\frac{1}{N} \sum_{i=1}^{N} 
				\Big\{
				\Big[\big(D_i - \hat{\pi}(X_{i})\big)U_i\Big]^{2}
				-
				\Big[\big(D_i - \pi(X_{i})\big)U_i\Big]^{2}
				\Big\}
				=C_{11N} + C_{12N}.
			\end{align*}
			
			For $C_{11N}$, we have:
			\begin{align*}
				&\|C_{11N} \|
				=
				O(1) 
				\Big\|
				\frac{1}{N} \sum_{i=1}^{N} 
				\big(D_i - \hat{\pi}(X_{i})\big)^{2} (\widehat{U}_i^2 - U_i^2)
				\Big\|
				\\
				&\leq 
				O(1)
				\Big\|
				\E
				\big(D_i - \hat{\pi}(X_{i})\big)^{2} (\widehat{U}_i^2 - U_i^2)
				\Big\| + o_P(1)
				\\
				&\leq 
				O(1)
				\Big\{
				\E
				\big\|
				D_i - \hat{\pi}(X_{i})
				\big\|^{4}
				\E
				\big\|
				\widehat{U}_i^2 - U_i^2
				\big\|^{2}
				\Big\}^{1/2} + o_P(1) =
				o_{P}(1),
			\end{align*}
			where the first inequality follows from \Cref{assumption as:2}(a), the second equality is due to Cauchy-Schwarz inequality, the third equality follows from the consistency of $\hat{\theta}$ such that $\widehat{U}_i = U_i +o_P(1)$ and the similar proofs as $A_{1N}$. For $C_{12N}$, we have:
			\begin{align*}
				\|
				C_{12N} 
				\|
				=
				&
				O(1) 
				\Big\|
				\frac{1}{N} \sum_{i=1}^{N} 
				\big(\pi(X_{i}) - \hat{\pi}(X_{i})\big)
				\big(D_{i} - \pi(X_{i}) + D_{i} - \hat{\pi}(X_{i}) \big) U_i^{2}
				\Big\|
				\\
				\leq
				&
				O(1) 
				\Big\|
				\E
				\big(\pi(X_{i}) - \hat{\pi}(X_{i})\big)
				\big(D_{i} - \pi(X_{i}) + D_{i} - \hat{\pi}(X_{i}) \big) U_i^{2}
				\Big\| + o_P(1)
				\\
				\leq
				&
				O(1) 
				\Big\|
				\E
				\big(\pi(X_{i}) - \hat{\pi}(X_{i})\big)
				\big(D_{i} - \pi(X_{i}) \big) U_i^{2}
				\Big\|
				\\
				&+
				O(1)
				\Big\|
				\E
				\big(\pi(X_{i}) - \hat{\pi}(X_{i})\big)
				\big( D_{i} - \hat{\pi}(X_{i}) \big) U_i^{2}
				\Big\| + o_P(1)
				\\
				=
				&
				O(1)C_{121N} + O(1)C_{122N} + o_P(1),
			\end{align*}
			where the first inequality comes from \Cref{assumption as:2}(a), and the second equality comes from the triangular inequality.  For $C_{121N}$, we have:
			\begin{align*}
				C_{121N} 
				&= \E
				\Big\|
				\big(\pi(X_{i}) - \hat{\pi}(X_{i})\big)
				\big(D_{i} - \pi(X_{i}) \big) U_i^{2}
				\Big\|
				\\
				&
				\leq 
				O(1) \E
				\Big\|
				\big(\pi(X_{i}) - \hat{\pi}(X_{i})\big)
				\big(D_{i} - \pi(X_{i}) \big)
				\Big\|
				\\
				&
				\leq 
				O(1)
				\Big\{
				\E
				\big\|
				\pi(X_{i}) - \hat{\pi}(X_{i})
				\big\|^2
				\E
				\big\|
				D_{i} - \pi(X_{i})
				\big\|^2
				\Big\}^{1/2}=o(1),
			\end{align*}
			where the first inequality comes from \Cref{assumption as:2}(a), the second inequality comes from Cauchy-Schwarz inequality, and the third equality comes from the similar proofs as $A_{2N}$. Similarly, we have $C_{122N} = o(1)$.
			
			For $C_{2N}$, we have:
			\bea
			&&\|
			C_{2N} 
			\|
			=
			\Big\|
			\big(
			\frac{1}{N} \sum_{i=1}^{N}
			\hat{\upsilon}(X_{i})
			\big)^{-2} 
			\frac{1}{N} \sum_{i=1}^{N}
			[\big(D_i - \pi(X_{i})\big)U_i]^{2}			
			\nonumber\\
			&&	
			-
			\big(\frac{1}{N} \sum_{i=1}^{N}
			\upsilon(X_{i})
			\big)^{-2} 
			\frac{1}{N} \sum_{i=1}^{N}
			[\big(D_i - \pi(X_{i})\big)U_i]^{2}	
			\Big\|
			\nonumber\\
			&&\leq 
			\Big\|
			\big\{
			\E
			\hat{\upsilon}(X_{i})
			\big\}^{-2} 
			\E
			\big[
			\big(D_i - \pi(X_{i})\big)U_i
			\big]^{2}			
			-
			\big\{\E
			\upsilon(X_{i})
			\big\}^{-2} 
			\E
			\big[
			\big(D_i - \pi(X_{i})\big)U_i
			\big]^{2}	
			\Big\|		+ o_P(1)
			=o_P(1),
			\nonumber
			\eea
			where the second inequality comes from \Cref{assumption as:2}(a), and the third equality comes from the consistency of $\hat{\theta}$ such that $\hat{\upsilon}(X_i) = \upsilon(X_i) +o_P(1)$ and the similar proofs as $A_{1N}$.
			
			For $C_{3N}$, by the Law of Large Numbers, we have
			\begin{align*}
				\big(\frac{1}{N} \sum_{i=1}^{N}
				\upsilon(X_{i})
				\big)^{-2} 
				\frac{1}{N} \sum_{i=1}^{N}
				\Big[\big(D_i - \pi(X_{i})\big)U_i\Big]^{2}		
				&
				= 
				\big(\E
				\upsilon(X_{i})
				\big)^{-2} 
				\E
				\Big[\big(D_i - \pi(X_{i})\big)U_i\Big]^{2}	 + o_{P}(1)
				\\
				&=
				\sigma_{\omega}^{2} + o_{P}(1).
			\end{align*}
			
			In sum, we have shown $\widehat{\sigma}_{\omega}^{2} -  \sigma_{\omega}^{2} = o_{P}(1)$.
			
			\textbf{(b)}
			We write
			\bea
			&&\hat{\sigma}^{2} - \sigma^{2} 
			=
			\frac{1}{N} \sum_{i=1}^{N}
			\Big[
			\hat{\upsilon}(X_{i})^{-1}\big(D_i - \hat{\pi}(X_{i})\big)\widehat{U}_i
			\Big]^{2}
			-
			\sigma^{2}
			\nonumber\\
			&& =
			\frac{1}{N} \sum_{i=1}^{N}
			\Big[\hat{\upsilon}(X_{i})^{-1}\big(D_i - \hat{\pi}(X_{i})\big)\widehat{U}_i\Big]^{2}
			-
			\frac{1}{N} \sum_{i=1}^{N}
			\Big[\hat{\upsilon}(X_{i})^{-1}\big(D_i - \pi(X_{i})\big)U_i\Big]^{2}
			\nonumber\\
			&& + 
			\frac{1}{N} \sum_{i=1}^{N}
			\Big[\hat{\upsilon}(X_{i})^{-1}\big(D_i - \pi(X_{i})\big)U_i\Big]^{2}			
			-	
			\frac{1}{N} \sum_{i=1}^{N}
			\Big[\upsilon(X_{i})^{-1}\big(D_i - \pi(X_{i})\big)U_i\Big]^{2}		
			\nonumber\\
			&&+
			\frac{1}{N} \sum_{i=1}^{N}
			[\upsilon(X_{i})^{-1}\big(D_i - \pi(X_{i})\big)U_i]^{2}		
			-
			\sigma^{2} 
			= D_{1N} + D_{2N} + D_{3N}.
			\nonumber
			\eea
			
			For $D_{1N}$, we have:
			\begin{align*}
				\|
				D_{1N}
				\|
				=
				&
				\Big\|
				\frac{1}{N} \sum_{i=1}^{N}
				\big[
				\hat{\upsilon}(X_{i})^{-1}\big(D_i - \hat{\pi}(X_{i})\big)\widehat{U}_i
				\big]^{2}
				-
				\frac{1}{N} \sum_{i=1}^{N}
				\big[
				\hat{\upsilon}(X_{i})^{-1}\big(D_i - \pi(X_{i})\big)U_i
				\big]^{2}
				\Big\|
				\\
				\leq 		
				&
				O(1)
				\Big\|
				\frac{1}{N} \sum_{i=1}^{N}
				\big[ 
				\big(D_i - \hat{\pi}(X_{i})\big)\widehat{U}_i
				\big]^{2}
				-
				\frac{1}{N} \sum_{i=1}^{N}
				\big[ 
				\big(D_i - \pi(X_{i})\big)U_i
				\big]^{2}
				\Big\|
				\leq
				o_{P}(1),
			\end{align*}
			where the first inequality comes from the boundness of $\upsilon(\cdot)$, and the second equality comes from the similar proofs as $C_{1N}$.

			For $D_{2N}$, we have:
			\begin{align*}
				\|
				D_{2N} 
				\|
				&=
				\Big\|
				\frac{1}{N} \sum_{i=1}^{N}
				[\hat{\upsilon}(X_{i})^{-1}\big(D_i - \pi(X_{i})\big)U_i]^{2}			
				-	
				\frac{1}{N} \sum_{i=1}^{N}
				[\upsilon(X_{i})^{-1}\big(D_i - \pi(X_{i})\big)U_i]^{2}		
				\Big\|
				\\
				&=
				\Big\|
				\frac{1}{N} \sum_{i=1}^{N} 
				[\hat{\upsilon}(X_{i})^{-1} - \upsilon(X_{i})^{-1}]
				[\big(D_i - \pi(X_{i})\big)U_i]^{2}	
				\Big\|		
				\\
				&
				\leq
				\Big\|		
				\E
				[\hat{\upsilon}(X_{i})^{-1} - \upsilon(X_{i})^{-1}]
				[\big(D_i - \pi(X_{i})\big)U_i]^{2}			
				\Big\|		+ o_P(1)
				\\
				&
				\leq
				\Big\{
				\E
				\big\|		
				\hat{\upsilon}(X_{i})^{-1} - \upsilon(X_{i})^{-1}
				\big\|^{2}			
				\E
				\big\|
				\big(D_i - \pi(X_{i})\big)U_i
				\big\|^{4}			
				\Big\}^{1/2}			+ o_P(1)
				=
				o_{P}(1),
			\end{align*}
			where the third inequality comes from \Cref{assumption as:2}(a), the fourth inequality comes from Cauchy-Schwarz inequality, and the last inequality comes from the consistency of $\hat{\theta}$ such that $\hat{\upsilon}(X_i) = \upsilon(X_i) +o_P(1)$ and the similar proofs as $C_{2N}$.
			
			For $D_{3N}$, by the Law of Large Numbers, we have:
			\begin{align*}
				\frac{1}{N} \sum_{i=1}^{N}
				\Big[
				\upsilon(X_{i})^{-1}\big(D_i - \pi(X_{i})\big)U_i
				\Big]^{2}		
				= 
				\E	
				\Big[
				\upsilon(X_{i})^{-1}\big(D_i - \pi(X_{i})\big)U_i
				\Big]^{2} + o_{P}(1)
				=
				\sigma^{2} + o_{P}(1).
			\end{align*}
			
			In sum, we have shown $\widehat{\sigma}^{2} -  \sigma^{2} = o_{P}(1)$.
			
		\end{proof}

		\setcounter{equation}{0}
		\setcounter{table}{0}
		\renewcommand{\theequation}{D.\arabic{equation}}
		\renewcommand{\thesubsection}{D.\arabic{subsection}}
		\renewcommand{\thefigure}{D.\arabic{figure}}
		\renewcommand{\thetable}{D.\arabic{table}}
		\renewcommand{\thelemma}{D.\arabi{lemma}}
		\renewcommand{\theremark}{D.\arabic{remark}}
		\renewcommand{\thecorollary}{D.\arabic{corollary}}
		\renewcommand{\theassumption}{D.\arabic{assumption}}
		\renewcommand{\theproposition}{D.\arabic{proposition}}				
		
		\noindent{\bf \large Appendix D: Additional Simulation Results}\label{section:addtional_simulation}
		\bigskip
		
		\noindent{\bf Appendix D.1: Estimating the Propensity Score Function}
		\medskip	
		
		We first conduct a Monte Carlo study to investigate the finite sample properties of our semiparametric estimator proposed in \Cref{subsec:ps} for the propensity score.
		We consider the following four functional forms (DGPs) with a vector of regressors $X=(X_{1}, X_{2})^{\prime} \sim N(0, I_{2})$:
		
		\begin{itemize}
			
			\item[] 1A. $g_0(X^{\prime}\theta_0) = \sin(X^{\prime}\theta_0)$,
			where $\theta_0  = (0.8,-0.6)^{\prime}$.
			
			\item[] 1B. $g_0(X^{\prime}\theta_0) = 0.5\{(X^{\prime}\theta_0)^3 - (X^{\prime}\theta_0)\}$,
			where $\theta_0$ is the same as in Setting 1A.
			
			\item[] 2A. $g_0(X^{\prime}\theta_0) = \sin(X^{\prime}\theta_0)$, where
			$\theta_0  = (0.5,-0.5)^{\prime}$.  
			
			\item[] 2B.
			$g_0(X^{\prime}\theta_0) = 0.5\{(X^{\prime}\theta_0)^3 - (X^{\prime}\theta_0)\}$,
			where $\theta_0$ is the same as in Setting 1B.
			
		\end{itemize}
		
		The propensity score function is defined as $D = \Lambda \big( g_0(X'\theta_{0}) \big)$. The DGPs in Settings 1A and 1B satisfy the $\|\theta_{0}\|=1$ for identification purpose,
		while the DGPs in Settings 2A and 2B violate $\|\theta_{0}\|=1$.
		
		Let $\hat{\theta}= (\hat{\theta}_{1}, \hat{\theta}_{2})'$
		be the estimator of $\theta_{0}$. We evaluate the estimators in terms of bias, standard deviation (std) and root mean-squared error (RMSE).
		We summarize the simulation results in \Cref{table: simulation_ps_ks}. For the DGPs in Settings 1A and 1B that satisfy $\|\theta_{0}\|=1$, the simulation results show that our estimator performs well. Moreover, by focusing on the bias and RMSE separately, it is clear that both terms vanish quickly as the sample size increases. The asymptotic normal approximation is accurate as the standard deviations become smaller as the sample size grows. Our estimation approach is valid for the polynomial single index structure or more complicated periodic single index form. More importantly, the convergence speed is $O(1/N)$ along the direction of true parameter  $\theta_{0}$, consistent with the results in \Cref{lemma:indentification2}.	
		
		{\small
			\begin{table}[!htbp]
				\centering
				\caption{\label{table: simulation_ps_ks} Simulation Results for the Estimators of Propensity Score Function of Settings 1A to 2B.}
				\begin{tabular}{lccrrcccrr}
					\toprule
					& Setting & $N$    & \multicolumn{1}{c}{$\widehat{\theta}_{1}$} & \multicolumn{1}{c}{$\widehat{\theta}_{2}$} & & Setting & $N$    & \multicolumn{1}{c}{$\widehat{\theta}_{1}$} & \multicolumn{1}{c}{$\widehat{\theta}_{2}$} \\ \hline
					Bias &         &      &          &          &  &         &      &          &          \\ \hline
					& 1A & 400  & -0.0030 & 0.0019 &  & 2A & 400  & -0.0114 & -0.0094 \\
					&     & 800  & 0.0011 & -0.0010 &  &     & 800  & -0.0051 & 0.0050  \\
					&     & 1600 & 0.0003 & 0.0003 &  &     & 1600 & 0.0019  & 0.0015  \\
					& 1B & 400  & -0.0010 & -0.0015 &  & 2B & 400  & 0.0108  & -0.0081 \\
					&     & 800  & -0.0003 & 0.0005 &  &     & 800  & -0.0052 & 0.0043  \\
					&     & 1600 & 0.0001  & 0.0003 &  &     & 1600 & -0.0009 & 0.0009  \\\hline
					Std  &     &      &         &        &  &     &      &         &         \\\hline
					&1A & 400  & 0.0356  & 0.0482 &  & 2A & 400  & 0.0439  & 0.0449  \\
					&     & 800  & 0.0232  & 0.0309 &  &     & 800  & 0.0305  & 0.0314  \\
					&     & 1600 & 0.0195  & 0.0249 &  &     & 1600 & 0.0215  & 0.0217  \\
					& 1B & 400  & 0.0222  & 0.0349 &  & 2B & 400  & 0.0471  & 0.0482  \\
					&     & 800  & 0.0138  & 0.0184 &  &     & 800  & 0.0317  & 0.0334  \\
					&     & 1600 & 0.0136  & 0.0167 &  &     & 1600 & 0.0268  & 0.0282 \\ \bottomrule
				\end{tabular}\\
				%		\begin{minipage}{0.95\textwidth}
					%			\vspace{0.5em}
					%			\footnotesize We report the simulation results of the semiparametric estimator proposed in \Cref{subsec:ps} for the propensity score function, including the sample mean deviation over the true value (bias) and the sample standard deviation (std) over 10,000 replications. The DGPs are Settings 1A to 2B. $N$ is the number of sample size.
					%		\end{minipage}
			\end{table}
		}

		For the DGPs in Settings 2A and 2B that violate $\|\theta_{0}\|=1$, the estimated value $\hat{\theta}_{1}$ and $\hat{\theta}_{2}$ converge to $1/\sqrt{2}$ and  -$1/\sqrt{2}$, respectively. For all sample sizes we studied, the estimators have relative small biases. The standard deviations of $\hat{\theta}_{1}$ and $\hat{\theta}_{2}$ become smaller as the sample size increases. The asymptotic normal approximation is accurate and the convergence speed is $O(1/\sqrt{N})$ along all other direction orthogonal to the true value $\theta_{0}$, in accordance with the results in \Cref{lemma:indentification2}.

		We further conduct the simulation study to investigate the performance of our estimators for six dimensions of covariates. We consider the following four DGPs for the propensity score with the vector of regressors $X=(X_{1}, X_{2},X_{3},X_{4},X_{5}, X_{6})^{\prime} \sim N(0,I_{6})$:
		
		\begin{itemize}
			\item 3A. $g_0(X^{\prime}\theta_0) = \sin(X^{\prime}\theta_0)$, where $\theta_0  = (\sqrt{0.2},\sqrt{0.3},\sqrt{0.25},-\sqrt{0.1},\sqrt{0.08},-\sqrt{0.07})^{\prime}$.
			
			\item 3B. $g_0(X^{\prime}\theta_0) = 0.5\{(X^{\prime}\theta_0)^3 - (X^{\prime}\theta_0)\} $, where $\theta_0$ is the same as in Setting 3A.

			\item
			4A. $g_0(X^{\prime}\theta_0) = \sin(X^{\prime}\theta_0)$, 
			where     $\theta_0 = (\sqrt{0.2},\sqrt{0.4},\sqrt{0.6},-\sqrt{0.25},\sqrt{0.1},-\sqrt{0.45})^{\prime}$.

			\item 4B.
			$g_0(X^{\prime}\theta_0) = 0.5\{(X^{\prime}\theta_0)^3 - (X^{\prime}\theta_0)\}$, where
			$\theta_0$ is the same as in Setting 4A.  
		\end{itemize}

		The propensity score function is defined as $D = \Lambda \big( g_0(X'\theta_{0}) \big)$. The DGPs in Settings 3A and 3B satisfy $\|\theta_{0}\|=1$,
		while the DGPs in Settings 4A and 4B violate $\|\theta_{0}\|=1$.

		We show the simulation results for Settings 3A and 3B in \Cref{table: simulation_ps_hd_ks} and the results for Settings 4A and 4B in \Cref{table: simulation_ps_mis_hd_ks}, respectively. Our estimation approach is valid for the six dimensions of covariates we considered. All estimators have small biases for the three sample sizes we considered. The biases vanish quickly when the sample size increases and RMSEs generally decrease as the sample size increases. The asymptotic normal approximation is accurate as the standard deviations become smaller as the sample size grows.  The convergence speed is consistent with the results in \Cref{lemma:indentification2}. 
		
		{\small
			\begin{table}
				\centering
				\caption{\label{table: simulation_ps_hd_ks}Simulation Results for the Estimators of Propensity Score Function of Settings 3A to 3B.}
				\begin{tabular}{lccrrrrrr}
					\toprule
					& Setting &  $N$    & \multicolumn{1}{c}{$\widehat{\theta}_{1}$} & \multicolumn{1}{c}{$\widehat{\theta}_{2}$} & \multicolumn{1}{c}{$\widehat{\theta_{3}}$} & \multicolumn{1}{c}{$\widehat{\theta_{4}}$} & \multicolumn{1}{c}{$\widehat{\theta_{5}}$} & \multicolumn{1}{c}{$\widehat{\theta_{6}}$} \\ \hline
					Bias &         &      &         &         &         &         &         &         \\ \hline
					& 3A & 400  & -0.0059 & -0.0047 & -0.0076 & 0.0040 & -0.0047 & 0.0042 \\
					&     & 800  & 0.0025 & 0.0024  & -0.0028 & -0.0017 & 0.0016 & -0.0017 \\
					&     & 1600 & 0.0011 & -0.0013 & 0.0011 & -0.0009 & -0.0005 & 0.0011 \\
					& 3B & 400  & 0.0018 & 0.0036  & 0.0024 & 0.0031 & -0.0014 & 0.0043 \\
					&     & 800  & -0.0008 & 0.0010  & -0.0009 & 0.0006 & -0.0004 & 0.0009 \\
					&     & 1600 & 0.0001 & -0.0001 & -0.0001 & 0.0001 & 0.0002  & 0.0001 \\\hline
					Std  &     &      &        &         &         &        &         &        \\\hline
					& 3A & 400  & 0.0524 & 0.0542  & 0.0629  & 0.0476 & 0.0384  & 0.0641 \\
					&     & 800  & 0.0340 & 0.0351  & 0.0392  & 0.0293 & 0.0230  & 0.0381 \\
					&     & 1600 & 0.0236 & 0.0244  & 0.0266  & 0.0199 & 0.0152  & 0.0253 \\
					& 3B & 400  & 0.0326 & 0.0387  & 0.0387  & 0.0339 & 0.0244  & 0.0511 \\
					&     & 800  & 0.0203 & 0.0210  & 0.0232  & 0.0173 & 0.0137  & 0.0267 \\
					&     & 1600 & 0.0142 & 0.0147  & 0.0161  & 0.0120 & 0.0100  & 0.0165\\ \bottomrule
				\end{tabular}\\
				%		\begin{minipage}{0.95\textwidth}
					%			\vspace{0.5em}
					%			\footnotesize We report the simulation results of the semiparametric estimator proposed in \Cref{subsec:ps} for the propensity score function, including the sample mean deviation over the true value (bias) and the sample standard deviation (std) over 10,000 replications. The DGPs are Settings 3A to 3B. $N$ is the number of sample size. 
					%		\end{minipage}
			\end{table}
			
			\begin{table}
				\centering
				\caption{\label{table: simulation_ps_mis_hd_ks}Simulation Results for the Estimators of Propensity Score Function of Settings 4A to 4B.}
				\begin{tabular}{lccrrrrrr}
					\toprule
					& Setting &  $N$    & \multicolumn{1}{c}{$\widehat{\theta}_{1}$} & \multicolumn{1}{c}{$\widehat{\theta}_{2}$} & \multicolumn{1}{c}{$\widehat{\theta_{3}}$} & \multicolumn{1}{c}{$\widehat{\theta_{4}}$} & \multicolumn{1}{c}{$\widehat{\theta_{5}}$} & \multicolumn{1}{c}{$\widehat{\theta_{6}}$} \\ \hline
					Bias &         &      &         &         &         &         &         &         \\ \hline    
					& 4A & 400  & 0.0134 & 0.0219  & -0.0234 & -0.0165 & -0.0092 & 0.0199  \\
					&     & 800  & 0.0123 & 0.0137  & -0.0214 & -0.0146 & 0.0086  & -0.0142 \\
					&     & 1600 & 0.0029 & 0.0114  & -0.0116 & -0.0103 & -0.0030 & -0.0114 \\
					& 4B & 400  & 0.0125 & -0.0265 & -0.0134 & 0.0195  & 0.0071  & 0.0189  \\
					&     & 800  & 0.0114 & -0.0055 & 0.0125  & 0.0056  & -0.0068 & 0.0070  \\
					&     & 1600 & 0.0033 & 0.0070  & -0.0064 & -0.0032 & -0.0046 & -0.0014 \\\hline
					Std  &     &      &        &         &         &         &         &         \\\hline
					& 4A & 400  & 0.0601 & 0.0467  & 0.0557  & 0.0526  & 0.0537  & 0.0677  \\
					&     & 800  & 0.0420 & 0.0323  & 0.0395  & 0.0371  & 0.0378  & 0.0472  \\
					&     & 1600 & 0.0293 & 0.0226  & 0.0273  & 0.0260  & 0.0268  & 0.0332  \\
					& 4B & 400  & 0.0432 & 0.0470  & 0.0487  & 0.0496  & 0.0426  & 0.0624  \\
					&     & 800  & 0.0337 & 0.0380  & 0.0374  & 0.0405  & 0.0319  & 0.0525  \\
					&     & 1600 & 0.0298 & 0.0271  & 0.0324  & 0.0287  & 0.0236  & 0.0375\\ \bottomrule
				\end{tabular}
				%		\begin{minipage}{0.95\textwidth}
					%			\vspace{0.5em}
					%			\footnotesize We report the simulation results of the semiparametric estimator proposed in \Cref{subsec:ps} for the propensity score function, including the sample mean deviation over the true value (bias) and the sample standard deviation (std) over 10,000 replications. The DGPs are Settings 4A to 4B. $N$ is the number of sample size.
					%		\end{minipage}	
			\end{table}
		}

		We further conduct a Monte Carlo study to investigate the finite sample properties of our semiparametric estimator proposed in \Cref{subsec:ate}, and examine the average treatment effect under potential misspecification of propensity score function and the relaxation of $\|\theta_{0}\|=1$ in \Cref{eq:ps1}. We consider the following eight DGPs for both the propensity score and potential outcome.
		\medskip

		\noindent{\bf Setting 7A}: $g_0(X^{\prime}\theta_0) = \sin(X^{\prime}\theta_0)$, where $\theta_0 = (\theta_{0,1}, \theta_{0,2})^{\prime} = (0.5,-0.5)^{\prime}$, the vector of covariates $X=(X_{1}, X_{2})^{\prime}$ are generated from standard normal distributions $N(0,1)$. The propensity score function is defined as $D = \Lambda \big( g_0(X'\theta_{0}) \big)$ and the outcome model is generated as $Y=\beta_d D + X_{1}+ X_{2} + \epsilon$, where $\epsilon \sim N(0,1)$ and $\beta_d = 1$.
		\smallskip
		
		\noindent{\bf Setting 7B}: $g_0(X^{\prime}\theta_0) = 0.5\{(X^{\prime}\theta_0)^3 - (X^{\prime}\theta_0)\}$,  where $\theta_0 = (\theta_{0,1}, \theta_{0,2})^{\prime} = (0.5,-0.5)^{\prime}$, the vector of covariates $X=(X_{1}, X_{2})^{\prime}$ are generated from standard normal distributions $N(0,1)$. The propensity score function is defined as $D = \Lambda \big( g_0(X'\theta_{0}) \big)$ and the outcome model is generated as $Y=\beta_d D + X_{1}+ X_{2} + \epsilon$, where $\epsilon \sim N(0,1)$ and $\beta_d = 1$.
		\smallskip
		
		\noindent{\bf Setting 8A}: $D=1\{X^{\prime}\theta_0+\varepsilon>0\}$, where $\varepsilon \sim N(0,1)$ is independent of $X$, $\theta_0 = (\theta_{0,1}, \theta_{0,2})^{\prime} = (0.8,-0.6)^{\prime}$,  the vector of covariates $X=(X_{1}, X_{2})^{\prime}$ are generated from standard normal distributions $N(0,1)$. The outcome model is generated as $Y=\beta_d D + X_{1}+ X_{2} + \epsilon$, where $\epsilon \sim N(0,1)$ and $\beta_d = 1$.
		\smallskip
		
		\noindent{\bf Setting 8B}: $D=1\{(X^{\prime}\theta_0)^3 + (X^{\prime}\theta_0)^2+ X^{\prime}\theta_0 + \varepsilon>0\}$, where $\varepsilon \sim N(0,1)$ is independent of $X$, $\theta_0 = (\theta_{0,1}, \theta_{0,2})^{\prime} = (0.8,-0.6)^{\prime}$,  the vector of covariates $X=(X_{1}, X_{2})^{\prime}$ are generated from standard normal distributions $N(0,1)$. The outcome model is generated as $Y=\beta_d D + X_{1}+ X_{2} + \epsilon$, where $\epsilon \sim N(0,1)$ and $\beta_d = 1$.
		\smallskip
		
		\noindent{\bf Setting 9A}: $g_0(X^{\prime}\theta_0) = \sin(X^{\prime}\theta_0)$, where $\theta_0 = (\theta_{0,1}, \theta_{0,2})^{\prime} = (0.8,-0.6)^{\prime}$ and the vector of covariates $X=(X_{1}, X_{2})^{\prime}$ are generated from standard normal distributions $N(0,1)$. The propensity score function is defined as $D = \Lambda \big( g_0(X'\theta_{0}) \big)$ and the outcome model is generated as $Y=\beta_d D + X_{1}+ X_{2} + \epsilon$, where $\epsilon$ is the chi-square with 1 degree of freedom minus its median and $\beta_d = 1$.
		\smallskip
		
		\noindent{\bf Setting 9B}: $g_0(X^{\prime}\theta_0) = 0.5\{(X^{\prime}\theta_0)^3 - (X^{\prime}\theta_0)\}$,  where $\theta_0 = (\theta_{0,1}, \theta_{0,2})^{\prime} = (0.8,-0.6)^{\prime}$ and the vector of covariates $X=(X_{1}, X_{2})^{\prime}$ are generated from the standard normal distribution $N(0,1)$. The propensity score function is defined as $D = \Lambda \big( g_0(X'\theta_{0}) \big)$ and the outcome model is generated as $Y=\beta_d D + X_{1}+ X_{2} + \epsilon$, where $\epsilon$ is Cauchy distribution and $\beta_d = 1$.
		\smallskip

		\noindent{\bf Setting 10A}: $g_0(X^{\prime}\theta_0) = 2X_{1}X_{2}$, where the vector of covariates $X=(X_{1}, X_{2})^{\prime}$ are generated from standard normal distributions $N(0,1)$. The propensity score function is defined as $D = \Lambda \big( g_0(X'\theta_{0}) \big)$ and the outcome model is generated as $Y=\beta_d D + X_{1}+ X_{2} + \epsilon$, where $\epsilon \sim N(0,1)$ and $\beta_d = 1$.
		\smallskip
		
		\noindent{\bf Setting 10B}: $D=1\{2X_{1}X_{2}+\varepsilon>0\}$, where $\varepsilon \sim N(0,1)$ is independent of $X$ and the vector of covariates $X=(X_{1}, X_{2})^{\prime}$ are generated from standard normal distributions $N(0,1)$. The outcome model is generated as $Y=\beta_d D + X_{1}+ X_{2} + \epsilon$, where $\epsilon \sim N(0,1)$ and $\beta_d = 1$.
		\medskip

		The DGPs in Settings 7A and 7B violate $\|\theta_{0}\|=1$ in \Cref{eq:ps1}, the DGPs in Settings 8A and 8B have probit propensity score function and are supposed to be handled by our semiparametric single-index models, DGPs in Settings 9A and 9B have heavy-tail error distribution of the outcome equations, and DGPs in Settings 10A and 10B impose additional misspecification of propensity score function with respect to the semiparametric single-index model.

		{\small
			%	\begin{landscape}
				\begin{table}
					\centering
					\caption{\label{table: simulation with ate2}Simulation Results the Estimators of Average Treatment Effect of Settings 7A to 10B.}
					\resizebox{\columnwidth}{!}{%
						\begin{tabular}{lccrccrccrccr}
							\toprule
							& Setting & $N$  & \multicolumn{1}{c}{$\beta_d$} & Setting & $N$  & \multicolumn{1}{c}{$\beta_d$} & Setting & $N$  & \multicolumn{1}{c}{$\beta_d$} & \multicolumn{1}{c}{Setting} & $N$                  & \multicolumn{1}{c}{$\beta_d$} \\ \hline
							Bias &         &      &                               &         &      &                               &         &      &                               &                             &  &                               \\ \hline
							& 7A      & 400  & 0.0044                        & 8A      & 400  & 0.0046                        & 9A      & 400  & 0.0039                        & 10A                         & 400                  & -0.0059                       \\
							&         & 800  & -0.0021                       &         & 800  & -0.0017                       &         & 800  & -0.0022                       &                             & 800                  & 0.0027                        \\
							&         & 1600 & 0.0002                        &         & 1600 & -0.0007                       &         & 1600 & -0.0002                       &                             & 1600                 & 0.0002                        \\
							& 7B      & 400  & 0.0028                        & 8B      & 400  & -0.0052                       & 9B      & 400  & -0.0026                       & 10B                         & 400                  & 0.0046                         \\
							&         & 800  & -0.0011                       &         & 800  & 0.0035                        &         & 800  & 0.0011                        &                             & 800                  & -0.0019                        \\
							&         & 1600 & -0.0003                       &         & 1600 & -0.0007                       &         & 1600 & -0.0003                       &                             & 1600                 &-0.0005                        \\ \hline
							Std  &         &      &                               &         &      &                               &         &      &                               &                             &   &                               \\ \hline
							& 7A      & 400  & 0.0402                        & 8A      & 400  & 0.0442                        & 9A      & 400  & 0.0370                        & 10A                         & 400                  & 0.0502                        \\
							&         & 800  & 0.0277                        &         & 800  & 0.0307                        &         & 800  & 0.0255                        &                             & 800                  & 0.0346                        \\
							&         & 1600 & 0.0192                        &         & 1600 & 0.0216                        &         & 1600 & 0.0175                        &                             & 1600                 & 0.0224                        \\
							& 7B      & 400  & 0.0366                        & 8B      & 400  & 0.0483                        & 9B      & 400  & 0.0337                        & 10B                         & 400                  & 0.0457                         \\
							&         & 800  & 0.0253                        &         & 800  & 0.0345                        &         & 800  & 0.0234                        &                             & 800                  & 0.0303                         \\
							&         & 1600 & 0.0177                        &         & 1600 & 0.0239                        &         & 1600 & 0.0163                        &                             & 1600                 & 0.0208                        \\ \bottomrule
						\end{tabular}
					}
					%		\begin{minipage}{0.95\textwidth}
						%			\vspace{0.5em}
						%			\footnotesize We report the simulation results of the semiparametric estimator proposed in \Cref{subsec:ate} for the average treatment effects, including the sample mean deviation over the true value (bias) and the sample standard deviation (std) over 10,000 replications. The DGPs are Settings 7A to 10B. $N$ is the number of sample size and $k$ is the truncation parameter.
						%		\end{minipage}	
				\end{table}
				%	\end{landscape}
		}

		We summarize the simulation results in \Cref{table: simulation with ate2}. Our estimator performs well under the relaxation of $\|\theta_{0}\|=1$ and misspecification of propensity score and outcome function. All estimators have relatively small biases, standard deviations, and RMSEs for the three sample sizes we studied, and all three terms generally decrease as the sample size increases. Our simulation results show that our estimators are valid for certain degree of misspecification in propensity score and outcome equation.
		
 	\bigskip
 	
	\noindent{\bf Appendix D.2: Comparison with Other Estimators in Propensity Score Approximation}
		\medskip 
		
		We employ a single-index structure as an approximation for the unknown propensity score function. In this section, we investigate its accuracy and compare the finite sample performance of our proposed method with recent approaches to propensity score estimation based on covariates balancing \cite[e.g.,][]{Anna2021JAE,imai2014covariate,Zubizarreta} and high-dimensional selection \cite[e.g.,][]{belloni2017a,Chernozhukov2018,SunTan2021}. To challenge these assumptions, we employ four data generating processes (DGPs) that feature unbounded link functions and unbounded support for propensity score functions.
		\medskip
		
		\noindent{\bf Setting 9A}: $g_0(X^{\prime}\theta_0) = \sin(X^{\prime}\theta_0)$, where $\theta_0 = (\theta_{0,1}, \theta_{0,2})^{\prime} = (0.8,-0.6)^{\prime}$ and the vector of covariates $X=(X_{1}, X_{2})^{\prime}$ are generated from standard normal distributions $N(0,1)$. The propensity score function is defined as $D = \Lambda \big( g_0(X'\theta_{0}) \big)$.
		\smallskip
		
		\noindent{\bf Setting 9B}: $g_0(X^{\prime}\theta_0) = 0.5\{(X^{\prime}\theta_0)^3 - (X^{\prime}\theta_0)\}$,  where $\theta_0 = (\theta_{0,1}, \theta_{0,2})^{\prime} = (0.8,-0.6)^{\prime}$ and the vector of covariates $X=(X_{1}, X_{2})^{\prime}$ are generated from the standard normal distribution $N(0,1)$. The propensity score function is defined as $D = \Lambda \big( g_0(X'\theta_{0}) \big)$.
		\smallskip

		\noindent{\bf Setting 10A}: $g_0(X^{\prime}\theta_0) = 2X_{1}X_{2}$, where the vector of covariates $X=(X_{1}, X_{2})^{\prime}$ are generated from standard normal distributions $N(0,1)$. The propensity score function is defined as $D = \Lambda \big( g_0(X'\theta_{0}) \big)$.
		\smallskip

		\noindent{\bf Setting 10B}: $D=1\{2X_{1}X_{2}+\varepsilon>0\}$, where $\varepsilon \sim N(0,1)$ is independent of $X$ and the vector of covariates $X=(X_{1}, X_{2})^{\prime}$ are generated from standard normal distributions $N(0,1)$.
		\medskip

		%	\begin{landscape}
			\begin{table} 
				\centering
				\caption{\label{table: simulation_ps_compare} Simulation Results for the Comparison between Different Estimators of Propensity Score Function of Settings 9A to 10B.}
				\resizebox{\columnwidth}{!}{%
					\begin{tabular}{lcccccccccc}
						\toprule
						& Setting & $N$  & SP    & CB & HD & Setting & $N$  & SP     & CB & HD \\ \hline
						Bias &         &      &         &            &            &         &      &         &            &            \\ \hline
						& 9A     & 400  & 0.0463 & -0.1019     & 0.0917    & 10A     & 400  & 0.0536 & -0.0802     & 0.0941    \\
						&         & 800  & -0.0255  & -0.0436     & -0.0462     &         & 800  & -0.0217  & 0.0448    & -0.0386     \\
						&         & 1600 & 0.0068 & -0.0176     & -0.0133     &         & 1600 & 0.0093 & -0.0125     & 0.0163    \\
						& 9B     & 400  & -0.0372  & -0.0958     & 0.0784    & 10B     & 400  & -0.0498  & -0.0691     & -0.0853     \\
						&         & 800  & -0.0196  & 0.0465    & -0.0399     &         & 800  & -0.0247  & 0.0366    & -0.0418     \\
						&         & 1600 & 0.0053 & 0.0073    & 0.0102    &         & 1600 & 0.0038 & 0.0074    & 0.0087    \\ \hline
						RMSE &         &      &         &            &            &         &      &         &            &            \\ \hline
						& 9A     & 400  & 0.0617  & 0.1054     & 0.0946     & 10A     & 400  & 0.0638  & 0.0826     & 0.0771     \\
						&         & 800  & 0.0402  & 0.0702     & 0.0627     &         & 800  & 0.0420  & 0.0546     & 0.0507     \\
						&         & 1600 & 0.0328  & 0.0538     & 0.0485     &         & 1600 & 0.0326  & 0.0435     & 0.0388     \\
						& 9B     & 400  & 0.0664  & 0.1136     & 0.1012     & 10B     & 400  & 0.0687  & 0.0894     & 0.0816     \\
						&         & 800  & 0.0459  & 0.0765     & 0.0689     &         & 800  & 0.0463  & 0.0614     & 0.0548     \\
						&         & 1600 & 0.0386  & 0.0646     & 0.0577     &         & 1600 & 0.0390  & 0.0519     & 0.0466     \\ \hline
						Std  &         &      &         &            &            &         &      &         &            &            \\ \hline
						& 9A     & 400  & 0.0676  & 0.1370     & 0.1168     & 10A     & 400  & 0.0718  & 0.1024     & 0.1016     \\
						&         & 800  & 0.0419  & 0.0751     & 0.0674     &         & 800  & 0.0436  & 0.0599     & 0.0557     \\
						&         & 1600 & 0.0324  & 0.0540     & 0.0491     &         & 1600 & 0.0329  & 0.0432     & 0.0395     \\
						& 9B     & 400  & 0.0708  & 0.1397     & 0.1173     & 10B     & 400  & 0.0758  & 0.1043     & 0.1046     \\
						&         & 800  & 0.0461  & 0.0824     & 0.0730     &         & 800  & 0.0489  & 0.0656     & 0.0605     \\
						&         & 1600 & 0.0387  & 0.0640     & 0.0583     &         & 1600 & 0.0391  & 0.0523     & 0.0468     \\ \bottomrule
					\end{tabular}
				}
			\end{table}
			 
		\Cref{table: simulation_ps_compare} displays the finite sample performance of the three approaches, where we denote our estimator as SP, the covariates balancing estimator as CB, and the high-dimensional selection estimator as HD. All three estimators exhibit near-zero bias in all cases. Our estimator also works well when the DGPs are non-single index and outperforms the covariates balancing \cite[e.g.,][]{Anna2021JAE,imai2014covariate,Zubizarreta} and high-dimensional selection \cite[e.g.,][]{belloni2017a,Chernozhukov2018,SunTan2021} in terms of standard deviation and root mean squared error (RMSE) for all considered sample sizes. Overall, our method is superior to the existing methods when dealing with DGPs that are non-single index.
		\medskip
		
		\noindent{\bf Appendix D.3: Comparison with Other Estimators in Treatment Effects Performance}
		
		In this section, we compare the finite sample performance of our approach with two alternative approaches in \cite{Liu2018} and \cite{sun2021estimation}.  Both of these papers assume boundedness of the link function and the function support, which imposes additional constraints on their asymptotic theory and numerical performance. We use the following four DGPs with unbounded link function and unbounded support for propensity score functions. 
		\medskip
		
		\noindent{\bf Setting 11A}: $g_0(X^{\prime}\theta_0) = 10\exp(X^{\prime}\theta_0)$, where $\theta_0 = (\theta_{0,1}, \theta_{0,2})^{\prime} = (0.8,-0.6)^{\prime}$ and the vector of covariates $X=(X_{1}, X_{2})^{\prime}$ are generated from independent standard normal distributions $N(0,1)$. The propensity score function is defined as $D = \Lambda \big( g_0(X'\theta_{0}) \big)$ and the outcome model is generated as $Y=\beta_d D + X_{1}+ X_{2} + \epsilon$, where $\epsilon \sim N(0,1)$ and $\beta_d = 1$.
		\smallskip
		
		\noindent{\bf Setting 11B}: $g_0(X^{\prime}\theta_0) = 10 \{(X^{\prime}\theta_0)^5 - (X^{\prime}\theta_0)^3\} + 10\exp(X^{\prime}\theta_0)$,  where $\theta_0 = (\theta_{0,1}, \theta_{0,2})^{\prime} = (0.8,-0.6)^{\prime}$ and the vector of covariates $X=(X_{1}, X_{2})^{\prime}$ are generated from independent standard normal distributions $N(0,1)$. The propensity score function is defined as $D = \Lambda \big( g_0(X'\theta_{0}) \big)$ and the outcome model is generated as $Y=\beta_d D + X_{1}+ X_{2} + \epsilon$, where $\epsilon \sim N(0,1)$ and $\beta_d = 1$.
		\smallskip
		
		\noindent{\bf Setting 12A}: $g_0(X^{\prime}\theta_0) = 10\exp(X^{\prime}\theta_0)$, where $\theta_0 = (\theta_{0,1}, \theta_{0,2})^{\prime} = (0.8,-0.6)^{\prime}$ and the vector of covariates $X=(X_{1}, X_{2})^{\prime}$ are generated from independent Cauchy distribution. The propensity score function is defined as $D = \Lambda \big( g_0(X'\theta_{0}) \big)$ and the outcome model is generated as $Y=\beta_d D + X_{1}+ X_{2} + \epsilon$, where $\epsilon \sim N(0,1)$ and $\beta_d = 1$.
		\smallskip
		
		\noindent{\bf Setting 12B}: $g_0(X^{\prime}\theta_0) = 10 \{(X^{\prime}\theta_0)^5 - (X^{\prime}\theta_0)^3\} + 10\exp(X^{\prime}\theta_0)$,  where $\theta_0 = (\theta_{0,1}, \theta_{0,2})^{\prime} = (0.8,-0.6)^{\prime}$ and the vector of covariates $X=(X_{1}, X_{2})^{\prime}$ are generated from independent Cauchy distribution. The propensity score function is defined as $D = \Lambda \big( g_0(X'\theta_{0}) \big)$ and the outcome model is generated as $Y=\beta_d D + X_{1}+ X_{2} + \epsilon$, where $\epsilon \sim N(0,1)$ and $\beta_d = 1$.
		\medskip
		
		%	\begin{landscape}
			\begin{table} 
				\centering
				\caption{\label{table: simulation_ate_compare} Simulation Results for the Comparison between Different Estimators of Average Treatment Effect of Settings 11A to 12B.}
					\begin{tabular}{lcccccccccc}
						\toprule
						& Setting & $N$  & SP    & Efficient & NP & Setting & $N$  & SP     & Efficient & NP \\ \hline
						Bias &         &      &         &            &            &         &      &         &            &            \\ \hline
						& 11A     & 400  & -0.0445 & 0.1003     & -0.0899    & 12A     & 400  & -0.0517 & 0.0816     & -0.0922    \\
						&         & 800  & 0.0237  & 0.0444     & 0.0480     &         & 800  & 0.0229  & -0.0436    & 0.0408     \\
						&         & 1600 & -0.0072 & 0.0168     & 0.0145     &         & 1600 & -0.0087 & 0.0131     & -0.0155    \\
						& 11B     & 400  & 0.0384  & 0.0944     & -0.0776    & 12B     & 400  & 0.0486  & 0.0705     & 0.0867     \\
						&         & 800  & 0.0204  & -0.0457    & 0.0411     &         & 800  & 0.0235  & -0.0374    & 0.0420     \\
						&         & 1600 & -0.0045 & -0.0081    & -0.0090    &         & 1600 & -0.0042 & -0.0082    & -0.0075    \\ \hline
						RMSE &         &      &         &            &            &         &      &         &            &            \\ \hline
						& 11A     & 400  & 0.0625  & 0.1060     & 0.0934     & 12A     & 400  & 0.0646  & 0.0838     & 0.0769     \\
						&         & 800  & 0.0410  & 0.0710     & 0.0613     &         & 800  & 0.0432  & 0.0550     & 0.0515     \\
						&         & 1600 & 0.0330  & 0.0546     & 0.0493     &         & 1600 & 0.0332  & 0.0443     & 0.0396     \\
						& 11B     & 400  & 0.0672  & 0.1142     & 0.1004     & 12B     & 400  & 0.0695  & 0.0902     & 0.0828     \\
						&         & 800  & 0.0467  & 0.0773     & 0.0697     &         & 800  & 0.0471  & 0.0626     & 0.0560     \\
						&         & 1600 & 0.0394  & 0.0654     & 0.0589     &         & 1600 & 0.0398  & 0.0529     & 0.0474     \\ \hline
						Std  &         &      &         &            &            &         &      &         &            &            \\ \hline
						& 11A     & 400  & 0.0684  & 0.1362     & 0.1176     & 12A     & 400  & 0.0726  & 0.1038     & 0.1024     \\
						&         & 800  & 0.0427  & 0.0769     & 0.0682     &         & 800  & 0.0448  & 0.0607     & 0.0565     \\
						&         & 1600 & 0.0332  & 0.0554     & 0.0499     &         & 1600 & 0.0335  & 0.0448     & 0.0403     \\
						& 11B     & 400  & 0.0716  & 0.1409     & 0.1185     & 12B     & 400  & 0.0766  & 0.1051     & 0.1054     \\
						&         & 800  & 0.0479  & 0.0836     & 0.0748     &         & 800  & 0.0487  & 0.0668     & 0.0613     \\
						&         & 1600 & 0.0395  & 0.0656     & 0.0591     &         & 1600 & 0.0399  & 0.0531     & 0.0476     \\ \bottomrule
					\end{tabular}
				
				%		\begin{minipage}{\textwidth}
					%			\vspace{0.5em}
					%			\footnotesize \emph{Note:} We report the sample mean deviation over true value (Bias), the sample standard deviation (Std), and the root mean-squared error (RMSE) of estimators over 10,000 replications. The DGPs are Setting 11A to 12B. $N$ is the number of sample size. 
					%		\end{minipage}
			\end{table}
			%	\end{landscape}

%		In \Cref{table: simulation_ate_compare}, we show the finite sample performance of these three approaches, where we denote our estimator as SP, the locally efficient estimator as Efficient \citep[][]{Liu2018}, and the nonparametric estimator using cross-validation least squares to select bandwidth as NP \citep[][]{sun2021estimation}. First, all three estimators are nearly unbiased in all cases, while our estimator has relatively smaller bias when either boundedness of the link function or boundedness of function support does not hold. Second, our estimator has smaller standard deviation and RMSE than the estimators by \cite{Liu2018} and \cite{sun2021estimation} for all sample sizes we considered. Overall, our proposed method dominates the existing methods for the data with unbounded link function and unbounded support.
		Similarly, \Cref{table: simulation_ate_compare} presents the finite sample performance of the three approaches under the same conditions, where we denote our estimator as SP, the locally efficient estimator as Efficient \citep[][]{Liu2018}, and the nonparametric estimator using cross-validation least squares to select bandwidth as NP \citep[][]{sun2021estimation}. Our estimator consistently exhibits a smaller bias compared to the alternative methods when the boundedness assumptions are not met. Furthermore, our estimator demonstrates a lower standard deviation and RMSE than those of \cite{Liu2018} and \cite{sun2021estimation} across all sample sizes. In conclusion, our proposed method proves to be more advantageous than the existing methods for analyzing data with unbounded link functions and unbounded support.
		\medskip 
		
		\noindent{\bf Appendix D.4: Additional Simulation with Various Choices of Truncation Parameters}
		\medskip
%		
%		In this section, we propose the algorithm for choosing the truncation parameter for empirical application. The truncation error becomes negligible
%		asymptotically under the regularity conditions. For the choice of the truncation parameter, we propose the leave-one-out cross-validation as follows: 
%		
%		
%		
%		\vspace{0.5cm}
%		\noindent 
%		\textbf{Algorithm 2.}
%		Let $\mathcal{K}$ be
%		a set of candidate values of the truncation parameter.
%		\medskip
%		
%		(i) For each  $k \in \mathcal{K}$, we apply Algorithm 1 for the observations $\{(Y_{i},X_{i},D_{i})\}$ without using $j$-th observation and
%		obtain the leave-one-out estimate,
%		denoted by $\widehat{Y}_{k, -j}$. 
%		
%		(ii) The optimal truncation parameter
%		minimizes the prediction mean of squares: 
%		\begin{equation*}
%			\min_{k \in \mathcal{K}} \sum_{j=1}^{N}(\widehat{Y}_{k,-j}-Y_{j})^{2}.
%		\end{equation*}
		
		%We further demonstrate that the estimation results are insensitive to the truncation parameter $k$  in our extensive simulation exercises. See \Cref{section:addtional_simulation}. 

		We show simulation results for the finite sample performance of the estimation approach with various choices of truncation parameter $k$. \Cref{table: simulation_ps_ks_different_k,table: simulation_ps2_ks_different_k} show the simulation results of the propensity score estimation with the choices of truncation parameter $k$ ranging from $k = 2$ to $k = 6$. \Cref{table: simulation with ate_different_k,table: simulation with ate2_different_k} show the simulation results of the average therapeutic effect with the choices of truncation parameter $k$ ranging from $k = 2$ to $k = 6$.
		Overall, the estimators are valid against different choices of truncation parameters.

		{\small	
			% 				\begin{landscape}
				\begin{table}[!htbp]
					\centering
					\caption{\label{table: simulation_ps_ks_different_k}Simulation Results for the Estimators of Propensity Score Function with Different Truncation Parameters for Setting 1A and 1B.}
					\resizebox{\columnwidth}{!}{%
						\begin{tabular}{lcrrrrrrrrrrr}
							\toprule
							&         &      & \multicolumn{2}{c}{$k = 2$} & \multicolumn{2}{c}{$k = 3$} & \multicolumn{2}{c}{$k = 4$} & \multicolumn{2}{c}{$k = 5$} & \multicolumn{2}{c}{$k = 6$} \\ \midrule
							& Setting & $N$    & $\widehat{\theta}_{1}$     & $\widehat{\theta}_{2}$      & $\widehat{\theta}_{1}$     & $\widehat{\theta}_{2}$     & $\widehat{\theta}_{1}$     & $\widehat{\theta}_{2}$     & $\widehat{\theta}_{1}$     & $\widehat{\theta}_{2}$     & $\widehat{\theta}_{1}$     & $\widehat{\theta}_{2}$     \\ \hline
							Bias &     &      &         &        &         &         &         &         &         &         &         &         \\
							& 1A & 400  & 0.0125  & 0.0077 & -0.0121 & 0.0075  & -0.0127 & 0.0078  & -0.0131 & 0.0080  & 0.0134  & 0.0083  \\
							&     & 800  & 0.0044  & 0.0042 & -0.0043 & -0.0041 & -0.0045 & -0.0043 & -0.0046 & 0.0044  & 0.0048  & 0.0045  \\
							&     & 1600 & -0.0006 & 0.0006 & 0.0006  & 0.0006  & -0.0006 & 0.0006  & 0.0006  & -0.0006 & -0.0007 & -0.0007 \\
							& 1B & 400  & -0.0040 & 0.0062 & -0.0039 & -0.0060 & -0.0041 & 0.0062  & -0.0042 & -0.0064 & -0.0043 & -0.0066 \\
							&     & 800  & -0.0011 & 0.0020 & -0.0011 & 0.0019  & -0.0012 & 0.0020  & -0.0012 & 0.0020  & 0.0012  & 0.0021  \\
							&     & 1600 & 0.0003  & 0.0006 & 0.0003  & 0.0006  & 0.0003  & 0.0006  & 0.0003  & 0.0006  & 0.0003  & 0.0007  \\
							Std  &     &      &         &        &         &         &         &         &         &         &         &         \\
							& 1A & 400  & 0.0411  & 0.0553 & 0.0403  & 0.0542  & 0.0420  & 0.0564  & 0.0428  & 0.0575  & 0.0437  & 0.0587  \\
							&     & 800  & 0.0266  & 0.0353 & 0.0261  & 0.0347  & 0.0271  & 0.0361  & 0.0277  & 0.0368  & 0.0282  & 0.0375  \\
							&     & 1600 & 0.0223  & 0.0284 & 0.0228  & 0.0290  & 0.0219  & 0.0279  & 0.0232  & 0.0296  & 0.0237  & 0.0302  \\
							& 1B & 400  & 0.0254  & 0.0399 & 0.0249  & 0.0391  & 0.0259  & 0.0407  & 0.0265  & 0.0416  & 0.0270  & 0.0424  \\
							&     & 800  & 0.0158  & 0.0210 & 0.0154  & 0.0206  & 0.0161  & 0.0214  & 0.0164  & 0.0219  & 0.0167  & 0.0223  \\
							&     & 1600 & 0.0155  & 0.0191 & 0.0158  & 0.0195  & 0.0152  & 0.0187  & 0.0161  & 0.0199  & 0.0164  & 0.0203   \\ \bottomrule
						\end{tabular}
					}
					%					\begin{minipage}{\textwidth}
						%						\vspace{0.5em}
						%						\footnotesize We report the simulation results of the semiparametric estimator proposed in \Cref{subsec:ps} for the propensity score function, including the sample mean deviation over the true value (bias) and the sample standard deviation (std) over 10,000 replications. The data generating process is defined in model setting 1A and 1B. $N$ is the number of sample size. 
						%					\end{minipage}	
				\end{table}
				%			\end{landscape}
			
			%				\begin{landscape}
				\begin{table}[!htbp]
					\centering
					\caption{\label{table: simulation_ps2_ks_different_k}Simulation Results for the Estimators of Propensity Score Function with Different Truncation Parameters for Setting 2A and 2B.}
					\resizebox{\columnwidth}{!}{%
						\begin{tabular}{lccrrrrrrrrrr}
							\toprule
							&         &      & \multicolumn{2}{c}{$k = 2$} & \multicolumn{2}{c}{$k = 3$} & \multicolumn{2}{c}{$k = 4$} & \multicolumn{2}{c}{$k = 5$} & \multicolumn{2}{c}{$k = 6$} \\ \midrule
							& Setting & $N$    & $\widehat{\theta}_{1}$     & $\widehat{\theta}_{2}$      & $\widehat{\theta}_{1}$     & $\widehat{\theta}_{2}$     & $\widehat{\theta}_{1}$     & $\widehat{\theta}_{2}$     & $\widehat{\theta}_{1}$     & $\widehat{\theta}_{2}$     & $\widehat{\theta}_{1}$     & $\widehat{\theta}_{2}$     \\ \hline
							Bias &     &      &         &         &         &         &         &         &         &         &         &         \\
							& 2A & 400  & -0.0512 & 0.0418  & -0.0473 & 0.0387  & 0.0502  & -0.0410 & 0.0492  & 0.0402  & -0.0482 & -0.0394 \\
							&     & 800  & 0.0226  & -0.0223 & 0.0209  & 0.0206  & -0.0222 & 0.0219  & 0.0218  & -0.0215 & 0.0213  & -0.0211 \\
							&     & 1600 & -0.0086 & 0.0067  & -0.0079 & -0.0062 & 0.0084  & -0.0066 & -0.0082 & -0.0065 & -0.0081 & 0.0063  \\
							& 2B & 400  & 0.0481  & 0.0361  & 0.0445  & -0.0334 & -0.0472 & -0.0354 & -0.0463 & -0.0347 & -0.0453 & -0.0341 \\
							&     & 800  & 0.0233  & -0.0192 & 0.0215  & -0.0177 & 0.0228  & 0.0188  & 0.0224  & 0.0184  & -0.0220 & 0.0181  \\
							&     & 1600 & -0.0042 & -0.0042 & -0.0038 & 0.0039  & -0.0041 & 0.0041  & -0.0040 & 0.0040  & 0.0039  & 0.0040  \\
							Std  &     &      &         &         &         &         &         &         &         &         &         &         \\
							& 2A & 400  & 0.0182  & 0.0159  & 0.0164  & 0.0145  & 0.0177  & 0.0155  & 0.0173  & 0.0152  & 0.0317  & 0.0300  \\
							&     & 800  & 0.0085  & 0.0085  & 0.0078  & 0.0078  & 0.0083  & 0.0083  & 0.0081  & 0.0081  & 0.0179  & 0.0180  \\
							&     & 1600 & 0.0055  & 0.0057  & 0.0052  & 0.0054  & 0.0055  & 0.0057  & 0.0054  & 0.0056  & 0.0129  & 0.0135  \\
							& 2B & 400  & 0.0181  & 0.0153  & 0.0164  & 0.0141  & 0.0176  & 0.0150  & 0.0172  & 0.0147  & 0.0328  & 0.0308  \\
							&     & 800  & 0.0092  & 0.0090  & 0.0085  & 0.0084  & 0.0090  & 0.0089  & 0.0088  & 0.0087  & 0.0194  & 0.0199  \\
							&     & 1600 & 0.0064  & 0.0067  & 0.0060  & 0.0064  & 0.0063  & 0.0066  & 0.0062  & 0.0065  & 0.0153  & 0.0161 		\\\bottomrule
						\end{tabular}
					}
					%					\begin{minipage}{\textwidth}
						%						\vspace{0.5em}
						%						\footnotesize We report the simulation results of the semiparametric estimator proposed in \Cref{subsec:ps} for the propensity score function, including the sample mean deviation over the true value (bias) and the sample standard deviation (std) of the semiparametric estimator over 10,000 replications. The data generating process is defined in model setting 2A and 2B. $N$ is the number of sample size.
						%					\end{minipage}	
				\end{table}
				%			\end{landscape}

			\begin{table}
				\centering
				\caption{\label{table: simulation with ate_different_k}Simulation Results the Estimators of Average Treatment Effect with Truncation Parameters for Setting 5A to 5D.}
				%					\footnotesize
				\begin{tabular}{lccrrrrr}
					\toprule
					&         &      & $k = 2$   & $k = 3$   & $k = 4$  & $k = 5$   & $k = 6$   \\ \hline
					& Setting & $N$    & $\widehat{\beta}_{d}$ &  $\widehat{\beta}_{d}$       &   $\widehat{\beta}_{d}$      &       $\widehat{\beta}_{d}$  &  $\widehat{\beta}_{d}$       \\\hline
					Bias &     &      &         &         &         &         &         \\ 
					& 5A & 400  & 0.0052  & 0.0049  & 0.0050  & 0.0051  & 0.0051  \\
					&     & 800  & -0.0003 & 0.0003  & -0.0003 & -0.0003 & -0.0003 \\
					&     & 1600 & 0.0001  & -0.0001 & 0.0001  & 0.0001  & -0.0001 \\
					& 5B & 400  & 0.0065  & 0.0062  & 0.0063  & 0.0063  & 0.0064  \\
					&     & 800  & -0.0006 & 0.0005  & -0.0006 & 0.0006  & -0.0006 \\
					&     & 1600 & -0.0001 & -0.0001 & 0.0001  & -0.0001 & 0.0001  \\
					Std  &     &      &         &         &         &         &         \\
					& 5A & 400  & 0.0527  & 0.0487  & 0.0497  & 0.0507  & 0.0517  \\
					&     & 800  & 0.0365  & 0.0337  & 0.0344  & 0.0351  & 0.0358  \\
					&     & 1600 & 0.0284  & 0.0262  & 0.0268  & 0.0273  & 0.0278  \\
					& 5B & 400  & 0.0616  & 0.0569  & 0.0580  & 0.0592  & 0.0604  \\
					&     & 800  & 0.0445  & 0.0411  & 0.0420  & 0.0428  & 0.0436  \\
					&     & 1600 & 0.0327  & 0.0302  & 0.0308  & 0.0314  & 0.0320  \\
					Bias &     &      &         &         &         &         &         \\
					& 5C & 400  & 0.0056  & -0.0054 & -0.0054 & -0.0055 & -0.0056 \\
					&     & 800  & 0.0003  & -0.0002 & 0.0002  & -0.0002 & 0.0002  \\
					&     & 1600 & -0.0002 & 0.0002  & -0.0002 & 0.0002  & -0.0002 \\
					& 5D & 400  & 0.0029  & -0.0028 & 0.0028  & -0.0029 & 0.0029  \\
					&     & 800  & 0.0012  & 0.0011  & 0.0012  & 0.0012  & -0.0012 \\
					&     & 1600 & -0.0003 & 0.0003  & -0.0003 & -0.0003 & -0.0003 \\
					Std  &     &      &         &         &         &         &         \\
					& 5C & 400  & 0.0508  & 0.0470  & 0.0479  & 0.0489  & 0.0499  \\
					&     & 800  & 0.0350  & 0.0323  & 0.0330  & 0.0337  & 0.0343  \\
					&     & 1600 & 0.0227  & 0.0210  & 0.0214  & 0.0218  & 0.0222  \\
					& 5D & 400  & 0.0463  & 0.0427  & 0.0436  & 0.0445  & 0.0454  \\
					&     & 800  & 0.0307  & 0.0284  & 0.0289  & 0.0295  & 0.0301  \\
					&     & 1600 & 0.0210  & 0.0194  & 0.0198  & 0.0202  & 0.0206  \\ \bottomrule
				\end{tabular}\\
				%					\begin{minipage}{0.85\textwidth}
					%						\vspace{0.5em}
					%						\footnotesize We report the simulation results of the semiparametric estimator proposed in \Cref{subsec:ate} for the average treatment effects, including the sample mean deviation over the true value (bias) and the sample standard deviation (std) of the semiparametric estimator over 10,000 replications. The data generating process is defined in model setting 5A to 6B. $N$ is the number of sample size and $k$ is the truncation parameter. 
					%					\end{minipage}	
			\end{table}
			
			\begin{table}
				\centering
				\caption{\label{table: simulation with ate2_different_k}Simulation Results the Estimators of Average Treatment Effect with Truncation Parameters for Setting 7A to 8B.}
				%					\footnotesize
				\begin{tabular}{lccrrrrr}
					\toprule
					&         &      & $k = 2$   & $k = 3$   & $k = 4$  & $k = 5$   & $k = 6$   \\ \hline
					& Setting & $N$    & $\widehat{\beta}_{d}$ &  $\widehat{\beta}_{d}$       &   $\widehat{\beta}_{d}$      &       $\widehat{\beta}_{d}$  &  $\widehat{\beta}_{d}$       \\\hline
					Bias &     &      &         &         &         &         &         \\
					& 7A & 400  & -0.0058 & 0.0054  & 0.0055  & -0.0056 & 0.0057  \\
					&     & 800  & 0.0026  & -0.0024 & -0.0025 & 0.0025  & 0.0026  \\
					&     & 1600 & 0.0002  & -0.0002 & -0.0002 & -0.0002 & -0.0002 \\
					& 7B & 400  & 0.0030  & -0.0028 & 0.0029  & 0.0029  & 0.0030  \\
					&     & 800  & -0.0012 & 0.0011  & -0.0012 & 0.0012  & 0.0012  \\
					&     & 1600 & -0.0004 & 0.0003  & 0.0003  & -0.0003 & -0.0004 \\
					Std  &     &      &         &         &         &         &         \\
					& 7A & 400  & 0.0497  & 0.0470  & 0.0476  & 0.0483  & 0.0490  \\
					&     & 800  & 0.0342  & 0.0324  & 0.0328  & 0.0333  & 0.0337  \\
					&     & 1600 & 0.0222  & 0.0210  & 0.0212  & 0.0215  & 0.0218  \\
					& 7B & 400  & 0.0452  & 0.0427  & 0.0433  & 0.0439  & 0.0446  \\
					&     & 800  & 0.0300  & 0.0284  & 0.0287  & 0.0292  & 0.0296  \\
					&     & 1600 & 0.0205  & 0.0194  & 0.0197  & 0.0200  & 0.0203  \\
					Bias &     &      &         &         &         &         &         \\
					& 8A & 400  & 0.0094  & -0.0086 & 0.0088  & -0.0090 & 0.0092  \\
					&     & 800  & -0.0019 & 0.0017  & -0.0018 & -0.0018 & 0.0019  \\
					&     & 1600 & -0.0008 & 0.0007  & -0.0008 & 0.0008  & -0.0008 \\
					& 8B & 400  & -0.0165 & -0.0152 & 0.0155  & 0.0159  & -0.0162 \\
					&     & 800  & 0.0092  & -0.0085 & -0.0087 & -0.0089 & 0.0091  \\
					&     & 1600 & 0.0008  & -0.0007 & -0.0008 & -0.0008 & -0.0008 \\
					Std  &     &      &         &         &         &         &         \\
					& 8A & 400  & 0.0549  & 0.0519  & 0.0526  & 0.0534  & 0.0541  \\
					&     & 800  & 0.0317  & 0.0299  & 0.0304  & 0.0308  & 0.0312  \\
					&     & 1600 & 0.0211  & 0.0200  & 0.0203  & 0.0206  & 0.0209  \\
					& 8B & 400  & 0.0606  & 0.0573  & 0.0581  & 0.0589  & 0.0598  \\
					&     & 800  & 0.0394  & 0.0372  & 0.0377  & 0.0383  & 0.0388  \\
					&     & 1600 & 0.0211  & 0.0200  & 0.0203  & 0.0206  & 0.0209 \\\bottomrule
				\end{tabular}\\
			\end{table}
		}

		\clearpage
		\setcounter{equation}{0}
		\setcounter{table}{0}
		\setcounter{figure}{0}
		\renewcommand{\theequation}{E.\arabic{equation}}
		\renewcommand{\thesubsection}{E.\arabic{subsection}}
		\renewcommand{\thefigure}{E.\arabic{figure}}
		\renewcommand{\thetable}{E.\arabic{table}}
		\renewcommand{\thelemma}{E.\arabi{lemma}}
		\renewcommand{\theremark}{E.\arabic{remark}}
		\renewcommand{\thecorollary}{E.\arabic{corollary}}
		\renewcommand{\theassumption}{E.\arabic{assumption}}
		\renewcommand{\theproposition}{E.\arabic{proposition}}

		\noindent{\bf \large Appendix E: Additional Empirical Results}\label{section:addtional_empirical}
		\bigskip
		
		We apply our semiparametric methods to study the effect of job training on future earnings. Our dataset is based on the classic National Supported Work (NSW) demonstration dataset, as analyzed by \cite{lalonde1986evaluating} and reconstructed by  \cite{dehejia1999a}. 
		The NSW experiment took place from March 1975 until June 1977 and randomly assigned participants to the treatment group who received a guaranteed job for 9 to 18 months and frequent counselor meetings or control groups who were left in the labor market by themselves.

		The outcome variable $Y$ is the participant's earnings in 1978. The binary treatment variable $D$ is the job training status ($D = 1$ indicates participant is assigned training, $D = 0$ indicates participant is not assigned training). The covariates $X$ include the education, ethnicity, age, and employment variables before treatment, including earnings in 1974 and 1975.  Following the \cite{imbens2015a}, we augment this dataset with observations from the Current Population Survey (CPS). We focus on estimating the average treatment effect of the program for participants using the comparison group from the CPS. Due to data availability for the randomly assigned control group, we can assess whether the non-experimental estimators are accurate. We provide the summary statistics for these two datasets in \Cref{table: summary_job} and see much larger differences between the two groups, suggesting that it is important to carefully adjust for these differences in estimating the average treatment effects.

		{\small
			\begin{table}[!htbp]
				\centering
				\caption{\label{table: summary_job}Summary Statistics for Job Training data.}
				\begin{tabular}{lcccccc}
					\toprule
					& \multicolumn{2}{c}{Treatment Group} & \multicolumn{2}{c}{\begin{tabular}[c]{@{}c@{}}Experimental \\ Control Group\end{tabular}} & \multicolumn{2}{c}{\begin{tabular}[c]{@{}c@{}}Nonexperimental\\ Control Group\end{tabular}} \\ \cline{2-7}
					& \multicolumn{2}{c}{$N = 185$} & \multicolumn{2}{c}{$N = 260$}                                                       & \multicolumn{2}{c}{$N = 15992$}                                                       \\ \cline{2-7}
					& Mean             & Std              & Mean                                        & Std                                         & Mean                                          & Std                                         \\ \hline
					Age             & 25.82            & 7.16             & 25.05                                       & 7.06                                        & 33.23                                         & 11.05                                       \\
					Education       & 10.35            & 2.01             & 10.09                                       & 1.61                                        & 12.03                                         & 2.87                                        \\
					Black           & 0.84             & 0.36             & 0.83                                        & 0.38                                        & 0.07                                          & 0.26                                        \\
					Hispanic        & 0.06             & 0.24             & 0.11                                        & 0.31                                        & 0.07                                          & 0.26                                        \\
					Married         & 0.19             & 0.39             & 0.15                                        & 0.36                                        & 0.71                                          & 0.45                                        \\
					Nodegree        & 0.71             & 0.46             & 0.83                                        & 0.37                                        & 0.3                                           & 0.46                                        \\
					Earnings (1974) & 2095.57          & 4886.62          & 2107.03                                     & 5687.91                                     & 14016.8                                       & 9569.8                                      \\
					Earnings (1975) & 1532.06          & 3219.25          & 1266.91                                     & 3102.98                                     & 13650.8                                       & 9270.4                                      \\
					Earnings (1978) & 6349.14          & 7867.4           & 4554.8                                      & 5483.84                                     & 14846.66                                      & 9647.39                                     \\ \bottomrule
				\end{tabular}
			\end{table}
			
		}

		As the dataset has 7,657 observations in the control group with propensity scores less than 1.00e-05, the propensity score-related estimators may not be readily applied in this case. We compare the results of average treatment effect estimation based on bias-corrected matching (MBC) estimator \citep[][]{abadie2006a,abadie2011a} and regression adjustment (RA) estimator \citep[][]{lane1982analysis}. We summarize the estimation results for the average treatment effect of job training on earnings in 1978 in \Cref{table: ate_job} along with the mean and standard deviation. We first show the average difference between treatment group and experiment control group is 1794.34 and statistically significant. The RA estimator suggests that the job training increases the further earning by 3680.60 but is statistically insignificant. The MBC estimator shows that the average effect of 6293.11 is statistically significant but quite different from the experimental effect of 1794.34. Our ATE estimator is 4778.24 while variance-weighted ATE estimator is 2914.72. Both estimators are statistical significant, and the estimate of variance-weighted ATE is closest to the difference between treatment and experiment control group among all estimators we considered, suggesting that the our estimators may provide relatively accurate estimation and inference for the dataset with limited overlap. 
		
		\begin{table} 
			\centering
			\caption{\label{table: ate_job}Average Treatment Effect Estimation for the Job Training Data.}
			\resizebox{\columnwidth}{!}{%
				\begin{tabular}{lccccccc}
					\toprule
					& ATE    & Std     & Z-statistics & $p$-value & \multicolumn{2}{c}{95\% CI} \\ \midrule
					Difference    & 1794.34 & 671.00  & 2.67         & 0.01    & 474.01        & 3114.67     \\ \hline
					%			Bootstrap Difference & 1794.34 & 676.16  & 2.65         & 0.01    & 469.10        & 3119.59     \\ 
					MBC                  & 6293.11 & 1532.69 & 4.11         & 0.00    & 328.00        & 9297.12     \\
					RA                   & 3680.60 & 2622.33 & 1.40         & 0.16    & -1459.07      & 8820.27     \\
					Our Estimator ($\Delta^{ATE}$)                & 4778.24 & 631.95  & 7.56         & 0.00    & 3539.62       & 6016.86     \\ 
					Our Estimator ($\Delta_{\omega}^{ATE}$)               & 2914.72 & 391.81  & 7.44         & 0.00    & 2146.77       & 3682.67     \\ \bottomrule
				\end{tabular}
			}
		\end{table}
		
		{\small
			\begin{figure} [h!]
				\begin{center}
					\caption{\label{figure: link_job}Estimated nonparametric link function $g(\omega)$ of the propensity score for the job training data.}
					{\includegraphics[width=0.60\textwidth]{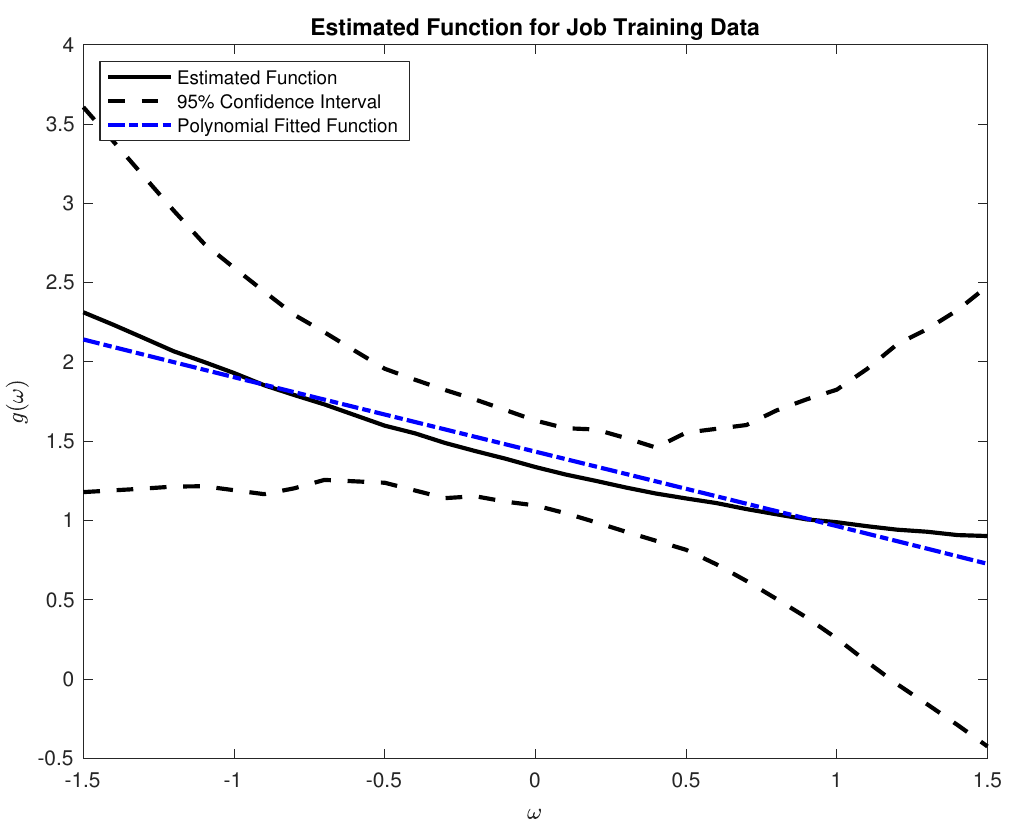}}
				\end{center}
			\end{figure}
		}
		
		We plot the estimated link function $\hat{g}$ of propensity score using the wild bootstrap simulation with 500 bootstrap repetitions in \Cref{figure: link_job}. The estimated link function $\hat{g}$ is shown in a solid line, the corresponding 95\% confidence interval is drawn in dashed lines. We calculate the optimal truncation parameter $k$ based on the minimization of  the prediction mean of squares. In this data example, the optimal truncation parameter $k = 2$. We estimate the approximated second-order polynomial fitted function using the ordinary least squares estimation approach as a reference as following (with standard errors in bracket):
		\begin{equation*}
			\widehat{g}_{k}(\omega_{i})=\underset{(0.098)}{1.34}
			-\underset{(0.072)}{0.47}\omega_{i}
			+
			\underset{(0.008)}{0.12}\omega_{i}^2,
		\end{equation*}
		where $\omega_{i}$ is the $x_{i}'\hat{\theta}$ for $1\leq i \leq N$.
		
		For reference, we also plot the approximated second-order polynomial fitted function using the ordinary least squares estimation approach. As shown, the estimated link functions $\hat{g}$ in the data sample of job training program are quite far away from the identity function, meaning that the logistic parametric assumption for propensity score may not be valid. One reason is that in the job training program data sample, there are 185 observations in treated group, 15,992 observations in control group, and 7,657 observations in the control group with propensity scores less than 1.00e-05. Therefore, most of the observations have the zero value for propensity score, making the regular parametric assumptions not feasible in this particular data sample. The estimated link function is very close to the approximated second-order polynomial fitted function, showing our model and method work well in this data sample.
		
}
		
\end{document}